\documentclass[aps, pre, preprint, longbibliography,floatfix, nofootinbib, superscriptaddress]{revtex4-1}
\usepackage{amsmath,amsfonts}
\usepackage{graphicx}
\usepackage{color}
\usepackage{subfigure}
\usepackage{tikz}
\usepackage{tikz-3dplot}
\usetikzlibrary{calc,quotes,angles}
\usepackage[colorlinks=true, citecolor=red, linkcolor=blue,urlcolor=blue ]{hyperref}
\usepackage{natbib}
 \def\di{\displaystyle}

\DeclareMathOperator\erf{erf}

\begin{document}
\title{Dynamics of a helical swimmer crossing viscosity gradients}
 \date{\today}
 \author{Christian \surname{Esparza L\'opez}}
 \affiliation{Department of Applied Mathematics and Theoretical Physics, University of Cambridge, Cambridge CB3 0WA, United Kingdom}
 \author{Jorge \surname{Gonzalez-Gutierrez}}
 \affiliation{Facultad de Ciencias en F\'isica y Matem\'aticas Universidad Aut\'onoma de Chiapas, Tuxtla Guti\'errez, Chiapas, M\'exico}
 \affiliation{Departamento de Termofluidos, Facultad de Ingenier\'ia, Universidad Nacional Autanoma de M\'exico, Av. Universidad 3000, Ciudad de M\'exico, 04510, M\'exico}
 \author{Francisco \surname{Solorio-Ordaz}}
 \affiliation{Departamento de Termofluidos, Facultad de Ingenier\'ia, Universidad Nacional Autanoma de M\'exico, Av. Universidad 3000, Ciudad de M\'exico, 04510, M\'exico}
 \author{Eric \surname{Lauga}}
	\email{e.lauga@damtp.cam.ac.uk}
 \affiliation{Department of Applied Mathematics and Theoretical Physics, University of Cambridge, Cambridge CB3 0WA, United Kingdom}
 \author{Roberto \surname{Zenit}}
\email{zenit@brown.edu}
 \affiliation{Instituto de Investigaciones en Materiales, Facultad de Ingenier\'ia, Universidad Nacional Autónoma de M\'exico, Av. Universidad 3000, Ciudad de M\'exico, 04510, M\'exico}
 \affiliation{School of Engineering, Brown University, 184 Hope St, Providence, RI 02912, USA}

\begin{abstract}
 We  experimentally and theoretically study the dynamics of a low-Reynolds number helical swimmer moving across viscosity gradients. Experimentally, a double-layer viscosity is generated by superposing two miscible fluids with similar densities but different dynamic viscosities. A synthetic helical magnetically-driven swimmer is then made to move across the viscosity gradients along four different configurations: either head-first (pusher swimmer) or tail-first (puller), and through either positive (i.e.~going from low to high viscosity) or negative viscosity gradients. We observe qualitative differences in the penetration dynamics for each case.  We find that the swimming speed can either increase or decrease while swimming across the viscosity interface, which results from the fact that the head and the tail of the swimmer can be in environments in which the local viscosity leads to different relative amounts of drag and thrust.  In order to rationalize the experimental measurements, we next develop a theoretical hydrodynamic model. We assume that the classical resistive-force theory of slender filaments is locally valid along the helical propeller and use it to calculate the swimming speed as a function of the position of the swimmer relative to the fluid-fluid interface. The predictions of the model agree well with experiments for the case of positive viscosity gradients.  When crossing across a negative  gradient, gravitational forces in the experiment become important, and  we modify the model  to include buoyancy, which agrees with experiments. In general our results show that it is harder for a pusher swimmer to cross from low to high viscosity, whereas for a puller swimmer it is the opposite. Our model is also extended to the case of a swimmer crossing a continuous viscosity gradient.
\end{abstract}

\maketitle
 
 \section{Introduction}
 
Taxis is the capability of biological cells to respond to an external stimulus, such a light or chemical gradients, and as a result move towards or away from it~\cite{dusenbery2009living}. In nature, the adaptability of microorganisms  to respond to a variety of cues has been demonstrated in gradients of light intensity (phototaxis)~\cite{jekely2008mechanism, bennett2015steering, giometto2015generalized, lozano2016phototaxis, dai2016programmable}, magnetic fields (magnetotaxis)~\cite{klumpp2016magnetotactic, rupprecht2016velocity, waisbord2016destabilization}, temperature (thermotaxis)~\cite{bahat2003thermotaxis, li2013serotonin, bickel2014polarization}, a gravitational potential (gravitaxis)~\cite{ten2014gravitaxis, campbell2013gravitaxis, campbell2017helical} and chemical stimuli (chemotaxis)~\cite{eisenbach2004chemotaxis}.

For many motile microorganisms, chemotaxis is a crucial method to escape from toxins (chemo-repulsion or negative chemotaxis) and to find sources of food (chemo-attraction or positive chemotaxis). Two illustrative examples are the well-studied  bacterium \emph{E.~coli}~\cite{berg2008coli}, whose study is at the heart of most of what we know about bacterial sensing and  information processing, and  spermatozoa looking for the ovum during fertilization~\cite{natr2003murray}. Beyond individual behaviour, microorganisms may also exhibit  collective dynamics through chemically-based communication. For example, when a \emph{Dictyostelium} cell (a type of mold) starves, it produces a chemical that induces a multicellular aggregation process, which  allows the cells to survive long starvation periods~\cite{eisenbach2004chemotaxis, bonner1947evidence, gerisch1968cell, eidi2017modelling}. The mechanism behind this phenomenon is captured in the classical Keller-Segel model~\cite{keller1970initiation, keller1971model}, and has been extended to describe some collective phenomena of \emph{E.~coli} bacteria showing chemo-attraction to self-produced autoinducers~\cite{ laganenka2016chemotaxis}.
 
A mechanical example of taxis, viscotaxis, emerges when a cell adapts its motion in response to viscosity gradients. Some microorganisms, such as \emph{Spiroplasma}~\cite{daniels1980aspects} and \emph{Leptospira interrogans}~\cite{kaiser1975enhanced, petrino1978viscotaxis, takabe2017viscosity},  have indeed the ability to respond to changes in viscosity.  
A particularly important example for human health is the colonization of the   stomach by the bacterium \emph{Helicobacter pylori}, which turns out to be another consequence of the  ability to move in viscosity gradients~\cite{illien2017fuelled, romanczuk2012active}.  Indeed, \emph{H.~pylori} is the only known bacteria to be capable of penetrating the intestinal mucus layer and reach the stomach wall~\cite{illien2017fuelled, romanczuk2012active}, thanks to an enzymatic degradation of the stomach mucosa~\cite{elgeti2015physics, Mirbagheri2016}. This leads to severe inflammation that can result in ulcerogenesis or neoplasia, and since the bacterium  infects about 50\% of the human population  it  is important to understand its pathogenesis~\cite{ottemann2002helicobacter}.  

 In this paper we focus on the mechanics of artificial bacteria in model systems displaying gradients in viscosity. In nature, the motion of helicoidal bacteria through a liquid environment is subject to a number of additional  physicochemical processes, including  screened electrostatics, the interactions with diffusing chemicals and biochemical noise~\cite{lauga2016bacterial}. From the point of view of continuum fluid mechanics, the dynamics of flagellated bacteria always takes place in the Stokesian regime  since the typical Reynolds numbers  range from $10^{-4}$ to $10^{-2}$. The hydrodynamics associated with the movement of such microorganisms is therefore dictated by the predominance of viscous forces and the absence of inertia. 

Some understanding already exists on the impact of viscosity gradients  on the dynamics of both passive and active (swimming) particles. For example, through cross-streamline migration in viscosity gradients, it is possible to sort soft passive particles in microflows~\cite{laumann2019focusing}. Heated particles create temperature gradients, which  induce local variations in viscosity in the surroundings of the particle~\cite{oppenheimer2016motion}. For simple swimmers composed of a small number of active spheres, viscotaxis has been recently shown to arise from a mismatch  in the viscous forces acting on the different parts of the swimmer, allowing both positive and negative viscotaxis in Newtonian fluids~\cite{liebchen2018viscotaxis}. Although that mechanism does not account for the possible existence of biological viscoreceptors~\cite{sherman1982viscosity}, the  positive viscotaxis in \emph{Spiroplasma}~\cite{daniels1980aspects} and \emph{Leptospira}~\cite{kaiser1975enhanced, petrino1978viscotaxis, takabe2017viscosity} can be explained in these terms. 

Using the classical squirmer model microswimmer~\cite{ lighthill1952squirming, blake1971spherical, pedley2016spherical}, work coupling the concentration of nutrients to the viscosity of the fluid showed  qualitative differences in the dynamics of swimming, in contrast to fluids with constant viscosity~\cite{eastham2019axisymmetric}. The squirmer model has also allowed to study theoretically the effect of  weak viscosity gradients on the motion of  general spherical swimmers, showing in particular how  the swimmer `mode' (i.e.~whether the swimmer is a pusher or a puller) is critical in setting the sign of the viscotaxis response~\cite{datt2019active}. However, and despite a  good  understanding of  locomotion of bacteria  in Newtonian fluids~\cite{laugabook}, a theory that explains how viscosity gradients affect the swimming of helical swimmers is currently not available.

Synthetic microswimmers have often been proposed as one modelling approach to study the motility of  microorganisms. Self-phoretic Janus colloids, for example, can be made to move through the generation of chemical, electrostatic or thermal  gradients~\cite{bechinger2016active}.  These systems have been shown to display similarity with biological chemotaxis, and the Keller-Segel equations for both, Janus colloids and chemotactic microorganisms, are similar~\cite{tindall2008overview,meyer2014active, saha2014clusters, pohl2015self, liebchen2015clustering}.  Chemotaxis also plays a significant role in the formation of dynamic clusters and patterns, synthetic colloid microswimmers suspensions~\cite{romanczuk2012active,pohl2014dynamic, saha2014clusters, elgeti2015physics,liebchen2015clustering, liebchen2016pattern,bechinger2016active,liebchen2017phoretic}.

Artificial helicoidal swimmers typically consist of a rigid magnetic head fixed to a metallic helical tail~\cite{zhang2009artificial} in which the whole body is  made to rotate by an external magnetic field. Propulsion arises as a result of the chirality of the helical tail, in close analogy to the  swimming of flagellated  bacteria, e.g.~\emph{E.~coli}~\cite{berg2008coli}. These types of synthetic swimmers preserve the basic physical characteristics that allow locomotion in low\textendash$\rm Re$ environments, namely the coupling  between rotation and translation for a  helical slender  filament  (here the  tail). Under this framework, many control parameters can be  explored experimentally   to quantify  the swimming motion in complex environments~\cite{gomez16, godinez2015}. 
 
Recently, inspired by the process through which \emph{H.~pylory} crosses the intestinal mucus layer, we conducted an experimental study on the  dynamics of helical swimmers moving through the interface between two immiscible fluids~\cite{gonzalez2019dynamics}. Depending on the orientation of the swimmer and the different stages of penetration (in particular whether the head or the tail reaches first the interface), the interface was shown to dramatically affect the swimmer. However, interfacial tension is not believed to play a significant role in the mucus zone, where instead high viscosity gradients are dominant. In this paper we therefore consider the case where the  helical swimmer crosses a viscosity interface. 

We construct  experimentally a stratified solution of two miscible fluids with different viscosities and study the motion of the artificial helical swimmer as it crosses the interface between the fluids. We show that the swimmer slows down as it crosses from a region of low to high viscosity head-first (i.e.~in the pusher mode) but that it speeds up when it approaches the interface with its tail forward (puller mode). In contrast, the swimmer always slows down when it moves down the gradient, regardless of its orientation. We then develop a hydrodynamic model to explain our observations. Inspired by a previous study on viscotaxis~\cite{liebchen2018viscotaxis}, we assume that the standard Newtonian Stokes drag laws are locally valid, and that the swimming behaviour is determined by an instantaneous balance between viscous propulsion and drag. For motion up the viscosity gradient, our model predicts a decrease (resp.~increase) in the swimming speed for pusher head-forward (resp.~puller tail-forward) orientation, which is consistent with the experimental observations. However, due to the reversibility of Stokes flows, our model would predict the opposite behaviour when the swimmer moves down the gradient, in contrast with   the experiments. Further analysis of our experiments reveal that  when the swimmer moves down the gradient it entrains a portion of the high-viscosity fluid into the low-viscosity region, regardless of its orientation. This drift volume increases the apparent density of the swimmer, thereby slowing it down due to gravitational forces. Including a buoyancy term in our model to account for this effect allow the theoretical  predictions  to come closer to the experimental observations.

The paper is organised as follows. In Section~\ref{sec:setup} we describe the synthetic swimmer, its characteristic geometrical parameters and the experimental setup. The experimental results for all   four   configurations are presented in Section~\ref{sec:results}, with a  focus on the swimming speed as a function of the swimmer position relative to the fluid interface. The mathematical model for a sharp viscosity gradient is developed in Section~\ref{sec:discrete} and its extension for a continuous viscosity profile is presented in Section~\ref{sec:cont_grad}. We next compare our  model with the experimental results in Section~\ref{sec:experiments}; a modified model that takes into account the fluid entrainment is discussed at the end of this section. Finally we discuss our results in Section~\ref{sec:conclusions}.
 
  \section{\label{sec:setup} Experimental setup and materials}
 
\begin{figure}
  \centering
  \subfigure[]{\includegraphics[height=10cm]{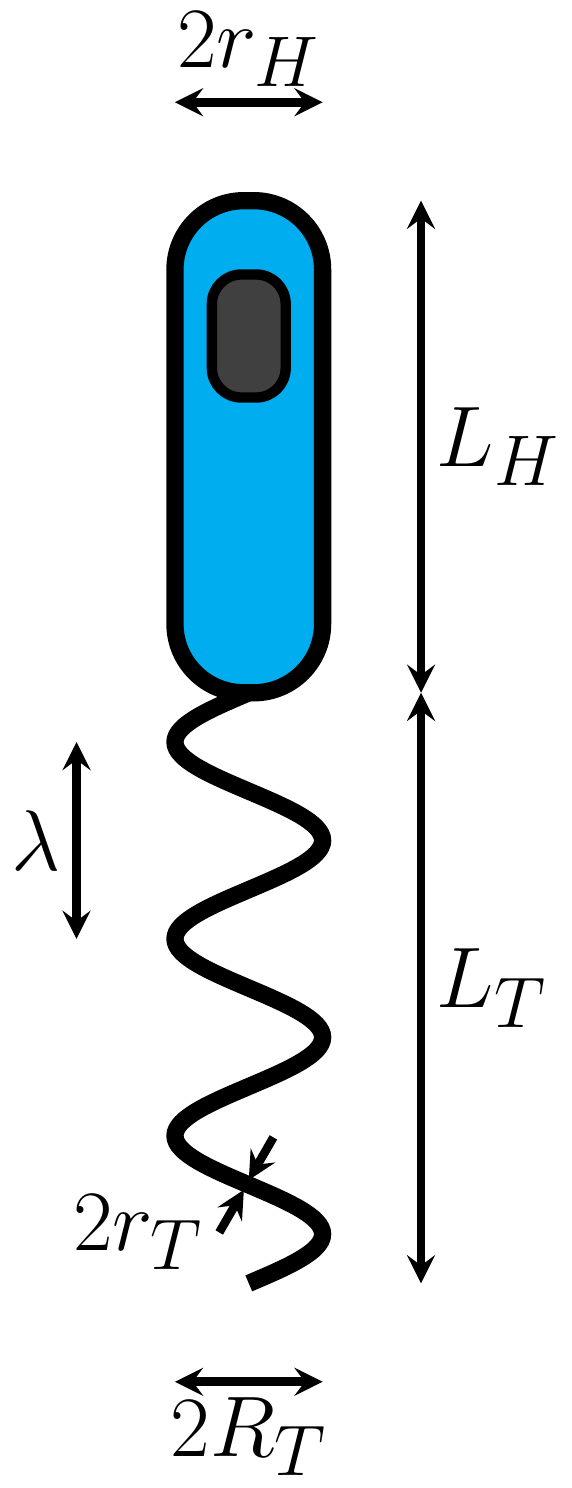}}\quad\quad 
  \subfigure[]{\includegraphics[height=10cm]{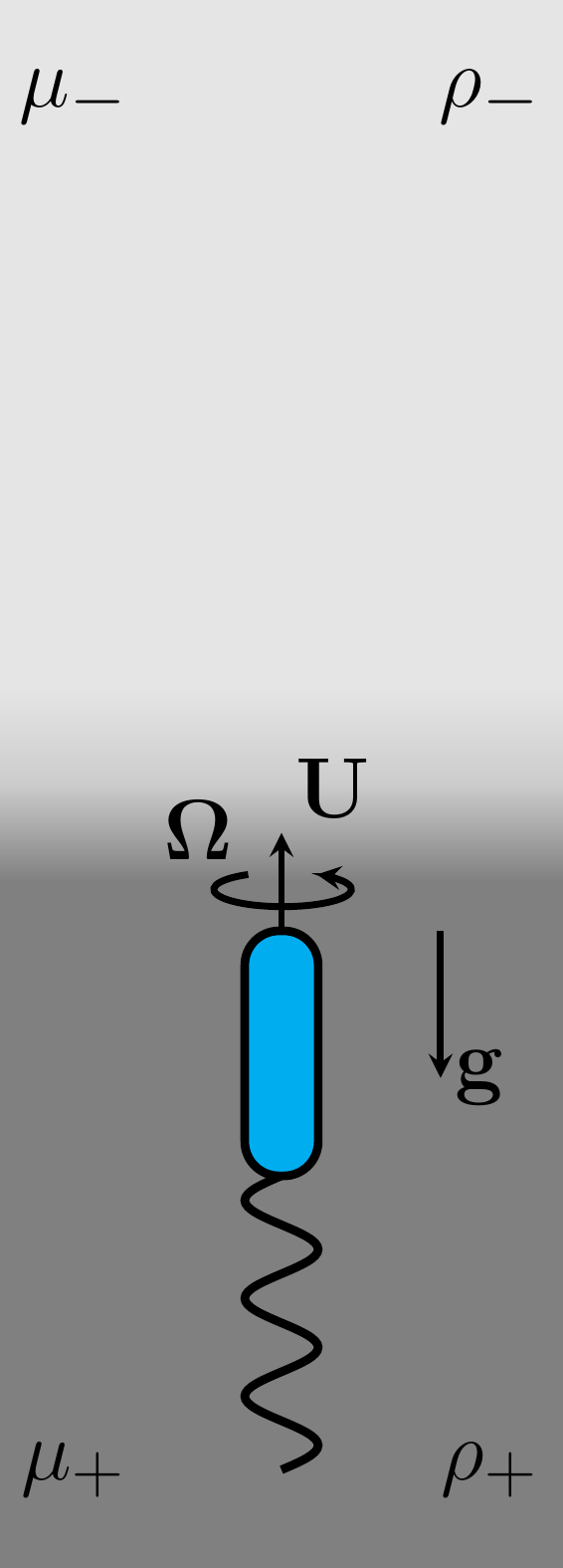}} 
    \caption{\label{fig:setup} Experimental setup. (a) Schematic representation of the helical swimmer. The dimensions of the device are: $2r_H= 4.5$~mm; $L_H=16$~mm; $L_T=16$~mm; $\lambda=5.3$~mm; $2R_T=4.5$~mm; and the pitch angle $\psi=45^{\circ}$. The thickness of the wire is $2r_T=0.9$~mm. (b) In this example, the swimmer moves head-first from a high-to-low viscosity fluid through a sharp gradient and here gravity is pointing downwards. Note that in all figures the dark and light gray denote high and low viscosity regions, respectively.}
 \end{figure}

To investigate the mechanics  of a synthetic swimmer crossing a layer of variable viscosity, we use the helical swimmer previously developed  by our group to study swimming in complex media~\cite{godinez2015, gomez16, gonzalez2019dynamics}. The helical swimmer, shown schematically in Fig.~\ref{fig:setup}(a), consists of a  cylindrical head and a right-handed helical tail, both of which are rigid; pictures of the swimmer are shown in Section \ref{sec:results}. 
The head of the swimmer contains a small magnet; since the entire setup is exposed to an external  magnetic field rotating below the step out frequency, the swimmer rotates at an imposed rate. Details on the setup can be found in Ref.~\cite{godinez2012}. With this setup, the speed of the swimmer  can be controlled by changing the rotation rate of the external magnetic field, and the   swimmer remains force-free throughout. In all experiments reported here, the swimmer moves  vertically in either the upwards or downwards direction. Furthermore, the swimming direction (head or tail first) can also be changed:  since the helical tail is chiral, reversing the rotation direction of the tail (by changing the rotation direction of the magnetic field) leads to the swimmer moving while either pushing or pulling the head.
 
 A viscosity gradient environment is produced by slowly superposing two miscible viscous liquids onto each other. They are placed, in sequence, in a transparent tank initially leading to a two-layer sharp viscosity gradient, as shown in Fig.~\ref{fig:setup}(b).  The bottom liquid is prepared by mixing glucose (530~ml) and water (100~ml), to have a viscosity of approximately \mbox{$\mu_+=2.74$~Pa.s}, at room temperature.  To ensure that the interface remains horizontal, a small amount of salt is added to this liquid (30~g. of NaCl) to increase its density slightly, $\rho_+=1367.4$~kg/m$^3$.
Note that the slight density stratification helps to maintain the layer stable to conduct  several experiments before replacing the fluids. Several combinations of the viscosity gradient are tested but we only report on one case. The viscosity and density of the top fluid are $\mu_-=0.55$~Pa.s and $\rho_-=1309.7$~kg/m$^3$, respectively. The fluid viscosities are measured with a viscometer (DV-III, Brokefield). The densities of the liquids are measured with a 25~ml pycnometer. 

The container, with dimensions 8.9$\times$8.9$\times$18~cm$^3$, with the swimmer inside is placed within the rotating Helmholtz coil, as in previous experiments~\cite{godinez2015, gomez16, gonzalez2019dynamics}. To reduce the crystallization of the glucose solutions at the free surface, the container is kept closed at all times. As explained above, the system is slightly density-stratified. Therefore, the swimmer cannot be neutrally buoyant in both top and bottom fluids. The density of the swimmer is  adjusted to make it as close as possible to that of the light fluid: $\rho_\text{swimmer}\approx 1270$~kg/m$^3$. Hence, the swimmer is slightly buoyant for both fluids. 

All experiments are conducted at a fixed rotation rate of the swimmer, $ {\Omega}/{2\pi}=2.92$~Hz and the swimmer moves at a constant terminal swimming speed, \mbox{$U_0\simeq 1.5-3.5$~mm/s} in one of the fluids. Due to the slight density mismatch the terminal speed is different for each fluid and for each direction of motion. The maximum Reynolds number is \mbox{$Re=0.035$}, using $\mu_-$, $r_H$ and the maximum swimming speed \mbox{$U_0=3.3$~mm/s}, as the characteristic viscosity, length and speed.

\subsection{\label{subsec:diffuse} Evolution of the viscosity gradient in time}

\begin{figure}
 \subfigure[]{\includegraphics[width=.25\linewidth]{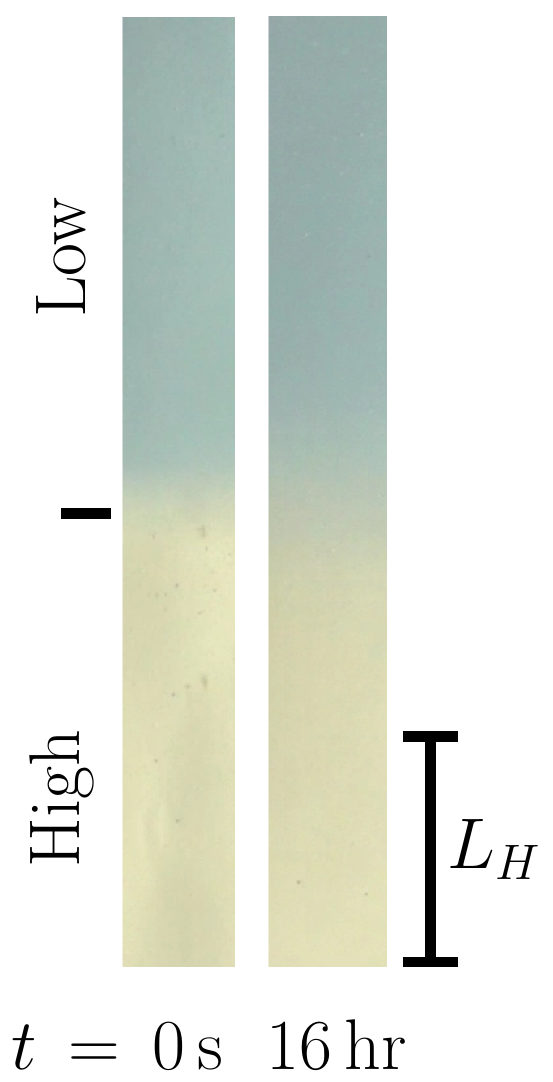}}\,\,\, 
 \subfigure[]{\includegraphics[width=.65\linewidth]{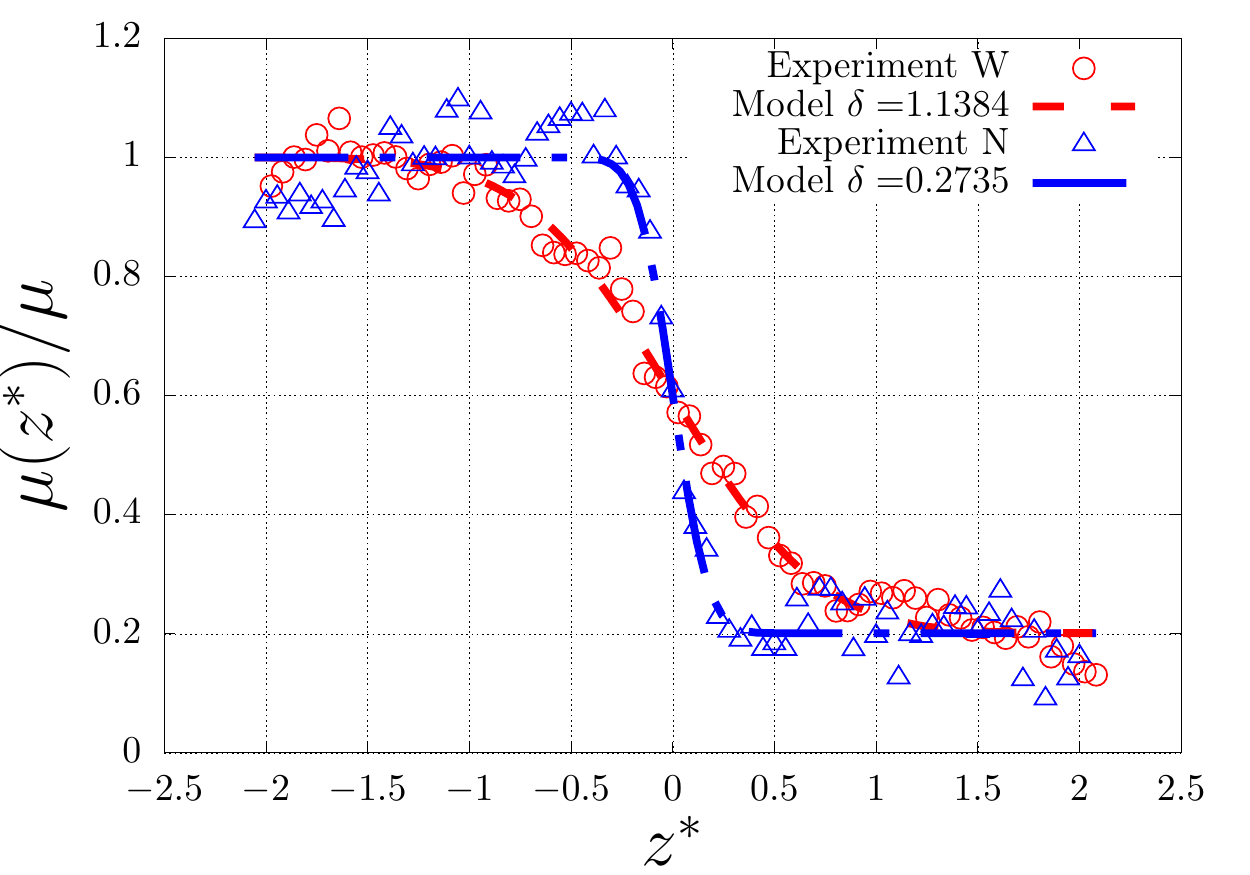}}
     \caption{\label{fig:setup2}  (a) The viscosity gradients at $t=0$ and 16 hours. (b) Normalized viscosity, $\mu(z)/\mu$, as a function of normalized distance from the interface, $z^*=z/L_H$. Data points represent the pixel intensity of the viscosity profile soon after the initial setup ($\delta=\Delta/L_H=0.2735$) and sixteen hours afterwards ($\delta=1.1384$). Solid line represents the best fit of Eq.~\eqref{eq:diff_2} replacing $\mu$ and $\mu'$ with the maximum and minimum intensities respectively.}
 \end{figure}

If left undisturbed, the two-fluid layer slowly mix, leading to a diffused non-sharp viscosity gradient. By conducting experiments at different times after the two-layer fluid is first prepared, we are  able to test the influence of the strength of the viscosity gradient on the swimming process. To quantify the thickness of the viscosity gradient, we apply the following procedure. Since the low viscosity fluid is dyed, it is possible to track the evolution of the gradient in time, using standard image processing. An example of how the gradient evolves from sharp to diffuse is shown in Fig.~\ref{fig:setup2}(a). By assigning a pixel-intensity value to a value of viscosity, it is then possible to measure how the viscosity evolves in space and time, as shown in Fig.~\ref{fig:setup2}(b). 
 
To obtain a quantitative assessment on how the viscosity gradient evolves in time, we can compare the experimental measurements with a classical diffusive model. We  assume that the viscosity of the fluid mixture is a Lagrangian function of some diffusive tracer with concentration $C(z,t)$, i.e.~\mbox{$\mu(z,t)=\mu[C(z,t)]$}. If $D$ is the diffusivity of the tracer $C$, then with a good approximation  the viscosity distribution $\mu(z,t)$ evolves according to the diffusion equation
 \begin{equation}
  \label{eq:diff_1}
  \frac{\partial\mu}{\partial t}=D\frac{\partial^2 \mu}{\partial z^2} .
 \end{equation}
 Using the Green's function method and the initial distribution \mbox{$\mu(z,0)=\mu'+\theta{(z)}(\mu-\mu')$}, where $\theta(z)$ is the Heaviside step function, we obtain the viscosity distribution as
 \begin{equation}
  \label{eq:diff_2}
  \mu(z,t)=\frac{\mu+\mu'}{2}+\frac{\mu-\mu'}{2}\erf{\left(\frac{z}{\sqrt{\Delta^2/2}}\right)},
 \end{equation}
where $\erf(x)$ is the error function and $\Delta=2\sqrt{2Dt}$ is the width of the transition region. We can then integrate this viscosity distribution, the result is
 \begin{equation}
  \label{eq:diff_3}
  \int_{h_1}^{h_2}{\frac{\mu(z)}{\mu'}\,\text{d}z}=\left[\frac{\mu+\mu'}{2\mu'}z+\frac{\mu-\mu'}{2\mu'}\left(z\erf{\left(\frac{z}{\sqrt{4Dt}}\right)+\sqrt{\frac{4Dt}{\pi}}e^{-z^2/4Dt}}\right)\right]_{h_1}^{h_2}.
 \end{equation}
 We may also assume that the mass density  fluid also evolves according to a diffusion equation but with a diffusivity $D'$ which may be different from $D$. Using the initial condition \mbox{$\rho(z)=\rho'+\theta{(z)}(\rho-\rho')$} we obtain a similar expression to Eq.~\eqref{eq:diff_3} for the integral of $\rho(z)$. Both expressions will be useful in Section~\ref{sec:cont_grad} where we develop a hydrodynamic model for the swimming motion in a continuous viscosity gradient.

 We show in Fig.~\ref{fig:setup2}(b) the viscosity profile obtained experimentally (data points) with  the pixel intensity corresponding to the viscositiy of each fluid. Here $z$ is the position along the gradient; the position of the fluid interface at $z=0$. The plot is presented in terms of the dimensionless distance  $z^*=z/L_H$, where $L_H$ is length of the head. Negative values of $z^*$ correspond to the bottom fluid which is more viscous while the high-viscosity fluid is located at $z^*>0$. By fitting Eq.~\eqref{eq:diff_2} to the experimental data (matching  $\mu$ and $\mu'$ with the maximum and the minimum intensities, respectively) we can find the thickness of the transition region $\delta=\Delta/L_H$ as a function of time. The experimental viscosity profile is then obtained by using this value of $\delta$ and the experimental values of the viscosities $\mu'=\mu_-$ and $\mu=\mu_+$ in Eq.~\eqref{eq:diff_2}. The fits are shown in Fig.~\ref{fig:setup2} as dashed lines. In the case where the measurement is conducted soon after the gradient is set up, referred to as a ``narrow gradient'' (N) in what follows, a value of $\delta=0.2735$ closely fits the data. For experiments conducted sixteen hours after the setup, which we will refer to as ``wide gradient'' (W), the value of $\delta=1.1384$ closely reproduces the experiments.

\subsection{Four different swimmer-viscosity interactions}

\begin{figure}[ht]
  \centering
  \subfigure[\,Case I]{\includegraphics[height=8cm]{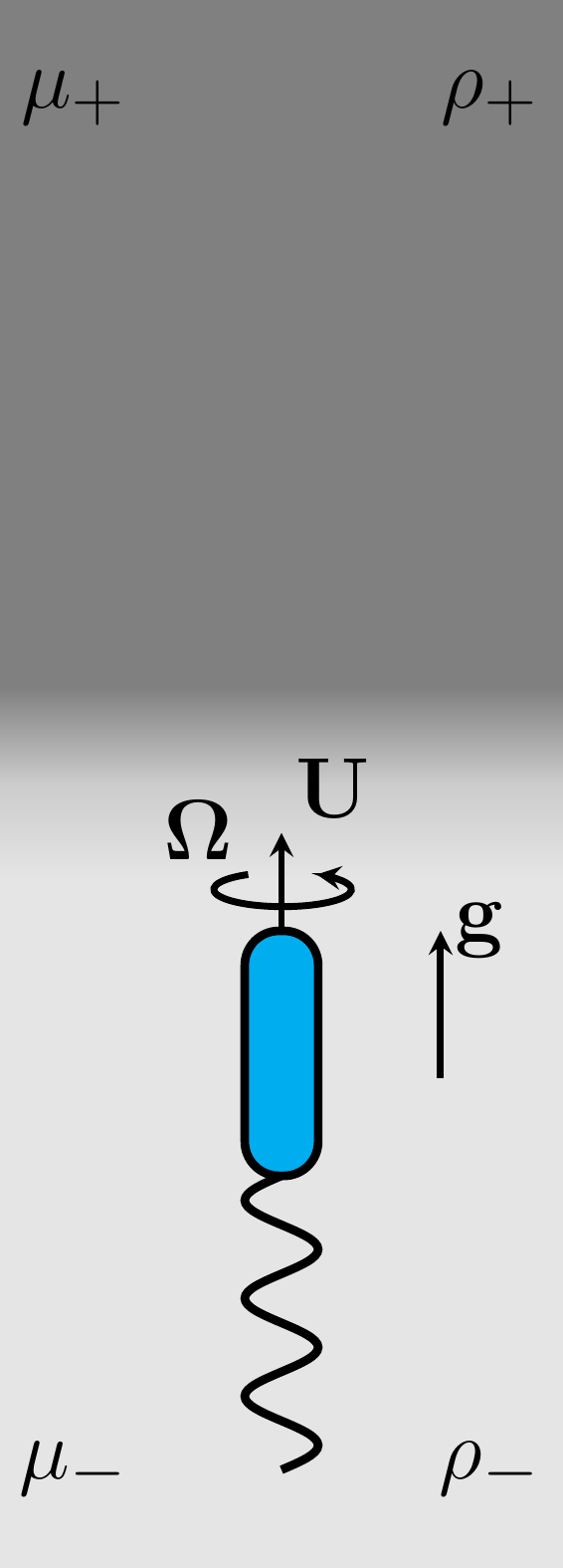}}\,\,
    \subfigure[\,Case II]{\includegraphics[height=8cm]{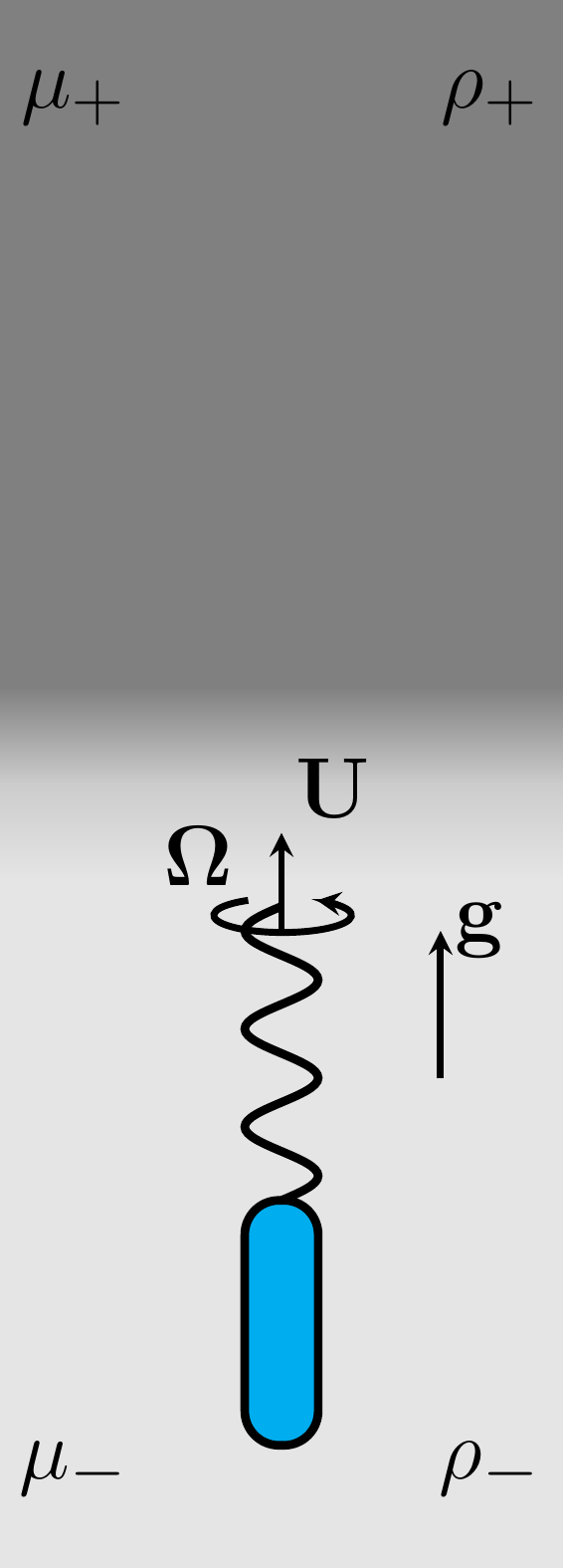}}\,\, 
  \subfigure[\,Case III]{\includegraphics[height=8cm]{figures/mswimmerc3.pdf}}\,\, 
  \subfigure[\,Case IV]{\includegraphics[height=8cm]{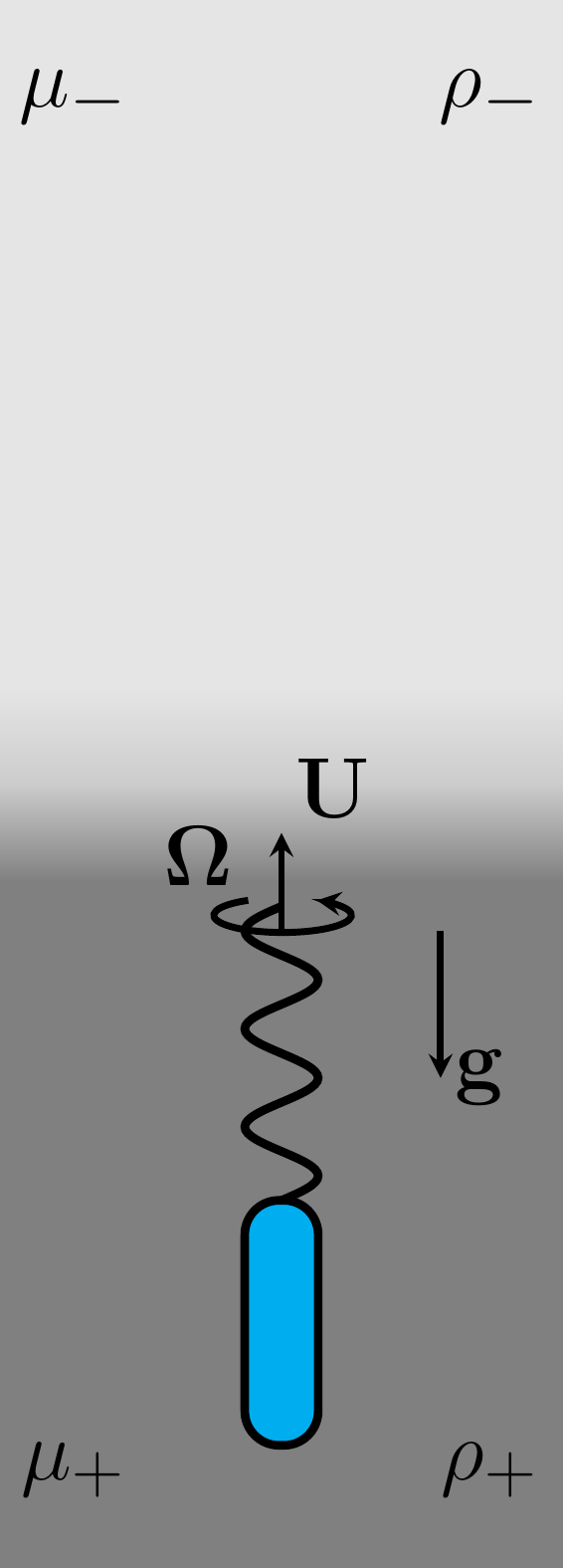}}
\caption{\label{fig:setup3} The four swimmer-viscosity interaction configurations for motion  across the viscosity gradient. In all cases $\mu_-<\mu_+$ and the motion is in the upwards direction; note that  gravity points upwards in (a) and (b), and downwards in (c) and (d). Notice as well the change in the sense of rotation of the tail, depending on its orientation. All conditions are described in Table~\ref{tab:2}.  }
 \end{figure}

With the setup described above, we can now consider  four distinct swimmer-viscosity interaction scenarios. First, the swimmer can move head-forward (pusher mode) or tail-forward (puller mode). Since the helical tail is chiral, reversing the rotation direction of the tail (by changing the rotation direction of the magnetic field) results, for the same swimmer, in a displacement in the opposite direction.  In a uniform fluid at small $Re$, the swimming speed is, as expected, unaffected by this change of direction (not shown). In addition to the swimming direction,  the swimmer can be made to swim across the interface from low to high viscosity (i.e.~from $\mu_-$ to $\mu_+$) or from high to low viscosity (i.e.~from $\mu_+$ to $\mu_-$). These four scenarios  are depicted schematically in Fig.~\ref{fig:setup3} and summarized in Table~\ref{tab:2}. In what follows, we will refer to the viscosity gradient as being positive when the swimmer moves from a low to a high viscosity region, or negative in the opposite case, from high to low. 
 \begin{table}[t]
     \centering
     \begin{tabular}{cccc}
         CASE     & DIRECTION  & GRADIENT \\
         \hline
         I      & head-first (pusher) & positive  \\
         II         & tail-first (puller) & positive \\
         III          & head-first (pusher) & negative  \\
         IV         & tail-first (puller) & negative  \\
   \hline
  \end{tabular}
   \caption{\label{tab:2} The four  possible swimmer-viscosity interactions  depicted in Fig.~\ref{fig:setup3}. }
 \end{table}
 
Before each experiment is conducted, the swimmer is slowly placed in the desired initial position and alignment, as far as possible from the interface, such that the viscosity gradient is not significantly disturbed. The motion is recorded with a video camera (920$\times$1080 pixels, Sony RX10II, 60 frames per second), using the same distance from the setup to the camera and lens magnification for all experiments. If the traveling time of the swimmer  across the two-fluid layer is smaller than the diffusion time, the viscosity gradient can be considered to be approximately constant. Although, in principle, it is possible to conduct experiments considering  different values of the viscosity gradient we only consider two cases here: a narrow gradient (N, $\delta$=0.274) and a wide gradient (W, $\delta$=1.138), as described above.

\section{\label{sec:results} Experimental results}

We now analyze the  crossing of the viscosity gradients by our swimmers along the  four configurations described in Table~\ref{tab:2}, for which we  find significantly contrasting behaviors. In all  experiments, the position $z$ measures the distance from the leading edge of the swimmer to the undisturbed interface; negative and positive $z$ values denote therefore locations before and after reaching the interface, respectively.

\subsection{\label{subsec:caseI} Case I: Head-first, positive viscosity gradient}

In this first case, the  swimmer is  placed initially at the upper part of the tank. The vertical displacement begins as soon as the rotating magnetic field forces the swimmer to rotate and swim downwards; it reaches quickly its steady-state speed, 
$U_0^{\mu_-}=1.75$~mm/s. After the interaction with the interface, the swimmer attains a new steady-state speed $U_0^{\mu_+}=U_{+}=2.5$~mm/s.

\begin{figure}
 \includegraphics[width=.9\linewidth]{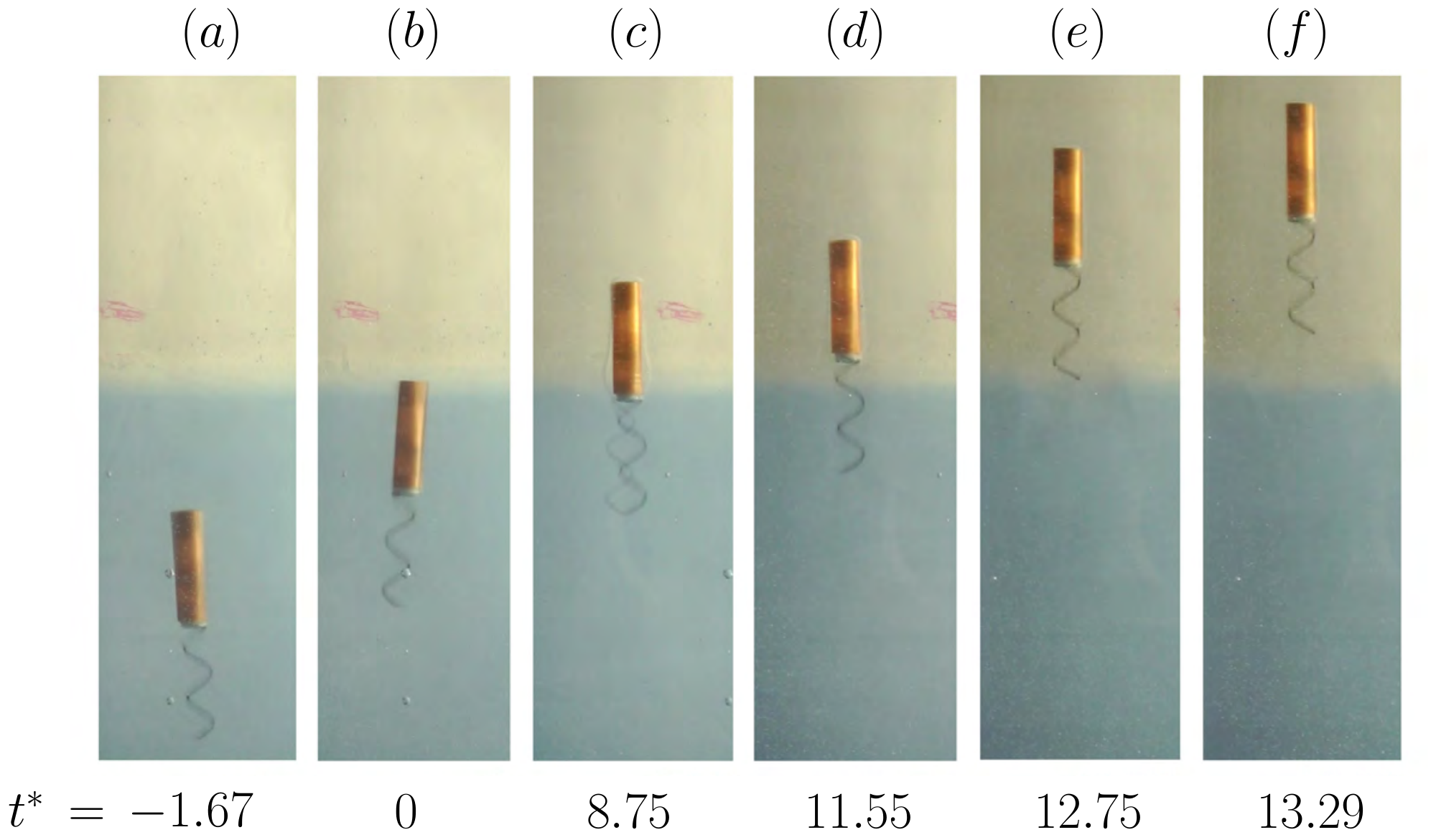}
     \caption{\label{fig:caseI_exp1} Case I: time sequence of the head-first (pusher) swimmer crossing a positive viscosity gradient, for $\delta$=0.274 (narrow gradient). Images have been  flipped so that the swimmer appears to move upwards.}
\end{figure}

\begin{figure}
     \subfigure[]{\includegraphics[width=.45\linewidth]{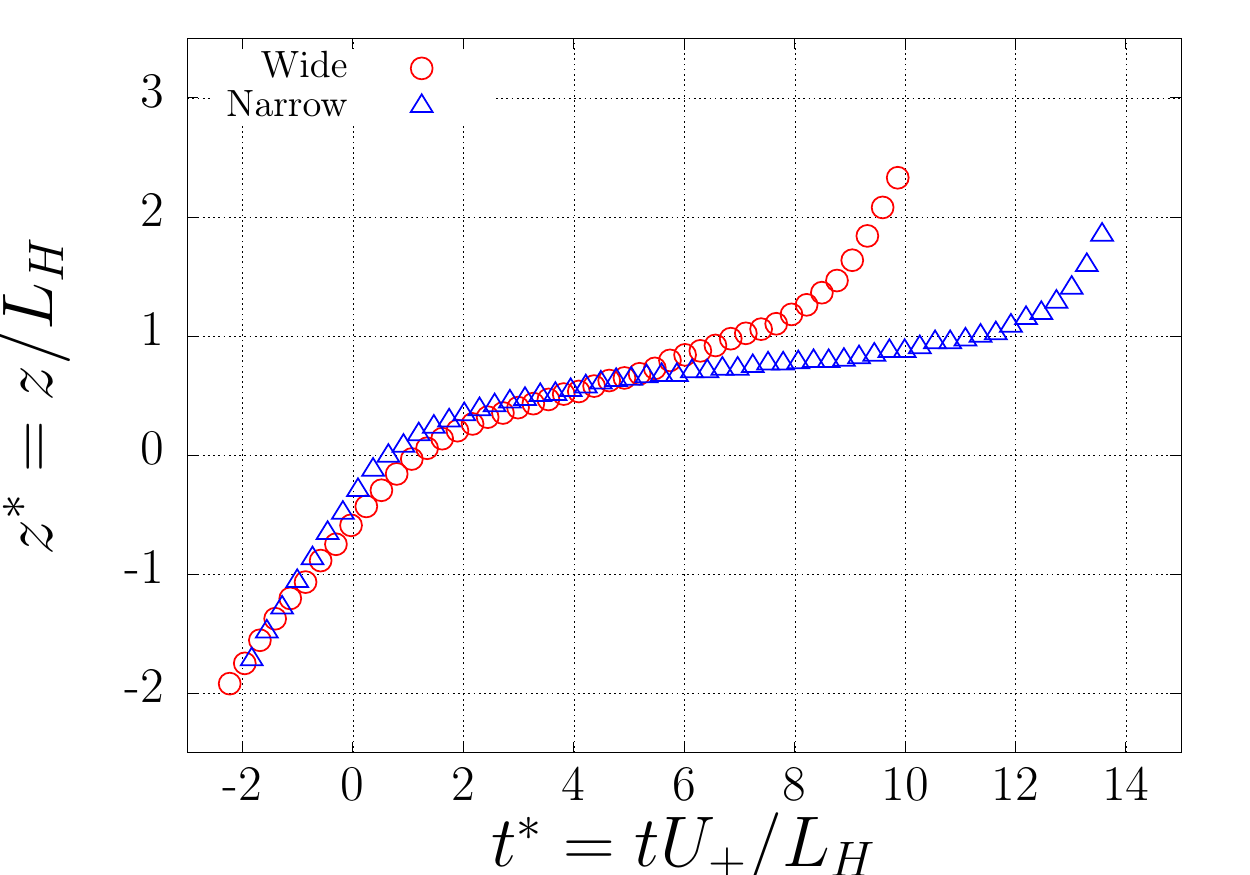}}\,
     \subfigure[]{\includegraphics[width=.45\linewidth]{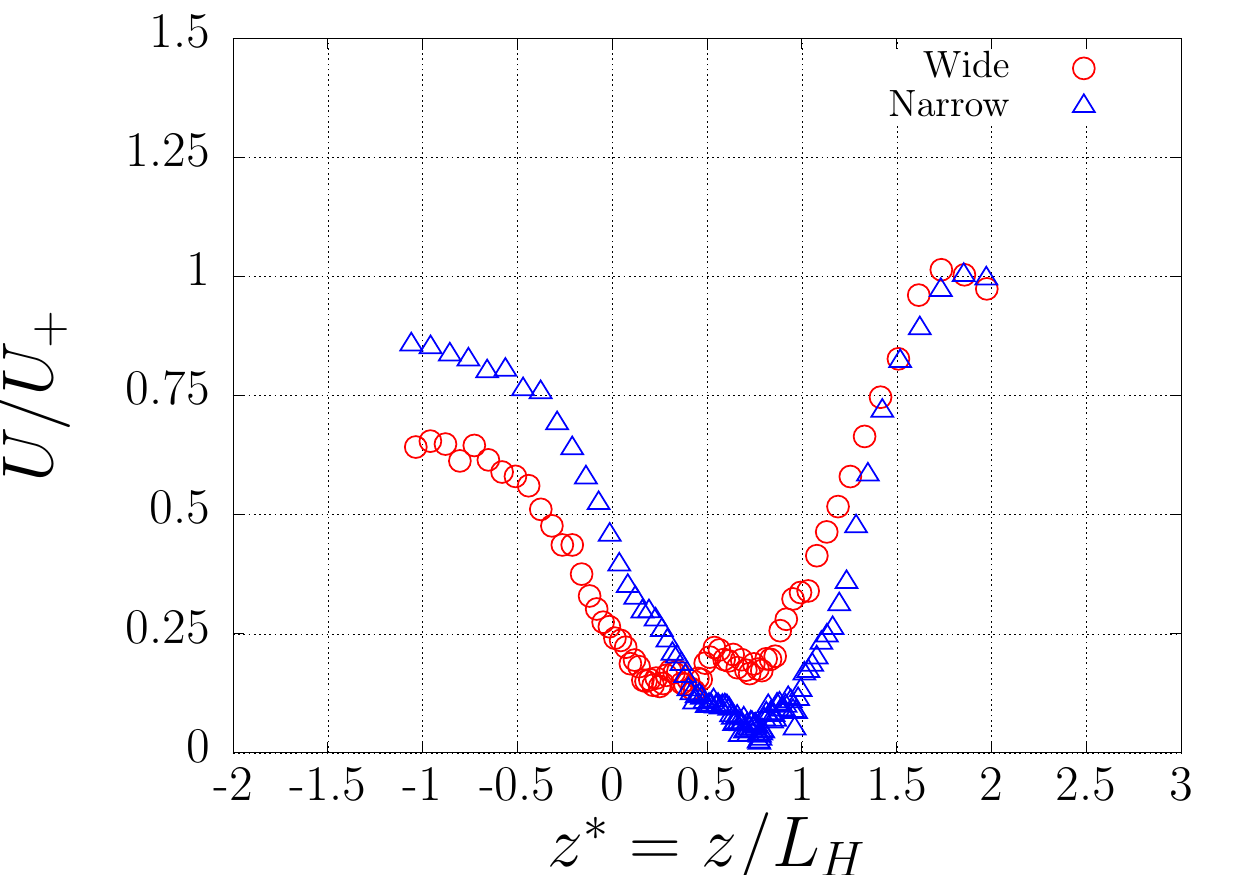}}
     \caption{\label{fig:caseI_exp2} Case I dynamics, head-first (pusher) swimmer crossing a positive viscosity gradient: (a) dimensionless position, $z^*$, as function of dimensionless time, $t^*$; (b) normalized speed  $U/U_{+}$ as a function of dimensionless position $z^*$. At $t^*\approx 0$ the swimmer reaches the interface, located at $z^*\approx 0$.}
\end{figure}

In Fig.~\ref{fig:caseI_exp1} we show a sequence of images illustrating  the crossing process. The time is given in dimensionless terms, $t^*=t U_{+}/L_H$, and $t^*=0$ represents the instant at which the swimmer (in this case the head) first reaches the interface. Along with the images, Fig.~\ref{fig:caseI_exp2}~(a) shows the normalized position of the swimmer, $z^*=z/L_H$, as a function of the normalized time, $t^*$ (note that the images have been  flipped so that the swimmer appears to move upwards);  Fig.~\ref{fig:caseI_exp2}~(b) shows the normalized speed $U/U_{+}$ as a function of $z^*$.

 
As the swimmer approaches the viscosity gradient, its speed progressively decreases, Fig.~\ref{fig:caseI_exp1}~(a-b). When the head of the swimmer begins to cross the interface [$z^*\approx 0$, Fig.~\ref{fig:caseI_exp1}~(b)], the speed decreases sharply reaching a minimum value at $z^*\approx 0.5$. Once the head has completely passed , the swimmer experiences  two different viscous environments simultaneously: the head is in the high-viscosity region while the tail is in the low-viscosity domain [Fig.~\ref{fig:caseI_exp1}~(c)]. Shortly after the head has crossed,  the swimmer rapidly increases its speed and the helical tail is progressively crossing the interface [Fig.~\ref{fig:caseI_exp1}~(d)]. Once the tail has completely gone through the interface, the swimmer attains its new steady-state speed, $z^*>2$ [Fig.~\ref{fig:caseI_exp1}~(e)\textendash(f)]. For the two values considered experimentally, the thickness of the viscosity gradient does not seem to affect the process significantly. However, we note that when the gradient is sharp, the swimmer spends a  longer time at the interface than  in the case of the wider viscosity gradient. 

 \subsection{\label{subsec:caseII} Case II: Tail-first, positive viscosity gradient}

\begin{figure}
 \includegraphics[width=.9\linewidth]{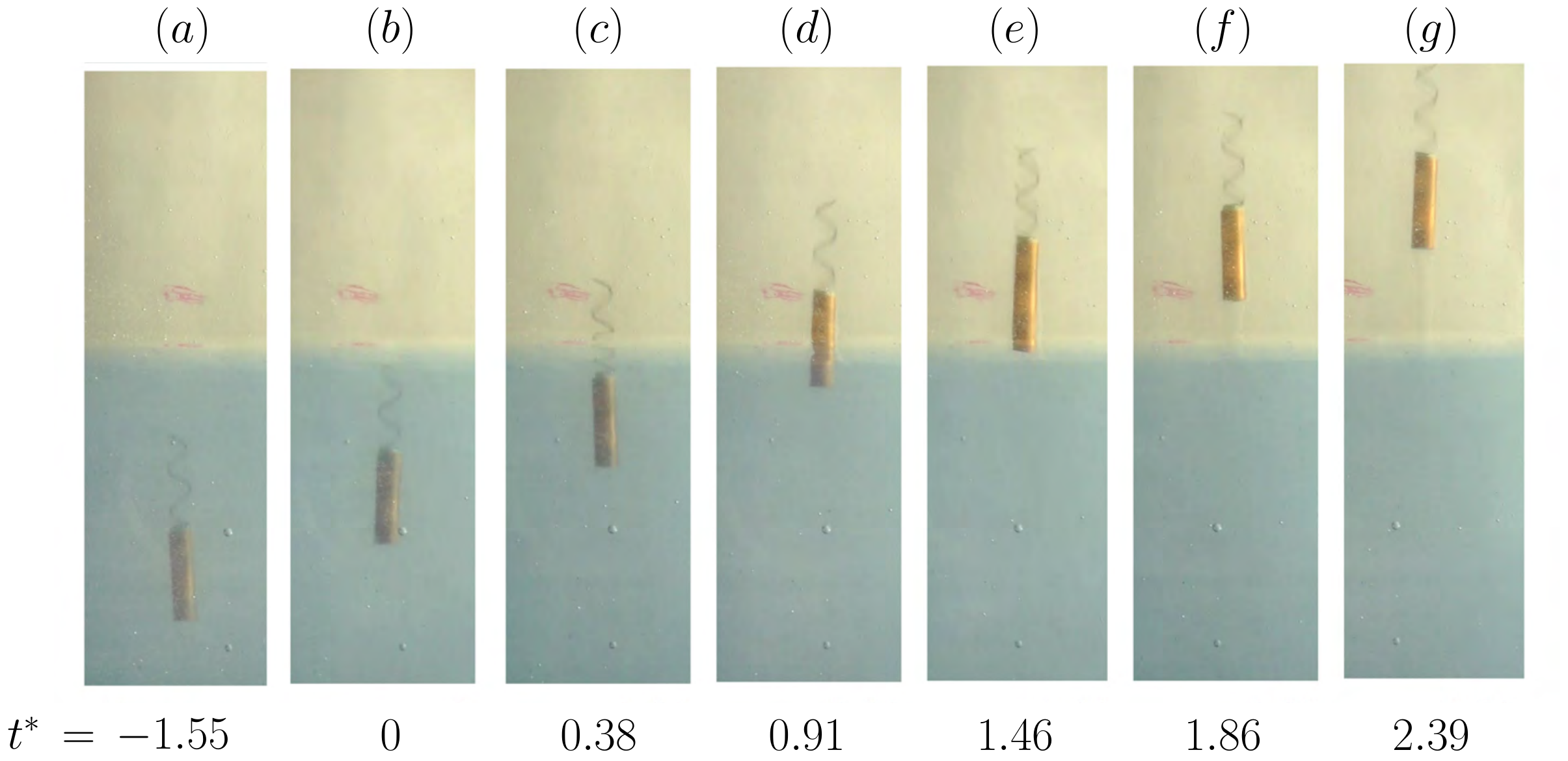}
     \caption{\label{fig:caseII_exp1} Case II: time sequence of the tail-first (puller) swimmer crossing a positive viscosity gradient, for $\delta$=0.274 (narrow gradient). Images have been  flipped so that the swimmer appears to move upwards.}
\end{figure}

\begin{figure}
     \subfigure[]{\includegraphics[width=.45\linewidth]{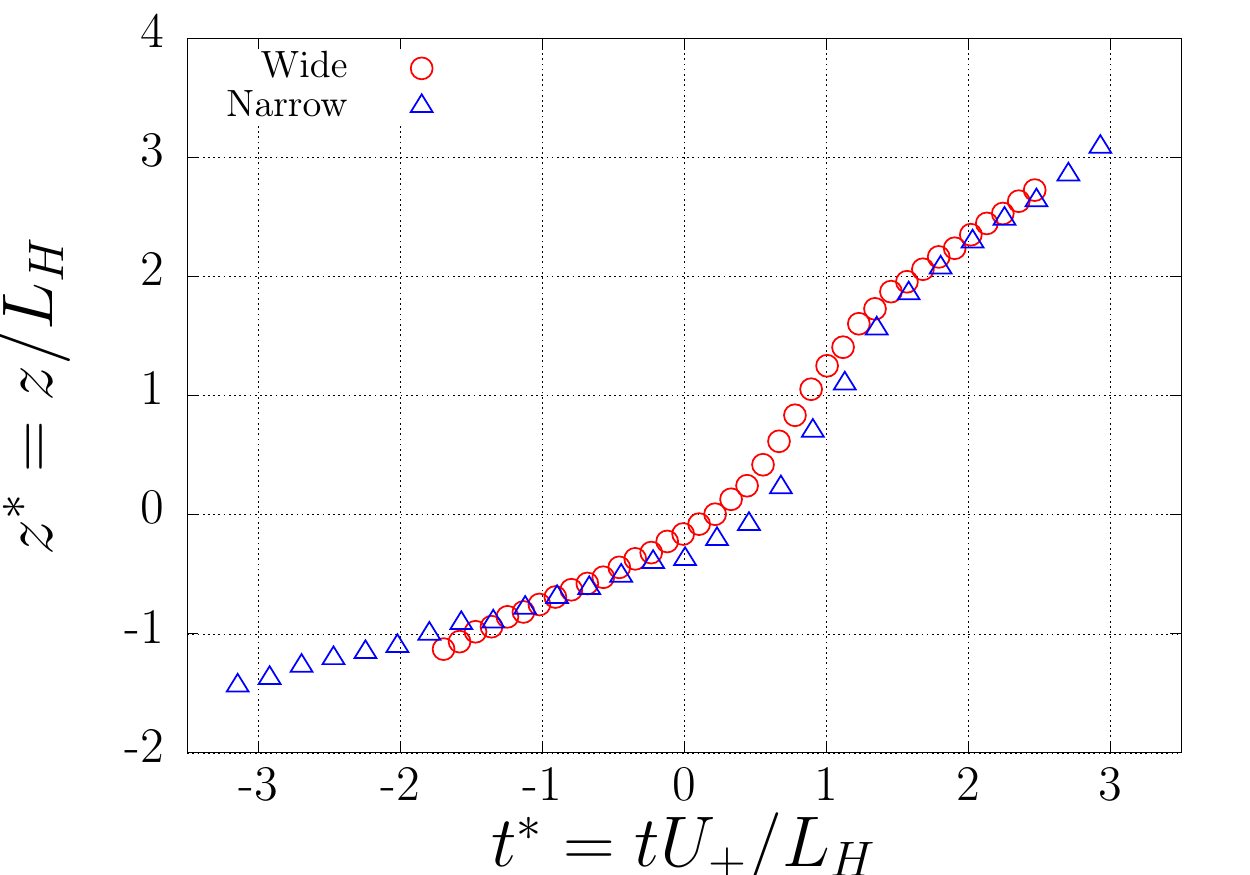}}\,
     \subfigure[]{\includegraphics[width=.45\linewidth]{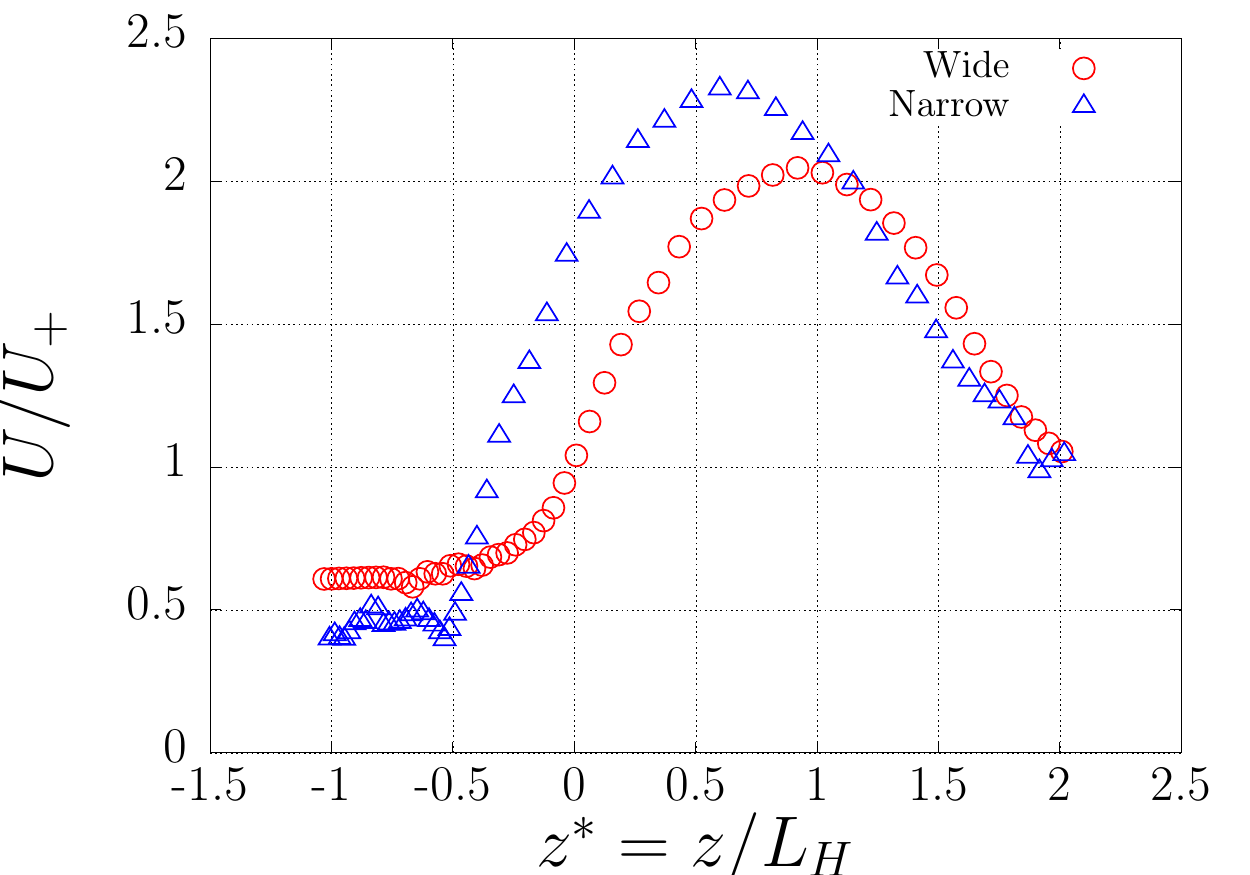}}
     \caption{\label{fig:caseII_exp2} Case II dynamics, tail-first (puller) swimmer crossing a positive viscosity gradient: (a) dimensionless position, $z^*$, as function of dimensionless time, $t^*$; (b) normalized speed  $U/U_{+}$ as a function of dimensionless position $z^*$. At $t^*\approx 0$ the swimmer reaches the interface, located at $z^*\approx 0$.}
\end{figure}

In this second case, the same swimmer is also placed near the top of the tank but the tail is oriented   towards the interface. By reversing the rotation direction, the   swimmer is then made to move tail-first (puller mode). In Fig.~\ref{fig:caseII_exp1} we show an image sequence of the process. The corresponding position and speed of the swimmer are plotted in Fig.~\ref{fig:caseII_exp2}.

As in the previous case, initially the swimmer moves at a constant speed when it is relatively far from the interface, see Fig.~\ref{fig:caseII_exp1}~(a). In contrast with the previous case, when the  tail of the swimmer reaches the interface, the swimming speed increases sharply, Fig.~\ref{fig:caseII_exp1}~(b). The swimming speed continues to increase until the head reaches the interface, Fig.~\ref{fig:caseII_exp1}~(c). The maximum speed reached is nearly twice that of the steady speed in the more viscous fluid. As the process progresses, the head crosses the interface and the swimming speed decreases from the maximum value to the free swimming value, see Fig.~\ref{fig:caseII_exp1}~(f). The process can be observed clearly in the two plots that show normalized position and speed in Fig.~\ref{fig:caseII_exp2}. As in the previous case, the thickness of the viscosity gradient does not significantly change the process.

\subsection{\label{subsec:caseIII} Case III: Head-first, negative viscosity gradient}

\begin{figure}
  \includegraphics[width=.9\linewidth]{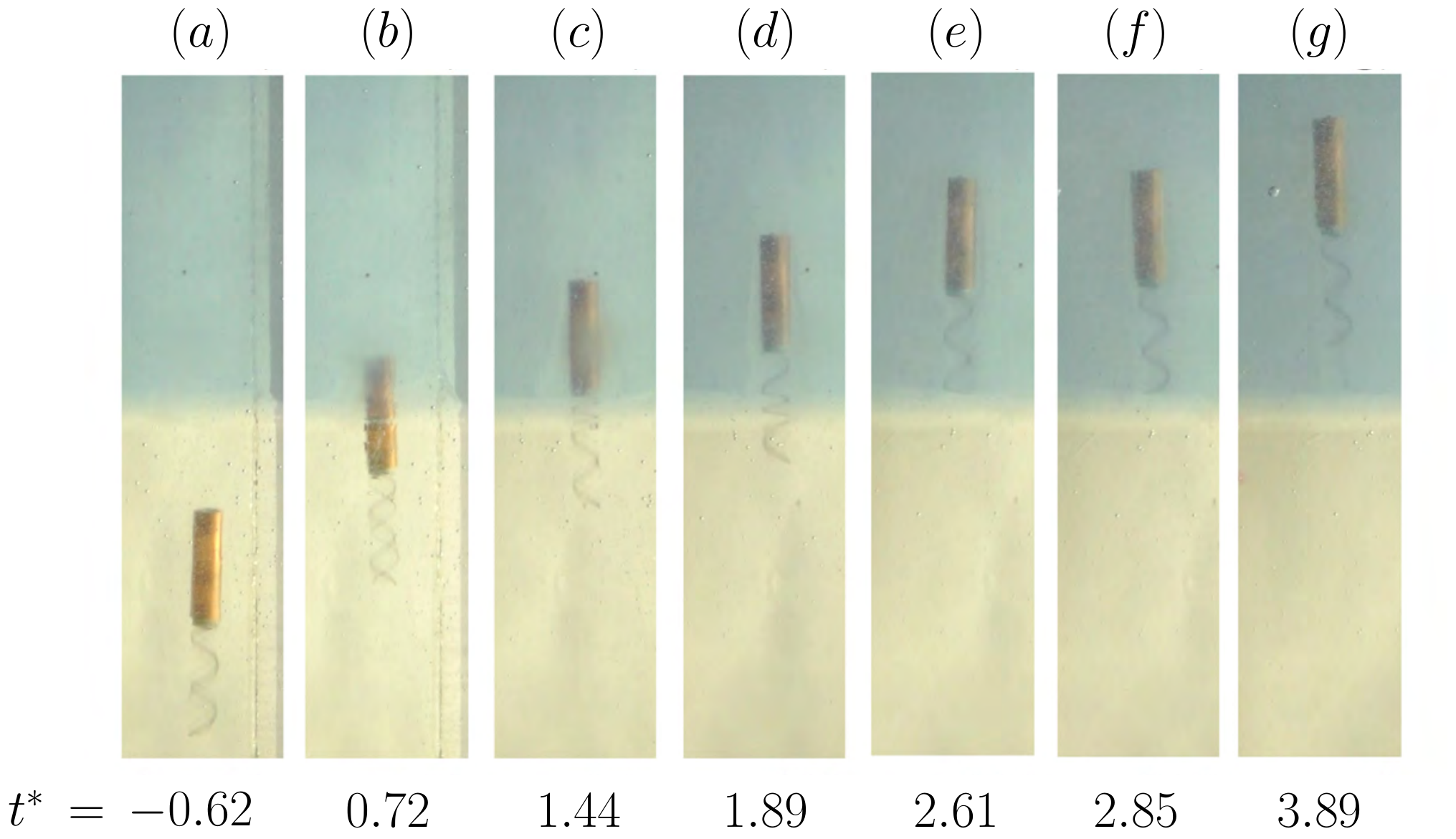}
\caption{\label{fig:caseIII_exp1} 
Case III: time sequence of the head-first (pusher) swimmer crossing a negative viscosity gradient, for $\delta$=0.274 (narrow gradient).}
\end{figure}

\begin{figure}
      \subfigure[]{\includegraphics[width=.45\linewidth]{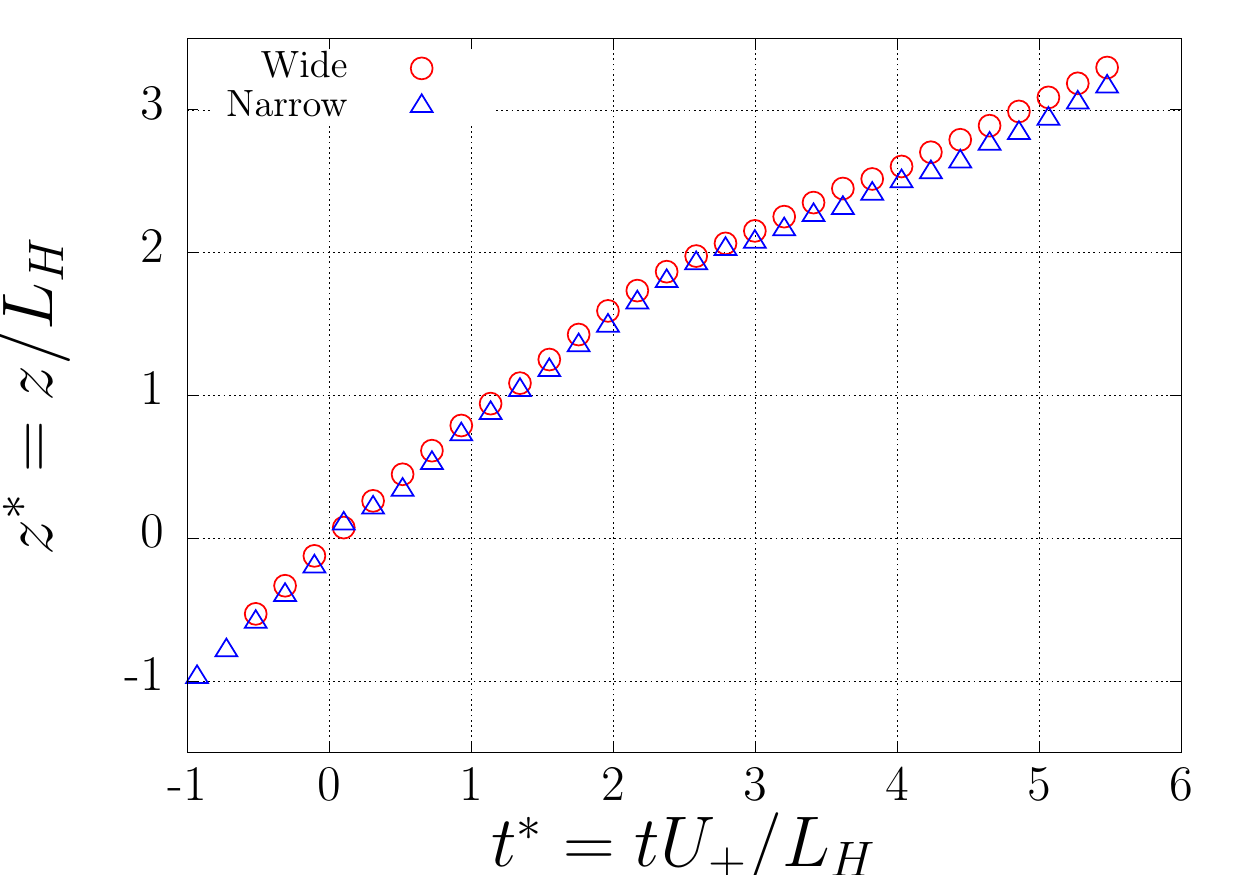}}\,
     \subfigure[]{\includegraphics[width=.45\linewidth]{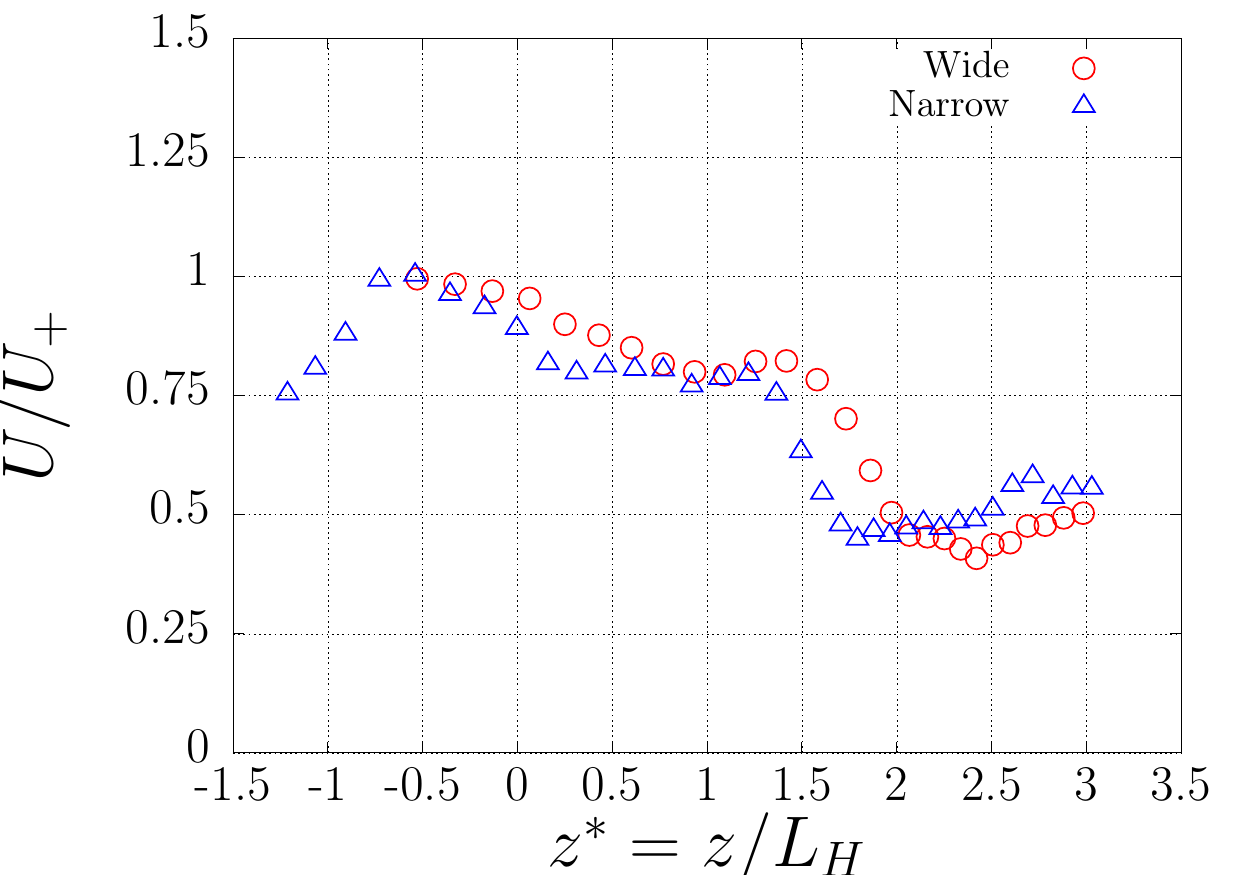}}
     \caption{\label{fig:caseIII_exp2} Case III dynamics, head-first swimmer (pusher) crossing a negative viscosity gradient: (a) dimensionless position, $z^*$, as function of dimensionless time, $t^*$; (b) normalized speed  $U/U_{+}$ as a function of dimensionless position $z^*$. At $t^*\approx 0$ the swimmer reaches the interface, located at $z^*\approx 0$.}
\end{figure}

In this third set of experiments, the swimmer  moves up from the bottom of the tank  head-first (i.e.~in pusher mode) across a negative viscosity gradient. Interestingly, we find some important differences with Case I. In Fig.~\ref{fig:caseIII_exp1} we show snapshots of the crossing process at different times while we plot in  Fig.~\ref{fig:caseIII_exp2}  the normalized position and the speed of the swimmer. 

The experiment starts with the swimmer moving in the high-viscosity fluid at constant speed, $U_0^{\mu_+}=U_{+}=3.2$ mm/s, see Fig.~\ref{fig:caseIII_exp1}~(a). As in the previous cases, the swimmer slows down as it approaches the interface, but the process is different from Case I since, although in both   cases the swimmer approaches the interface  head-first, the viscosity gradients are  in opposite directions. In the case of a negative viscosity gradient, the swimmer appears to entrain some of the  fluid with it as it crosses the interface, as can be seen in Fig.~\ref{fig:caseIII_exp1}~(b) and (c). This results in the swimming speed staying relatively constant during the crossing of the head. Once the head of the swimmer has completely crossed the interface, the tail remains in the more viscous fluid, Fig.~\ref{fig:caseIII_exp1}~(d), and the speed decreases sharply. At later times, the speed of the swimmer slowly increases until it reaches the free swimming speed $U_0^{\mu_-}$, see Fig.~\ref{fig:caseIII_exp1}~(f)\textendash(g). We note that the thickness of the viscosity gradient does not seem to affect the crossing process significantly.
\newpage

\subsection{\label{subsec:caseIV} Case IV: Tail-first, negative viscosity gradient}

The final case considers the dynamics of the swimmer moving tail-first from the high to low viscous fluid (negative gradient). In Case II, when the swimmer also moved tail-first, a significant increase of the swimming speed was observed during the interface crossing process. As shown below, the behaviour in this case is quite different. 

\begin{figure}[hp!]
 \includegraphics[width=.9\linewidth]{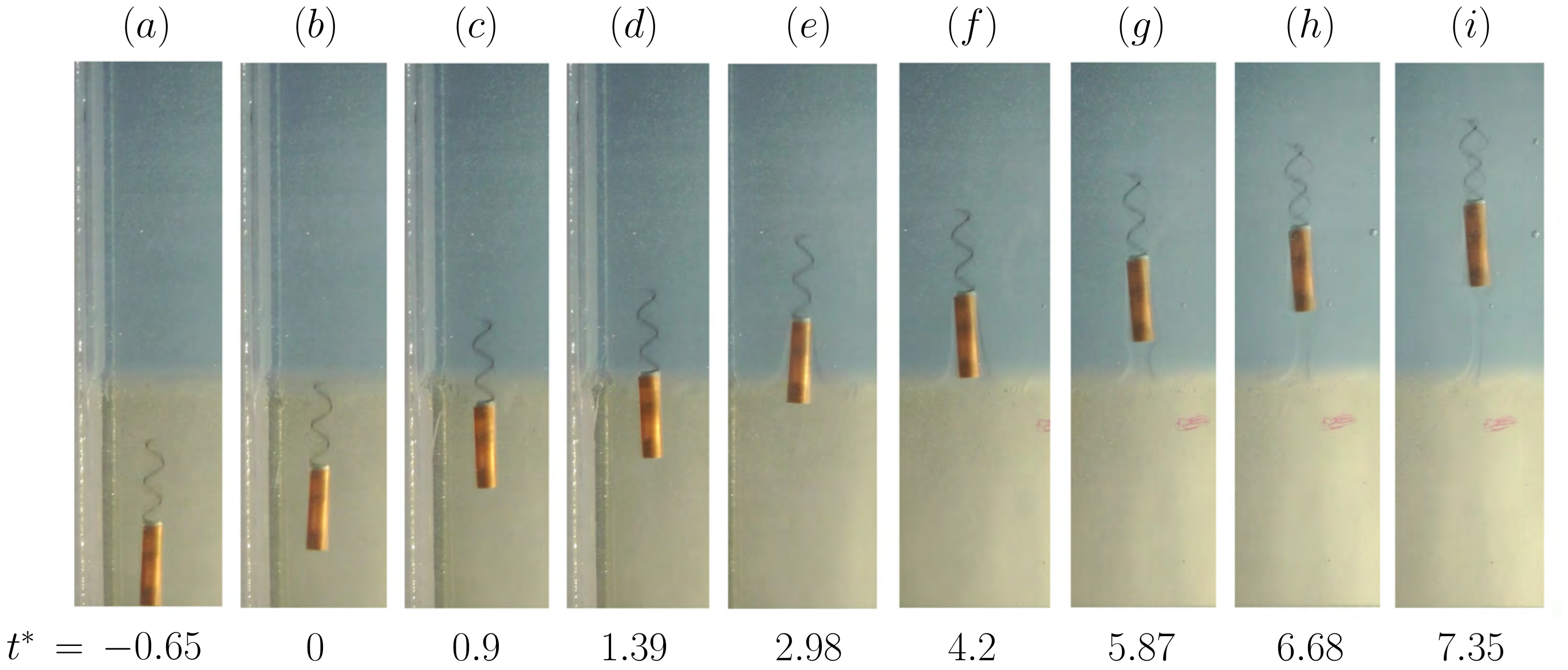}
     \caption{\label{fig:caseIV_exp1} Case IV: Time sequence of the tail-first swimmer crossing the viscosity gradients from high to low viscosity.}
\end{figure}

\begin{figure}[t]
     \subfigure[]{\includegraphics[width=.45\linewidth]{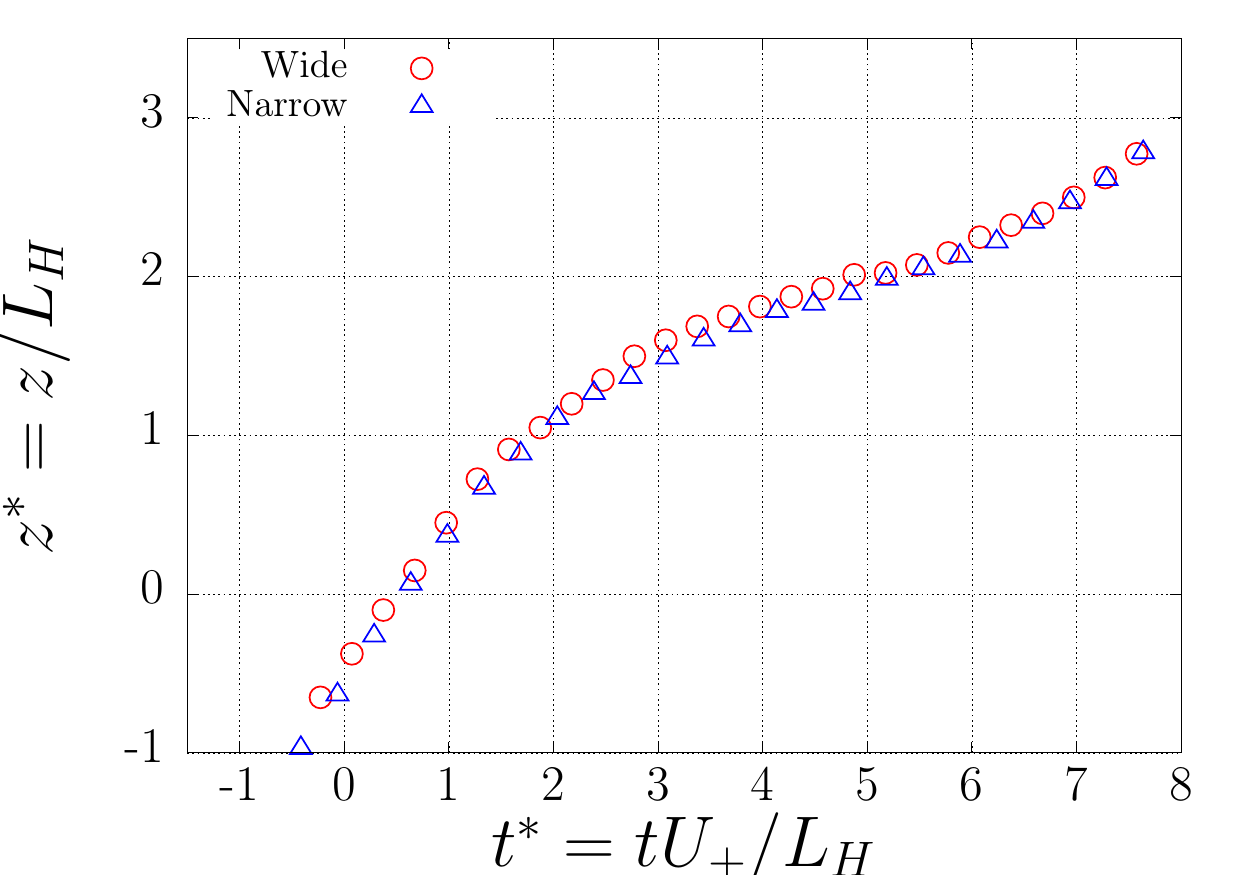}}\,
     \subfigure[]{\includegraphics[width=.45\linewidth]{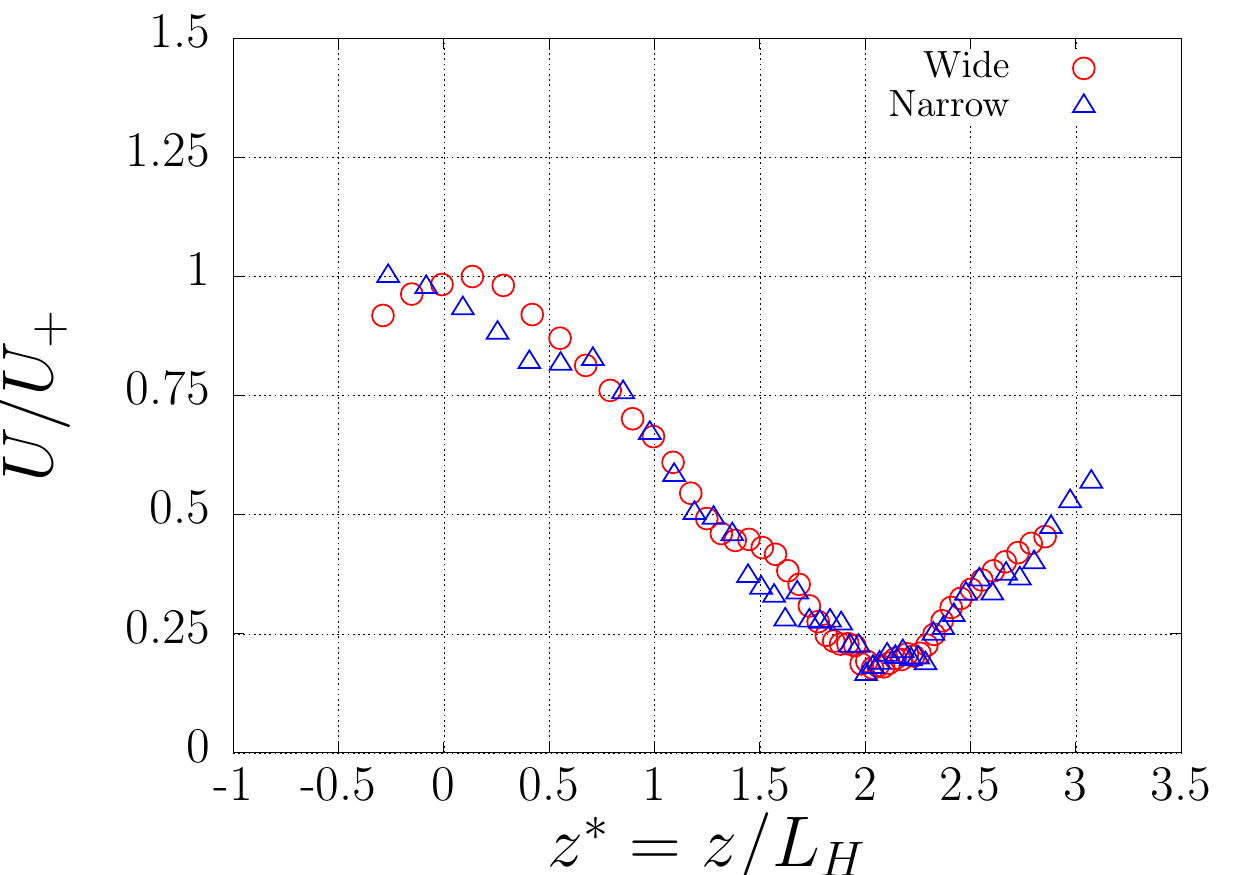}}
     \caption{\label{fig:caseIV_exp2} Case IV dynamics, tail-first (puller) swimming crossing  negative viscosity gradient: (a) dimensionless position, $z^*$, as function of dimensionless time, $t^*$; (b) normalized speed  $U/U_{+}$ as a function of dimensionless position $z^*$. At $t^*\approx 0$ the swimmer reaches the interface, located at $z^*\approx 0$.}
\end{figure}

In this case, the swimmer moves upwards from the bottom of the tank, approaching the interface tail-forward. A sequence of images of the dynamics are shown in Fig.~\ref{fig:caseIV_exp1} and we display in Fig.~\ref{fig:caseIV_exp2}  the corresponding position and speed of the swimmer for the two experiments with narrow and wide viscosity gradients (as in all previous cases,  Fig.~\ref{fig:caseIV_exp1} only illustrates the motion in the case of a narrow viscosity gradient).

The swimmer starts to move  from the bottom of the tank  towards the viscosity interface at a constant speed, see Fig.~\ref{fig:caseIV_exp1}~(a). When the tail reaches the viscosity interface, the speed increases slightly,  Fig.~\ref{fig:caseIV_exp1}~(b), but then continuously decreases as the swimmer crosses the interface, see Fig.~\ref{fig:caseIV_exp1}~(c)\textendash(f), reaching a minimum speed when the head finally reaches $z^*\approx 0$. As in the previous case, as the swimmer crosses the  gradient it is seen to entrain some of the more viscous fluid. The speed of the swimmer increases finally reaching its steady speed for $z^*>3$, see Fig.~\ref{fig:caseIV_exp1}~(h-i), while the more viscous fluid entrained by the  swimmer progressively return to its location at the bottom of the tank. 

 \section{\label{sec:discrete}Discrete interface model}

As shown in the four cases studied experimentally above, a rich dynamical process is observed, resulting from the intricate balance between drag and thrust for a swimmer straddling two domains of different viscosities. Since the two-fluid arrangement is naturally stratified (the  more-viscous fluid is denser), buoyancy may also play an important role in the process.  Based on these observations, we now propose a model to describe the motion of a swimmer immersed in a fluid of non-homogeneous viscosity. Following the experimental setup, the rigid swimmer consists of a cylindrical head and a helical tail.  It rotates at a fixed angular speed $\boldsymbol{\Omega}$, causing the tail to rotate and push on the surrounding fluid, thus propelling the swimmer forward with velocity $\mathbf{U}$. The size of the swimmer, velocity of motion and the viscosity of the fluid media are such that inertia can be neglected and we are in the creeping flow conditions.   Assuming that the classical resistive-force theory of slender filaments~\cite{Cox1970} remains applicable locally at each point along the swimmer,  the force per unit length acting on the swimmer is given by
 \begin{equation}
     \label{eq:resistive}
     \mathbf{f}=-\zeta_\perp\mathbf{u}+\left(\zeta_\perp-\zeta_\parallel\right)(
     \mathbf{u}
     \cdot\boldsymbol{\tau})\boldsymbol{\tau},
 \end{equation}
 \noindent where $\zeta_{\perp}$ and $\zeta_{\parallel}$ are,  respectively, the perpendicular and parallel drag coefficients per unit length given by~\cite{Doi1996}
  \begin{equation}
   \label{eq:drag_coefficients}
   \zeta_{\parallel}^{H,T}\approx \frac{2\pi \mu}{\ln{\left(L_{H,T}/r_{H,T}\right)}},
  \end{equation}
   where the superscripts denote head ($H$) and tail ($T$).    
   Using force balance we can then relate the swimming velocity to the angular velocity  with a linear relationship, \mbox{$\mathbf{U}=S\boldsymbol{\Omega}$}, with a prefactor $S$ that can be determined for different viscosity profiles. We  start below with a sharp (step) function, which is a good approximation to a  mixture of two miscible fluids of different viscosities at early times. We will  next  generalise to a continuous profile. We then complete the model by adding the effect of gravity to our calculations. The predictions of the model are finally compared against experimental data. 

\subsection{\label{subsec:sharp} Sharp viscosity gradient}
 We start by analysing the motion at early times when the gradient  in viscosity is sharp. In this case we can model the fluid as two semi-infinite domains with viscosity
 \begin{equation}
  \label{eq:viscosity}
  \mu(z)=\left\{\begin{array}{ccc}
                 \mu_1=\mu'	&\quad	&z\leq 0\\
                 \mu_2=\mu	&\quad	&0\leq z
                \end{array}\right.,
 \end{equation}
   where $z=0$ denotes the location of the  interface (see Fig.~1). The analysis below will be valid for any viscosity distribution and any orientation. Indeed, if we express the swimming speed as a function of the distance from the head to the fluid interface, instead of the vertical coordinate $z$, we can describe the dynamics in cases I and III (head-first negative and positive gradient) by taking $\mu'<\mu$ and $\mu<\mu'$, respectively. As discussed in Section~\ref{subsec:tail_first}, the tail-first dynamics then follows from the reversibility of Stokes flow.
 
\subsection{\label{subsec:head_first} Head-first interaction}
 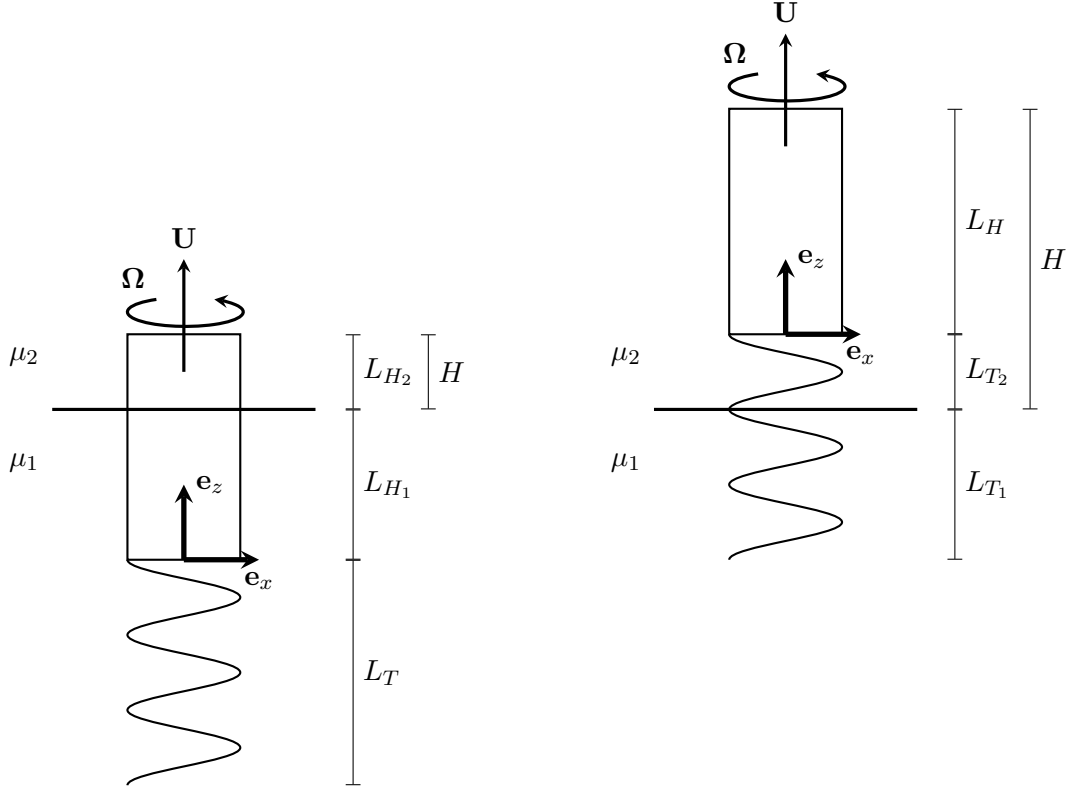
\begin{figure}
  \begin{tikzpicture}
   
   \draw[very thick] (-1,2)--(2.5,2);
   \path (-1,2)--++(135:1) node[right]{$\mu_2$};
   \path (-1,2)--++(225:1) node[right]{$\mu_1$};
   \draw[thick] (0,0)--(1.5,0)--(1.5,3)--(0,3)--cycle;
   \draw[very thick, -stealth] (0.75,2.5)--++(90:1.5) node[above]{$\mathbf{U}$};
   \draw[x=.5em,y=2em,very thick,-stealth,rotate=-90] (-18,0.5) node[anchor=south east] {$\boldsymbol{\Omega}$} arc (-150:150:1 and 1);
   \draw[thick, rotate=270] (0,0) cos (0.25,0.75) sin (0.5,1.5) cos (0.75,0.75) sin (1,0) cos (1.25,0.75) sin (1.5,1.5) cos (1.75,0.75) sin (2,0) cos (2.25,0.75) sin (2.5,1.5) cos (2.75,0.75) sin (3,0);
   \draw[|-|] (3,0)--node[right]{$L_{H_1}$}++(90:2);
   \draw[|-|] (3,2)--node[right]{$L_{H_2}$}++(90:1);
   \draw[|-|] (4,2)--node[right]{$H$}++(90:1);
   \draw[|-|] (3,-3)--node[right]{$L_T$}++(90:3);
   \draw[line width=2, -stealth] (0.75,0)--++(0:1) node[below]{$\mathbf{e}_x$};
   \draw[line width=2, -stealth] (0.75,0)--++(90:1) node[right]{$\mathbf{e}_z$};
   
   \draw[very thick] (7,2)--(10.5,2);
   \path (7,2)--++(135:1) node[right]{$\mu_2$};
   \path (7,2)--++(225:1) node[right]{$\mu_1$};
   \begin{scope}[shift={(8,3)}]
   \draw[thick] (0,0)--(1.5,0)--(1.5,3)--(0,3)--cycle;
   \draw[very thick, -stealth] (0.75,2.5)--++(90:1.5) node[above]{$\mathbf{U}$};
   \draw[x=.5em,y=2em,very thick,-stealth,rotate=-90] (-18,0.5) node[anchor=south east] {$\boldsymbol{\Omega}$} arc (-150:150:1 and 1);
   \draw [thick, rotate=270] (0,0) cos (0.25,0.75) sin (0.5,1.5) cos (0.75,0.75) sin (1,0) cos (1.25,0.75) sin (1.5,1.5) cos (1.75,0.75) sin (2,0) cos (2.25,0.75) sin (2.5,1.5) cos (2.75,0.75) sin (3,0);
   \draw[|-|] (3,0)--node[right]{$L_H$}++(90:3);
   \draw[|-|] (4,-1)--node[right]{$H$}++(90:4);
   \draw[|-|] (3,-3)--node[right]{$L_{T_1}$}++(90:2);
   \draw[|-|] (3,-1)--node[right]{$L_{T_2}$}++(90:1);
   \draw[line width=2, -stealth] (0.75,0)--++(0:1) node[below]{$\mathbf{e}_x$};
   \draw[line width=2, -stealth] (0.75,0)--++(90:1) node[right]{$\mathbf{e}_z$};
   \end{scope}
  \end{tikzpicture}
  \caption{\label{fig:experimental_setup}  Helical swimmer crossing a sharp viscosity gradient.}
 \end{figure}
 \subsubsection{Head crossing}
When the head crosses the viscosity gradient first (Fig.~\ref{fig:experimental_setup}, left), we model the drag exerted on each part of the head, as that experienced in an infinite fluid of viscosity $\mu_1=\mu'$ or $\mu_2=\mu$. The total viscous drag on the head is therefore given by
 \begin{equation}
  \label{eq:drag_h-f_1}
  \mathbf{F}_D=-\left[\zeta_{\parallel_1}^H L_{H_1}+\zeta_{\parallel_2}^H L_{H_2}\right]U\mathbf{e}_z=-\zeta_{\parallel_1}^H\left[L_{H_1}+\frac{\mu_2}{\mu_1}L_{H_2}\right]U\mathbf{e}_z,
 \end{equation}
  where $L_{H_{1,2}}$ are the lengths of the portions of the head above and below the interface (which therefore change in time as the swimmer moves through the interface). The resistance coefficients $\zeta_{\parallel_{1,2}}^H$ are proportional to the corresponding local viscosities and   depend on the
 geometry of the head~\cite{kimbook}. The tail is  modelled as a  right-handed helix of radius $R$ and pitch angle $\psi$.  We parametrise it using the arc-length $s=R\varphi/\sin{\psi}$, where $\varphi$ is the azimuthal coordinate. The position $  \mathbf{x}(s,t)$ of a
material point on the tail is therefore given by (see Fig.~\ref{fig:experimental_setup})
 \begin{equation}
   \label{eq:helix_pos}
  \mathbf{x}(s,t)=R\cos{\left(2\pi s/\ell+\Omega t\right)}\mathbf{e}_x+R\sin{\left(2\pi s/\ell+\Omega t\right)}\mathbf{e}_y+\left(Ut+bs\right)\mathbf{e}_z,
 \end{equation}
  where $\ell=2\pi R/\sin{\psi}$ is the arc-length per helical turn and $b=\cos{\psi}$. The tangent ($  \boldsymbol{\tau}$) and velocity vectors ($  \mathbf{u}$) are obtained by differentiation with respect to $s$ and $t$ respectively
 \begin{align}
  \label{eq:helix_tangent}
  \boldsymbol{\tau}(s)&=-\sin{\psi}\sin{\left(2\pi s/\ell+\Omega t\right)}\mathbf{e}_x+\sin{\psi}\cos{\left(2\pi s/\ell+\Omega t\right)}\mathbf{e}_y+\cos{\psi}\mathbf{e}_z ,\\
  \label{eq:helix_vel}
  \mathbf{u}(s)&=-R\Omega\sin{\left(2\pi s/\ell+\Omega t\right)}\mathbf{e}_x+R\Omega\cos{\left(2\pi s/\ell+\Omega t\right)}\mathbf{e}_y+U\mathbf{e}_z.
 \end{align}
 The total hydrodynamic force exerted on the helix is then obtained using resistive-force theory as  
 \begin{align}
  \label{eq:propulsion_h-f_1}
  \mathbf{F}_p&=-\int_0^{L_T/\cos{\psi}}{\zeta_{\perp_1}^T\mathbf{u}+\left(\zeta_{\parallel_1}^T-\zeta_{\perp_1}^T\right)(  \mathbf{u}\cdot  \boldsymbol{\tau})\boldsymbol{\tau}
  \,\text{d}s}\nonumber\\
  &=-\zeta_{\perp_1}^T\int_0^{L_T/\cos{\psi}}{\mathbf{u}+\left(\beta^{T}-1\right)(\mathbf{u}\cdot  \boldsymbol{\tau})\boldsymbol{\tau}\,\text{d}s},
 \end{align}
 \noindent where $\zeta_{\perp_1}^{T}$, $\zeta_{\parallel_1}^{T}$ are the perpendicular and parallel drag coefficients for the helix center line filament and $\beta^{T}=\zeta_{\parallel_1}^{T}/\zeta_{\perp_1}^{T}$, approximately equal to $1/2$ in the slender limit~\cite{Cox1970}.  
  Ignoring end effects, the propulsive force acts mainly in the $z$ direction, therefore we evaluate 
 \begin{align}
  \label{eq:propulsion_h-f_2}
\mathbf{e}_z\cdot  \mathbf{F}_p&=-\zeta_{\perp_1}^{T}\int_0^{L_T/\cos{\psi}}{\left[U+\left(\beta^{T}-1\right)b(\Omega R\sin{\psi}+Ub)\right]\,\text{d}s}\nonumber\\
  &=-\frac{\zeta_{\perp_1}^{T}L_T}{\cos{\psi}}\left[U\left(1+\left(\beta^{T}-1\right)\cos^2{\psi}\right)+\left(\beta^{T}-1\right)\Omega R\sin{\psi}\cos{\psi}\right].
 \end{align}
 We obtain the swimming velocity in terms of the angular velocity, by imposing the free swimming condition $\mathbf{e}_z\cdot (\mathbf{F}_p+\mathbf{F}_D)=0$. The swimming speed is then given by
 \begin{equation}
  \label{eq:swimming_vel_f-h_1}
  U=\frac{\zeta_{\perp_1}^{T}L_T\left(1-\beta^{T}\right)R\sin{\psi}\cos{\psi}\Omega}{\zeta_{\parallel_1}^H\cos{\psi}\left(L_{H_1}+\frac{\mu_2}{\mu_1}L_{H_2}\right)+\zeta_{\perp_1}^{T}L_T\left(1+(\beta^{T}-1)\cos^2{\psi}\right)}.
 \end{equation}
 Using the condition \mbox{$L_{H_1}+L_{H_2}=L_H$} and defining \mbox{$\lambda\equiv L_T/L_H$}, \mbox{$\xi\equiv\zeta_{\perp_1}^{T}/\zeta_{\parallel_1}^H$} and \mbox{$\ell_H\equiv L_{H_2}/L_H$}, we can write the swimming speed as a function of $\ell_H=h$ as
 \begin{equation}
  \label{eq:swimming_vel_f-h_2}
  U_1(h)=\frac{\xi\lambda\left(1-\beta^{T}\right)R\sin{\psi}\cos{\psi}\Omega}{\cos{\psi}\left(1+\di\frac{\mu_2-\mu_1}{\mu_1}h\right)+\xi\lambda\left(1+(\beta^{T}-1)\cos^2{\psi}\right)},
 \end{equation}
 \noindent where $h\equiv H/L_H$ is the dimensionless position of the swimmer's head.

  \subsubsection{Tail crossing}

 After the head has crossed the interface completely ($h=1$) the force balance changes. In this case, the head is completely immersed in the fluid of viscosity $\mu_2$ and the drag on the head is given by
 \begin{equation}
  \label{eq:drag_h-f_2}
  \mathbf{F}_D=-\zeta_{\parallel_2}^HL_HU\mathbf{e}_z.
 \end{equation}
 On the other hand, the propulsive force from the tail as it crosses the interface (Fig.~\ref{fig:experimental_setup}, right) 
 is now given by
 \begin{equation}
  \label{eq:propulsion_h-f_3}
 \mathbf{e}_z\cdot  \mathbf{F}_p=-\frac{\zeta_{\perp_1}^{T}L_{T_1}+\zeta_{\perp_2}^{T}L_{T_2}}{\cos{\psi}}\left[U\left(1+\left(\beta^{T}-1\right)\cos^2{\psi}\right)+\left(\beta^{T}-1\right)\Omega R\sin{\psi}\cos{\psi}\right],
 \end{equation}
 where $\beta^{T}$ is independent of the viscosities. 
 Applying the free swimming condition along $z$, we then obtain the swimming speed
 \begin{equation}
  \label{eq:swimming_vel_f-h_3}
  U=\frac{\di \xi\lambda\left(\frac{\mu_2}{\mu_1}+\frac{\mu_1-\mu_2}{\mu_1}\ell_{T}\right)\left(1-\beta^{T}\right)R\sin{\psi}\cos{\psi}\Omega}{\di \cos{\psi}\frac{\mu_2}{\mu_1}+\xi\lambda\left(\frac{\mu_2}{\mu_1}+\frac{\mu_1-\mu_2}{\mu_1}\ell_{T}\right)\left(1+(\beta^{T}-1)\cos^2{\psi}\right)},
 \end{equation}
 \noindent where we have used the fact that \mbox{$\zeta_{\perp_2}^{T}/\zeta_{\parallel_2}^H=\zeta_{\perp_1}^{T}/\zeta_{\parallel_1}^H=\xi $} and   defined $\ell_{T}\equiv L_{T_1}/L_T$. The position of the top part of the head is now \mbox{$H=L_H[1+\lambda(1-\ell_{T})]$}, so we can rewrite
 Eq.~\eqref{eq:swimming_vel_f-h_3} in terms of $h$ as follows
 \begin{equation}
  \label{eq:swimming_vel_f-h_4}
  U_2(h)=\frac{\di \xi\left(\lambda+\frac{\mu_2-\mu_1}{\mu_1}(h-1)\right)\left(1-\beta^{T}\right)R\sin{\psi}\cos{\psi}\Omega}{\di \cos{\psi}\frac{\mu_2}{\mu_1}+\xi\left(\lambda+\frac{\mu_2-\mu_1}{\mu_1}(h-1)\right)\left(1+(\beta^{T}-1)\cos^2{\psi}\right)}.
 \end{equation}
 \subsubsection{Summary}
In the calculations above we obtained  that $U_1(\ell_H=0)=U_2(\ell_{T}=0)=U_0$, so the swimming speeds are identical when the swimmer is completely immersed in either of the two fluids.  The final speed of the swimmer is then given parametrically by
 \begin{equation}
  \label{eq:swimming_vel_f-h_5}
  U(h)=\left\{\begin{array}{ccc}
                        U_0	&\quad	&h<0,\\
                        U_1(h)	&\quad	&0\leq h\leq1,\\
                        U_2(h)	&\quad	&1< h <1+\lambda,\\
                        U_0	&\quad	&1+\lambda<h.
                       \end{array}\right.
 \end{equation}
 
 We further note that the information about the direction of motion is only embedded in the values of the viscosities, hence Eq.~\eqref{eq:swimming_vel_f-h_5} is the swimming speed in Case I when \mbox{$\mu_1=\mu'<\mu_2=\mu$}, and Case III for the choice $\mu<\mu'$.

 \subsection{\label{subsec:position} Swimmer position}
  When the head crosses the interface, the swimming speed is of the form
  \begin{equation}
    \label{eq:vel_1_l_h}
    U_1(h)=\frac{A_1}{B_1+C_1 h} \Omega,
  \end{equation}
  \noindent where
  \begin{align}
    \label{eq:vel_1_A1}
    A_1&=\xi\lambda(1-\beta^{T})R\sin{\psi}\cos{\psi},\\
    \label{eq:vel_1_B1}
    B_1&=\cos{\psi}+\xi\lambda(1+(\beta^{T}-1)\cos^2{\psi}),\\
    \label{eq:vel_1_C1}
    C_1&=\frac{\mu_2-\mu_1}{\mu_1}\cos{\psi}.
\end{align}
On the other hand, when the tail crosses the interface, the speed is of the form
\begin{equation}
    \label{eq:vel_2}
    U_2(h)=\frac{A_2+D_2 h}{B_2+C_2 h}\Omega,
\end{equation}
\noindent where
\begin{align}
    \label{eq:vel_1_A2}
    A_2&=\xi\lambda'(1-\beta^{T})R\sin{\psi}\cos{\psi},\\
    \label{eq:vel_1_D2}
    D_2&=\xi\frac{\mu_2-\mu_1}{\mu_1}(1-\beta^{T})R\sin{\psi}\cos{\psi},\\
    \label{eq:vel_1_B2}
    B_2&=\cos{\psi}\frac{\mu_2}{\mu_1}+\xi\lambda'(1+(\beta^{T}-1)\cos^2{\psi}),\\
    \label{eq:vel_1_C2}
    C_2&=\xi\frac{\mu_2-\mu_1}{\mu_1}(1+(\beta^{T}-1)\cos^2{\psi}),
\end{align}
\noindent where $\lambda'=\lambda-(\mu_2-\mu_1)/\mu_1$. 

The swimming speed is the derivative with respect to time of the position of the head, therefore to find the position we   need to integrate the ordinary differential equation
\begin{equation}
    \label{eq:pos_i_l_h}
    L_H\frac{\text{d}h_i}{\text{d}t}=U_i(h).
\end{equation}
That equation is separable and can be integrated  to obtain   $h_1(t)$ and $h_2(t)$ as solutions of
\begin{align}
    \label{eq:h1_1_l_h}
    B_1 h_1+C_1 \frac{h_1^2}{2}&=\frac{A_1}{L_H}\Omega t,\\
    \label{eq:h2_1_l_h}
    L_H\left[\frac{D_2B_2-A_2C_2}{D_2^2}\ln{\left(\frac{A_2+D_2h_2}{A_2+D_2}\right)}+\frac{C_2}{D_2}(h_2-1)\right]&=\Omega(t-T_1),
\end{align}
  where $T_1=L_H(2B_1+C_1)/(2A_1\Omega)$ satisfies $h(T_1)=1$, that is, $T_1$ is the time at which the head has fully crossed the interface. Taking the condition $h(0)=0$, we then chose the positive branch of Eq.~\eqref{eq:h1_1_l_h} so the position of the head for $0\leq t\leq T_1$ is given by
\begin{equation}
    \label{eq:h1_2_l_h}
    h_1=\frac{B_1}{C_1}\left[\left(1+2\frac{A_1C_1\Omega t}{L_HB_1^2}\right)^{1/2}-1\right].
\end{equation}
In the second period,  $T_1\leq t\leq T_2$, the position of the head is given implicitly as the solution of
\begin{equation}
    \label{eq:h2_2_l_h}
    \cos{\psi}\ln{\left(1+\frac{\mu_2-\mu_1}{\lambda\mu_1}(h_2-1)\right)}+C_2(h_2-1)=\frac{D_2}{L_H}\Omega(t-T_1),
\end{equation}
  where we used Eqs.~\eqref{eq:vel_1_A2}-\eqref{eq:vel_1_C2} while \mbox{$T_2=L_H\left[\cos{\psi}\ln{(\mu_2/\mu_1)}+C_2\lambda\right]/(D_2\Omega)+T_1$} is the time at which the tail has completely crossed the interface, solution to $h_2(T_2)=1+\lambda$. As the swimming speed is constant, $U=U_0$, for $h\leq0$ and $1+\lambda\leq h$, or equivalently $t\leq0$ and $T_2\leq t$,  the position of the head as a function of time is given by
\begin{equation}
    h(t)=\left\{\begin{array}{ccc}
                 U_0 t  &\quad   &t\leq0,\\
                 h_1(t) &\quad   &0\leq t\leq T_1,\\
                 h_2(t) &\quad   &T_1\leq t\leq T_2,\\
                 1+\lambda+U_0(t-T_2)  &\quad   &T_2\leq t.
                \end{array} \right.
\end{equation}

\subsection{\label{subsec:buoyancy} Buoyancy}
\def\e{{\bf e}}
In order to maintain  a stable two-fluid configuration, the   fluids must have different densities. Experimentally, salt was added to the high-viscosity fluid to slightly increase its density. The effect on the swimmer is to add a buoyancy term in the force balance equation $(\mathbf{F}_D+\mathbf{F}_p+\mathbf{F}_g)\cdot\e_z=0$, where the buoyancy term is given by
\begin{equation}
 \label{eq:buoyancy_d1}
 \mathbf{F}_{g1}=\left[\pi R^2\left(L_{H_1}(\rho_H-\rho_1)+L_{H_2}(\rho_H-\rho_2)\right)+\pi a^2\frac{L_T}{\cos{\psi}}(\rho_T-\rho_1)\right]\mathbf{g},
\end{equation}
\noindent for $0\leq h\leq 1$ and
\begin{equation}
 \label{eq:buoyancy_d2}
 \mathbf{F}_{g2}=\left[\pi R^2L_{H}(\rho_H-\rho_2)+\frac{\pi a^2}{\cos{\psi}}\left(L_{T_1}(\rho_T-\rho_1)+L_{T_2}(\rho_T-\rho_2)\right)\right]\mathbf{g},
\end{equation}
\noindent for $1\leq h\leq 1+\lambda$. Here $\rho_H$ and $\rho_T$ are the effective densities of the head and the helical tail respectively, the density of fluid $i=1,2$ is denoted by $\rho_i$ and $\mathbf{g}$ is the gravitational acceleration. As none of these expressions contain  the swimming speed $U$ explicitly, we can modify Eq.~\eqref{eq:swimming_vel_f-h_2} and Eq.~\eqref{eq:swimming_vel_f-h_4} to include an extra buoyancy term $U_{gi}$ on the right hand side, given by
\begin{align}
 \label{eq:buoyancy_d3}
 U_{g1}&=\frac{(\mathbf{F}_{g1}\cdot\mathbf{e}_z)\cos{\psi}}{\zeta_{\parallel_1}^H\cos{\psi}\left(L_{H_1}+\frac{\mu_2}{\mu_1}L_{H_2}\right)+\zeta_{\perp_1}^{T}L_T\left(1+(\beta^{T}-1)\cos^2{\psi}\right)},\\
 \label{eq:buoyancy_d4}
 U_{g2}&=\frac{\mathbf{F}_{g2}\cdot\mathbf{e}_z\cos{\psi}}{\zeta_{\parallel_2}^H\cos{\psi}L_{H}+\left(\zeta_{\perp_1}^{T}L_{T_1}+\zeta_{\perp_1}^{T}L_{T_2}\right)\left(1+(\beta^{T}-1)\cos^2{\psi}\right)}.
\end{align}
Defining the typical buoyancy speed $u_{gi}^H\equiv\pi R^2(\rho_H-\rho_i)\mathbf{g}\cdot\mathbf{e}_z/\zeta_{\parallel_i}^ H$ we can then write
\begin{align}
 \label{eq:buoyancy_d5}
 U_{g1}&=\frac{u_{g1}^H}{B_1+C_1 h}\left[\left(1+\frac{\rho_1-\rho_2}{\rho_H-\rho_1}h\right)\cos{\psi}+\lambda\frac{a^ 2}{R^2}\frac{\rho_T-\rho_1}{\rho_H-\rho_1}\right],\\
 \label{eq:buoyancy_d6}
 U_{g2}&=\frac{u_{g2}^H}{B_2+C_2 h}\left[\cos{\psi}+(h-1)\frac{a^ 2}{R^2}\frac{\rho_1-\rho_2}{\rho_H-\rho_2}+\lambda\frac{a^2}{R^2}\frac{\rho_T-\rho_1}{\rho_H-\rho_2}\right],
\end{align}
\noindent where $B_i$ and $C_i$ are as given in Eqs.~\eqref{eq:vel_1_B1}, \eqref{eq:vel_1_C1}, \eqref{eq:vel_1_B2} and \eqref{eq:vel_1_C2}.

\section{\label{sec:cont_grad} Continuous interface model}
 At short times after depositing the fluids in the tank, the interface between the two fluid mixture is sharp and the analysis of Section~\ref{subsec:sharp} is appropriate. At later times however, the components responsible for the  increase in the  viscosity of the fluid   mixture diffuse, and therefore so does the viscosity profile (see our measurements in Section~\ref{subsec:diffuse}). We need therefore to include the case where the  distributions of viscosity  is continuous into our calculations for the swimming speed. This is the goal of this section.  In order to do this, we   first consider the case of a three-layer configuration with viscosities $\mu_1$, $\mu_2$ and $\mu_3$ where the layer of viscosity $\mu_2$ has thickness $\Delta$ smaller than both $L_T$ and $ L_H$ (see Fig.~\ref{fig:3fluid}). We   then generalise the result to an $n$-fluid mixture, keeping the overall thickness of the layer between the fluids of viscosity $\mu_1$ and $\mu_3$ fixed. Taking the thicknesses of each fluid layer to be infinitesimally small, we can then obtain the swimming speed for a continuous distribution of viscosity. Finally, we illustrate the impact of some specific distributions on the swimming speed.
 
 \subsection{\label{subsec:3fluid} 3-fluid configuration}
 \begin{figure}
  \begin{tikzpicture}
   
   \draw[very thick] (-1,2)--(2.5,2);
   \draw[very thick] (-1,1)--(2.5,1);
   \path (-1,2.5)--++(180:-0.25) node[right]{$\mu_3$};
   \path (-1,1.5)--++(180:-0.25) node[right]{$\mu_2$};
   \path (-1,0.5)--++(180:-0.25) node[right]{$\mu_1$};
   \draw[line width=2] (0,0)--(1.5,0)--(1.5,3)--(0,3)--cycle;
   \path (0.75,3)--++(90:2.5) node[above]{$\delta\leq h\leq1$};
   \draw[line width=2, rotate=270] (0,0) cos (0.25,0.75) sin (0.5,1.5) cos (0.75,0.75) sin (1,0) cos (1.25,0.75) sin (1.5,1.5) cos (1.75,0.75) sin (2,0) cos (2.25,0.75) sin (2.5,1.5) cos (2.75,0.75) sin (3,0);
   \draw[|-|] (3,0)--node[right]{$L_{H_1}$}++(90:1);
   \draw[|-|] (3,1)--node[right]{$L_{H_2}$}++(90:1);
   \draw[|-|] (3,2)--node[right]{$L_{H_3}$}++(90:1);
   \draw[|-|] (4,1)--node[right]{$H$}++(90:2);
   \draw[|-|] (3,-3)--node[right]{$L_T$}++(90:3);
   \draw[|-|] (-1.5,1)--node[left]{$\Delta$}++(90:1);
   
   \draw[very thick] (7,2)--(10.5,2);
   \draw[very thick] (7,1)--(10.5,1);
   \path (7,2.5)--++(180:-0.25) node[right]{$\mu_3$};
   \path (7,1.5)--++(180:-0.25) node[right]{$\mu_2$};
   \path (7,0.5)--++(180:-0.25) node[right]{$\mu_1$};
   \begin{scope}[shift={(8,3)}]
   \draw[line width=2] (0,0)--(1.5,0)--(1.5,3)--(0,3)--cycle;
   \draw [line width=2, rotate=270] (0,0) cos (0.25,0.75) sin (0.5,1.5) cos (0.75,0.75) sin (1,0) cos (1.25,0.75) sin (1.5,1.5) cos (1.75,0.75) sin (2,0) cos (2.25,0.75) sin (2.5,1.5) cos (2.75,0.75) sin (3,0);
   \path (0.75,-3)--++(90:-2.5) node[below]{$1+\delta\leq h\leq1+\lambda$};
   \draw[|-|] (3,0)--node[right]{$L_H$}++(90:3);
   \draw[|-|] (4,-2)--node[right]{$H$}++(90:4);
   \draw[|-|] (3,-3)--node[right]{$L_{T_1}$}++(90:1);
   \draw[|-|] (3,-2)--node[right]{$L_{T_2}$}++(90:1);
   \draw[|-|] (3,-1)--node[right]{$L_{T_3}$}++(90:1);
   \draw[|-|] (-1.5,-2)--node[left]{$\Delta$}++(90:1);
   \end{scope}
  \end{tikzpicture}
  \caption{\label{fig:3fluid} Helical swimmer crossing a $3$-fluid configuration at different times.}
 \end{figure}
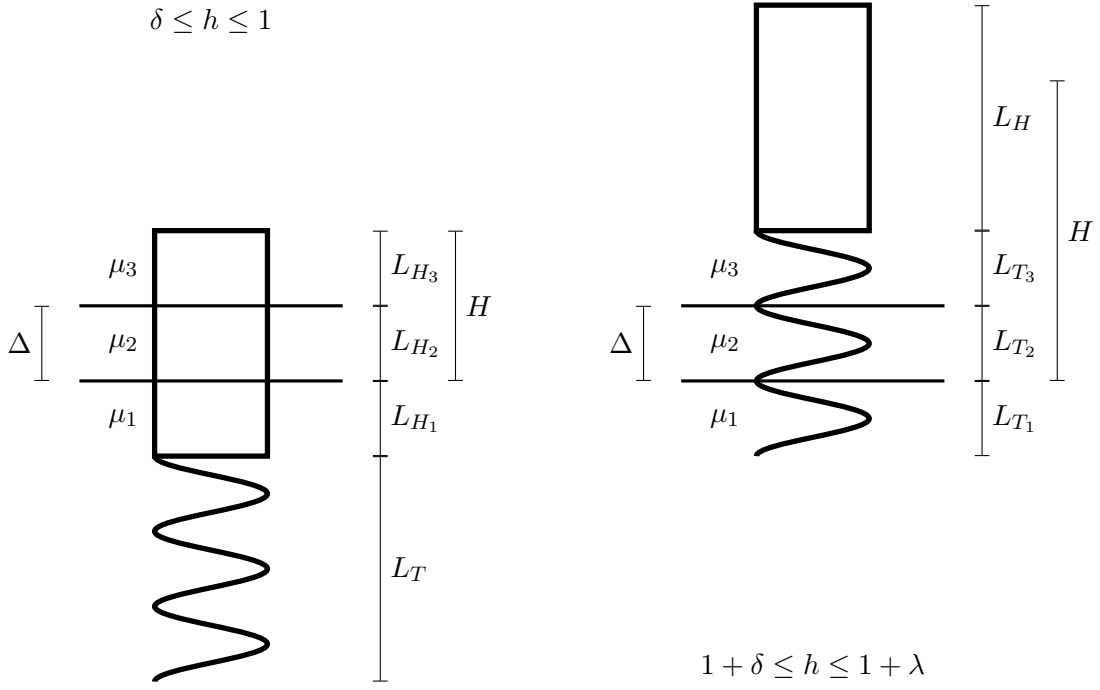
 We first analyse  the 3-fluid case which has a viscosity distribution
 \begin{equation}
  \label{eq:3f_1}
  \mu(z)=\left\{\begin{array}{ccc}
                 \mu_1=\mu'	&\quad	&z\leq0,\\
                 \mu_2	&\quad	&0\leq z\leq\Delta,\\
                 \mu_3=\mu	&\quad	&\Delta\leq z.
                \end{array}\right.
 \end{equation}
 In the head-first approach, the situation is initially the same as in the 2-fluid case until the head reaches the interface between the fluids $2$ and $3$. Hence, the swimming speed  is $U_0$ for $h\leq0$ and $U_1(h)$ for $0\leq h\leq \delta$, where $\delta\equiv\Delta/L_{H} \leq 1$ is the dimensionless thickness of the middle fluid layer (see Fig.~\ref{fig:3fluid}). After the swimmer crosses, its propulsion remains the same   for \mbox{$\delta\leq h\leq1$} whereas the drag has now three contributions
 \begin{equation}
  \label{eq:3f_2}
  \mathbf{F}_D=-\zeta_{\parallel_1}^HL_H\left[\ell_{H_1}+\frac{\mu_2}{\mu_1}\ell_{H_2}+\frac{\mu_3}{\mu_1}\ell_{H_3}\right]U\mathbf{e}_z,
 \end{equation}
   where $\ell_{H_i}\equiv L_{H_i}/L_H$. Continuing, for $1\leq h\leq1+\delta$, the head is now crossing the interface between fluids $2$ and $3$, while the tail is crossing the interface between fluids $1$ and $2$. In this case, the propulsion is the same as in the previous calculation, but the drag is now
 \begin{equation}
 \label{eq:3f_3}
  \mathbf{F}_D=-\zeta_{\parallel_1}^HL_H\left[\frac{\mu_2}{\mu_1}\ell_{H_2}+\frac{\mu_3}{\mu_1}\ell_{H_3}\right]U\mathbf{e}_z.
 \end{equation}
 Once the head has fully crossed the viscosity gradient, the drag is the same as in the previous calculation after, replacing $\mu_2$ by $\mu_3$. On the other hand, the tail is now partially immersed in the three fluids, so  the propulsion is now given by
 \begin{equation}
  \label{eq:3f_4}
\mathbf{e}_z\cdot   \mathbf{F}_p=\frac{-\zeta_{\perp_1}^{T}L_T}{\cos{\psi}}\left[\ell_{T_1}+\frac{\mu_2}{\mu_1}\ell_{T_2}+\frac{\mu_3}{\mu_1}\ell_{T_3}\right]\left[U\left(1+\left(\beta^{T}-1\right)\cos^2{\psi}\right)+\left(\beta^{T}-1\right)\Omega R\sin{\psi}\cos{\psi}\right],
 \end{equation}
 where $\ell_{T_i}\equiv L_{T_i}/L_T$. As $h$ continues increasing, the swimmer reaches the situation in which the tail is crossing only the interface between fluid $2$ and $3$, this is for \mbox{$1+\delta\leq h\leq1+\lambda+\delta$}. In this case the swimming speed is the same as $U_2$ in Eq.~\eqref{eq:swimming_vel_f-h_4}, replacing $\mu_1$, $\mu_2$ by $\mu_2$, $\mu_3$ respectively. Finally for $1+\lambda+\delta\leq h$, the swimming speed $U_0$ is constant. Putting all together, the swimming speed for $h\leq0$ is $U_0$ and $U_1(h)$ as given in Eq.\eqref{eq:swimming_vel_f-h_2} for $0\leq h\leq \delta$. For $\delta\leq h\leq1$ the swimming speed is
 \begin{equation}
  \label{eq:3f_5}
  U_2(h)=\frac{\xi\lambda(1-\beta^{T})\sin{\psi}\cos{\psi}R\Omega}{\cos{\psi}\left(\di1+\frac{\mu_2-\mu_3}{\mu_1}\delta+\frac{\mu_3-\mu_1}{\mu_1}h\right)+\xi\lambda(1+(\beta^{T}-1)\cos^2{\psi})},
 \end{equation}
  where we have used the normalisation condition $\sum{\ell_{H_i}=1}$. For $1\leq h\leq 1+\delta$ the swimming speed is obtained as
 \begin{equation}
  \label{eq:3f_6}
  U_3(h)=\frac{\di \xi\left(\lambda+\frac{\mu_2-\mu_1}{\mu_1}(h-1)\right)(1-\di \beta^{T})\sin{\psi}\cos{\psi}R\Omega}{\di \cos{\psi}\left(\frac{\mu_2}{\mu_1}+\frac{\mu_3-\mu_2}{\mu_1}(h-\delta)\right)+\xi\left(\lambda+\frac{\mu_2-\mu_1}{\mu_1}(h-1)\right)(1+(\beta^{T}-1)\cos^2{\psi})}.
 \end{equation}
 Continuing, for $1+\delta\leq h\leq 1+\lambda$ we have
 \begin{equation}
  \label{eq:3f_7}
  U_4(h)=\frac{\di \xi\left(\lambda+\frac{\mu_2-\mu_3}{\mu_1}\delta+\frac{\mu_3-\mu_1}{\mu_1}(h-1)\right)(1-\beta^{T})\sin{\psi}\cos{\psi}R\Omega}{\cos{\psi}\di\frac{\mu_3}{\mu_1}+\xi\left(\lambda+\frac{\mu_2-\mu_3}{\mu_1}\delta+\frac{\mu_3-\mu_1}{\mu_1}(h-1)\right)(1+(\beta^{T}-1)\cos^2{\psi})},
 \end{equation}
while for $1+\lambda\leq h\leq1+\lambda+\delta$ we have
 \begin{equation}
  \label{eq:3f_8}
  U_5(h)=\frac{\di \xi\left(\lambda\frac{\mu_2}{\mu_1}+\frac{\mu_3-\mu_2}{\mu_1}(h-1-\delta)\right)(1-\beta^{T})\sin{\psi}\cos{\psi}R\Omega}{\di \cos{\psi}\frac{\mu_3}{\mu_1}+\xi\left(\lambda\frac{\mu_2}{\mu_1}+\frac{\mu_3-\mu_2}{\mu_1}(h-1-\delta)\right)(1+(\beta^{T}-1)\cos^2{\psi})}.
 \end{equation}
 Finally, the swimming speed reaches its terminal value $U_0$ for $1+\lambda+\delta\leq h$.
  
\subsection{\label{subsec:n-fluid} $n$-fluid configuration}
\begin{figure}
  \begin{tikzpicture}
   \foreach \x in {1,...,6}{
   \draw[very thick] (-1,-1.5+0.5*\x)--++(0:3.5);
   }
   \foreach \x in {2,...,6}{
   \path (-1,-1.75+0.5*\x)--++(180:-0.25) node[right]{$\mu_\x$};
   }
   \path (-1,1.75)--++(180:-0.25) node[right]{$\mu$};
   \path (-1,-1.25)--++(180:-0.25) node[right]{$\mu'$};
   \draw[line width=2] (0,0)--(1.5,0)--(1.5,3)--(0,3)--cycle;
   \draw[line width=2, rotate=270] (0,0) cos (0.25,0.75) sin (0.5,1.5) cos (0.75,0.75) sin (1,0) cos (1.25,0.75) sin (1.5,1.5) cos (1.75,0.75) sin (2,0) cos (2.25,0.75) sin (2.5,1.5) cos (2.75,0.75) sin (3,0);
   \draw[|-|] (3,0)--node[right]{$L_{H_4}$}++(90:0.5);
   \draw[|-|] (3,0.5)--node[right]{$L_{H_5}$}++(90:0.5);
   \draw[|-|] (3,1)--node[right]{$L_{H_6}$}++(90:0.5);
   \draw[|-|] (3,1.5)--node[right]{$L_{H_7}$}++(90:1.5);
   \draw[|-|] (4,-1)--node[right]{$H$}++(90:4);
   \draw[|-|] (3,-3)--node[right]{$L_{T_1}$}++(90:2);
   \draw[|-|] (3,-1)--node[right]{$L_{T_2}$}++(90:0.5);
   \draw[|-|] (3,-0.5)--node[right]{$L_{T_3}$}++(90:0.5);
   \draw[|-|] (-1.5,-1)--node[left]{$\Delta$}++(90:2.5);
   
   \draw[|-stealth] (4.75,0)--node[below]{$n\rightarrow\infty$}++(0:1);
   \begin{scope}[shift={(8,0)}]
   \draw[very thick, dashed] (-1,1.5)--++(0:3.5);
   \draw[very thick, dashed] (-1,-1)--++(0:3.5);
   \path (-1,0.25)--++(180:0) node[right]{$\tilde{\mu}(z)$};
   \draw[very thick] (2,-3.5)--node[right]{$\mu'$}(2,-1)--(-0.5,1.5)--node[left]{$\mu$}(-0.5,3.5);
   \draw[line width=2] (0,0)--(1.5,0)--(1.5,3)--(0,3)--cycle;
   \draw[line width=2, rotate=270] (0,0) cos (0.25,0.75) sin (0.5,1.5) cos (0.75,0.75) sin (1,0) cos (1.25,0.75) sin (1.5,1.5) cos (1.75,0.75) sin (2,0) cos (2.25,0.75) sin (2.5,1.5) cos (2.75,0.75) sin (3,0);
   \draw[|-|] (3,-1)--node[right]{$H$}++(90:4);
   \draw[|-|] (-1.5,-1)--node[left]{$\Delta$}++(90:2.5);
   \end{scope}
  \end{tikzpicture}
  \caption{\label{fig:nfluid} The $n$-fluid configuration and a linear gradient of length $\Delta$.}
 \end{figure}
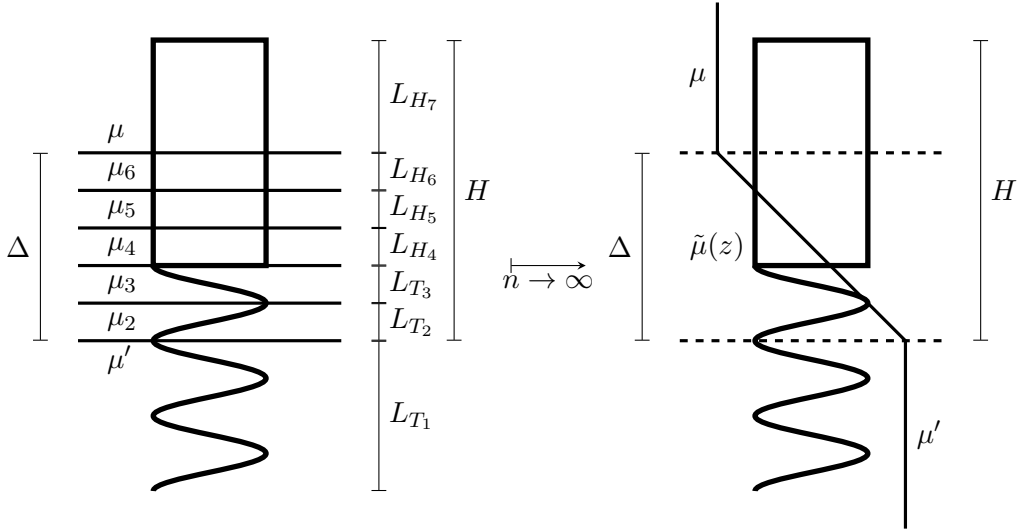

 From the calculations above, we observe that the propulsion and drag involve terms of the form \mbox{$\sum{(\mu_i)/{\mu_1}}\ell_{H_i,T_i}$}. Let us next consider an $n$-layer viscosity profile, given by 
 \begin{align}
  \label{eq:nf_1}
  \mu(z)=\left\{\begin{array}{ccc}
                 \mu_1=\mu'	&\quad	&z\leq0,\\
                 \tilde{\mu}(z)	&\quad	&0\leq z\leq\Delta,\\
                 \mu_n=\mu	&\quad	&\Delta\leq z,
                \end{array}\right.
 \end{align}
 \noindent where $\tilde{\mu}(z)=\mu_j$ for \mbox{$L_{j-2}\leq z \leq L_{j-1}$}, with
 \begin{equation}
  \label{eq:nf_2}
  L_j=\sum_{i=1}^{j}{\Delta_i}.
 \end{equation}
 Here $\Delta_i$ represents the thickness of the $(i+1)$-th layer of fluid, satisfying \mbox{$\sum{\Delta_i}=\Delta \leq L_H$}, with  \mbox{$i=1,2,\ldots,n-1$} (see Fig.~\ref{fig:nfluid}). Using the normalisation conditions $\sum{\ell_{H_i}}=1$ and $\sum{\ell_{T_i}}=\lambda$, and defining the constants \mbox{$b=(1-\beta^{T})\sin{\psi}\cos{\psi}R\Omega$} and \mbox{$c=1+(\beta^{T}-1)\cos^2{\psi}$} we can then  generalise the  expressions  for the swimming speed  in the previous section as follows.

 For $0\leq h\leq \delta$
 \begin{equation}
  \label{eq:nf_3}
  U_1(h)=\frac{\xi\lambda b}{\cos{\psi}\left(\di 1-h+\sum_{i=2}^{k_h}{\frac{\mu_i}{\mu'}\ell_{H_i}}\right)+\xi\lambda c},
 \end{equation}
   where $k_h$ is the lowest integer such that $H<L_{k_h}$. For $\delta\leq h\leq1$ we have
 \begin{equation}
  \label{eq:nf_4}
  U_2(h)=\frac{\xi\lambda b}{\di \cos{\psi}\left(1-h+\sum_{i=2}^{n-1}{\frac{\mu_i}{\mu'}\ell_{H_i}}+\frac{\mu}{\mu'}(h-\delta)\right)+\xi\lambda c},
 \end{equation}
   while  the swimming speed is next obtained for $1\leq h\leq1+\delta$  as
 \begin{equation}
  \label{eq:nf_5}
  U_3(h)=\frac{\di \xi\left(1+\lambda-h+\sum_{i=1}^{k_{h-1}}{\frac{\mu_i}{\mu'}\lambda\ell_{T_i}}\right) b}{\di \cos{\psi}\left(\sum_{i=k_{h-1}}^{n-1}{\frac{\mu_i}{\mu'}\ell_{H_i}}+\frac{\mu}{\mu'}(h-\delta)\right)+\xi\left(1+\lambda-h+\sum_{i=1}^{k_{h-1}}{\frac{\mu_i}{\mu'}\lambda\ell_{T_i}}\right) c}.
 \end{equation}
 Continuing, for $1+\delta\leq h\leq 1+\lambda$ we have
 \begin{equation}
  \label{eq:nf_6}
  U_4(h)=\frac{\di \xi\left(1+\lambda-h+\sum_{i=1}^{n-1}{\frac{\mu_i}{\mu'}\lambda\ell_{T_i}}+\frac{\mu}{\mu'}(h-1-\delta)\right) b}{\di \cos{\psi}\frac{\mu}{\mu'}+\xi\left(1+\lambda-h+\sum_{i=1}^{n-1}{\frac{\mu_i}{\mu'}\lambda\ell_{T_i}}+\frac{\mu}{\mu'}(h-1-\delta)\right) c},
 \end{equation}
  and finally, for $1+\lambda\leq h\leq1+\lambda+\delta$ we obtain the relationship
 \begin{equation}
  \label{eq:nf_7}
  U_5(h)=\frac{\di \xi\left(\sum_{i=k_{h-1-\lambda}}^{n-1}{\frac{\mu_i}{\mu'}\lambda\ell_{T_i}}+\frac{\mu}{\mu'}(h-1-\delta)\right) b}{\di \cos{\psi}\frac{\mu}{\mu'}+\xi\left(\sum_{i=k_{h-1-\lambda}}^{n-1}{\frac{\mu_i}{\mu'}\lambda\ell_{T_i}}+\frac{\mu}{\mu'}(h-1-\delta)\right) c}.
 \end{equation}

 As above, the swimming speed for $h\leq0$ and $1+\lambda+\delta\leq h$ is constant and equal to $U_0$. In the next section we consider the limit $n\rightarrow\infty$ with vanishing thicknesses $\Delta_i$ and fixed sum, $\sum{\Delta_i}=\Delta$. This will give the swimming speed in the case of a continuous distribution of viscosity. 
 
 \subsection{\label{subsec:continuous} The continuous limit}
\begin{figure}
  \begin{tikzpicture}
   \foreach \x in {1,...,7}{
   \draw[very thick] (-1,-1.5+0.5*\x)--++(0:3.5);
   }
   \foreach \x in {1,2,3}{
   \path (-1,0.25+0.5*\x)--++(180:0.25) node[left]{$\mu_\x$};
   \path (-1,0.75-0.5*\x)--++(180:0.25) node[left]{$\mu_{-\x}$};
   }
   \path (-1,2.75)--++(180:0.25) node[left]{$\vdots$};
   \path (-1,-1.25)--++(180:0.25) node[left]{$\vdots$};
   \draw[line width=2] (0,0)--(1.5,0)--(1.5,3)--(0,3)--cycle;
   \draw[line width=2, rotate=270] (0,0) cos (0.25,0.75) sin (0.5,1.5) cos (0.75,0.75) sin (1,0) cos (1.25,0.75) sin (1.5,1.5) cos (1.75,0.75) sin (2,0) cos (2.25,0.75) sin (2.5,1.5) cos (2.75,0.75) sin (3,0);
   \foreach \x in {1,2,3}{
   \draw[|-|] (3,0+0.5*\x)--node[right]{$\Delta_{\x}$}++(90:0.5);
   \draw[|-|] (3,1-0.5*\x)--node[right]{$\Delta_{-\x}$}++(270:0.5);
   }
   \path (3,2.75)--++(180:0.25) node[right]{$\vdots$};
   \path (3,-1.25)--++(180:0.25) node[right]{$\vdots$};
   
   \draw[|-stealth] (4.75,0.5)--node[below]{$n\rightarrow\infty$}++(0:1);
   
   \begin{scope}[shift={(8,0)}]
   \draw[very thick, dashed] (-1,2)--++(0:3.5);
   \draw[very thick, dashed] (-1,-1)--++(0:3.5);
   \draw[very thick] plot [smooth] coordinates{(2,-3) (2,-1) (1.5,0) (0.75,0.5) (0,1) (-0.5,2) (-0.5,4)};
   \path (2,-3)node[right]{$\mu'$}--(2,-1);
   \path (-0.5,2)--(-0.5,4)node[left]{$\mu$};
   \path (-1,0.5)--++(180:0) node[above]{$\mu(z)$};
   \draw[line width=2] (0,0)--(1.5,0)--(1.5,3)--(0,3)--cycle;
   \draw[line width=2, rotate=270] (0,0) cos (0.25,0.75) sin (0.5,1.5) cos (0.75,0.75) sin (1,0) cos (1.25,0.75) sin (1.5,1.5) cos (1.75,0.75) sin (2,0) cos (2.25,0.75) sin (2.5,1.5) cos (2.75,0.75) sin (3,0);
   \draw[dashed] (-1,0.5)--++(0:3.5);
   \path (2,0.5)--++(270:0.25)node[below]{$z=0$};
   \draw[|-|] (3,0.5)--node[right]{$H$}++(90:2.5);
   \draw[|-|] (-1.5,-1)--node[left]{$\Delta$}++(90:3);
   \end{scope}
  \end{tikzpicture}
  \caption{\label{fig:continuous} Continuous viscosity gradient. In this case $\Delta$ is the length scale for the transition region. For a diffuse layer for example, $\Delta\sim\sqrt{Dt}$, where $D$ is the diffusivity of $\mu(z)$.}
 \end{figure}
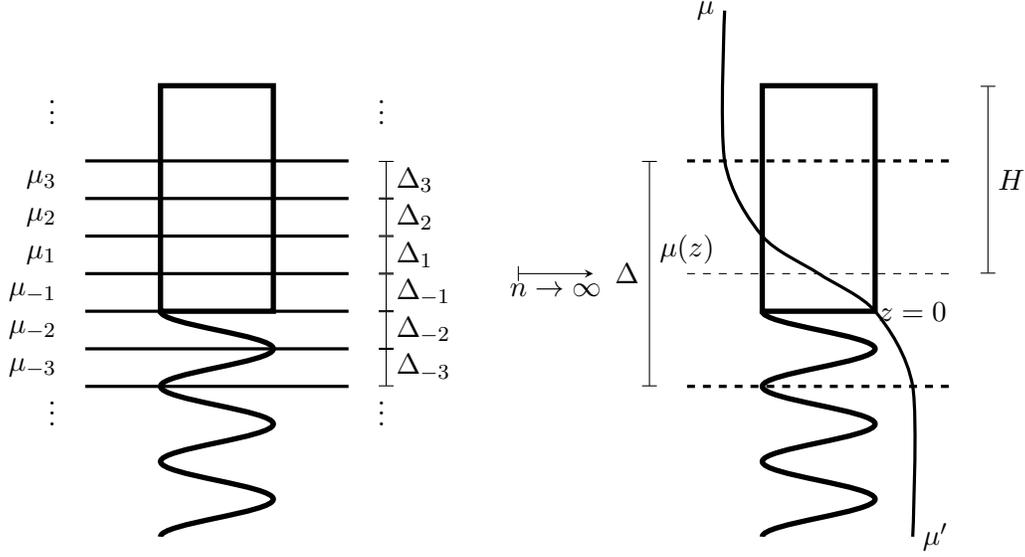
We consider now the limit in which the number of fluid layers tend to infinity while keeping fixed the total thickness of the layer between the semi-infinite fluids of viscosity $\mu$ and $\mu'$. In this case, the sums in Eqs.~\eqref{eq:nf_3}-\eqref{eq:nf_7} may be replaced by continuous integrals as 
 \begin{equation}
  \label{eq:cont_1}
  \sum_{i=k_{h_1}}^{k_{h_2}}{\frac{\mu_i}{\mu'}\ell_{H_i}}\rightarrow\int_{h_1}^{h_2}{\frac{\tilde{\mu}(z)}{\mu'}\,\text{d}z},
 \end{equation}
  where $\tilde{\mu}(z)$ represents any continuous viscosity distribution such that $\tilde{\mu}(0)=\mu'$ and $\tilde{\mu}(\delta)=\mu$;  similarly we may also replace $\lambda\ell_{T_i}\rightarrow \,\text{d}z$ in the corresponding sums. 
  
   Substituting  Eq.~\eqref{eq:cont_1} into Eqs.~\eqref{eq:nf_3}-\eqref{eq:nf_7} gives the swimming speed for a swimmer crossing from a region of viscosity $\mu'$ to a region of different viscosity $\mu$, with a continuous transition region of thickness $\Delta\leq L_H, L_T$. In order to generalise  this result for any continuous viscosity profile, we consider the configuration in Fig.~\ref{fig:continuous}. Repeating the calculation steps of the previous section and taking the continuous limit, we obtain the swimming speed as
 \begin{equation}
  \label{eq:cont_2}
  U(h)=\frac{\di \xi\left(\int_{h-1-\lambda}^{h-1}{\frac{\mu(z)}{\mu'}\,\text{d}z}\right)(1-\beta^{T})\sin{\psi}\cos{\psi}R\Omega+(\mathbf{F}_g\cdot\mathbf{e}_z\cos{\psi}/\zeta_{\parallel}^{H}L_H)}{\di \cos{\psi}\left(\int_{h-1}^{h}{\frac{\mu(z)}{\mu'}\,\text{d}z}\right)+\xi\left(\int_{h-1-\lambda}^{h-1}{\frac{\mu(z)}{\mu'}\,\text{d}z}\right)(1+(\beta^{T}-1)\cos^2{\psi})},
 \end{equation}
 \noindent where now $\mu(z)$ stands for an arbitrary continuous viscosity distribution satisfying the condition \mbox{$\mu(z\rightarrow-\infty)\rightarrow\mu'$} and \mbox{$\mu(z\rightarrow\infty)\rightarrow\mu$}. Notice that the transition between $\mu'$ and $\mu$ does not have to be monotonic, but in order to apply this model to the experimets presented in Section~\ref{sec:results}, we will chose a monotonic viscosity profile as the linear and diffuse distributions depicted in Figures~\ref{fig:nfluid} and~\ref{fig:continuous}, respectively. Again, the dynamics in cases I and III are obtained by setting $\mu'<\mu$ and $\mu<\mu'$, respectively. Notice as well that we included the buoyancy term in Eq.~\eqref{eq:cont_2}, which in the continuous limit is given by
 \begin{equation}
  \label{eq:cont_3}
  \mathbf{F}_g=\pi R^2L_H\left\{\int_{h-1}^{h}{[\rho_H-\rho(z)]\,\text{d}z}+\frac{a^2}{R^2\cos{\psi}}\int_{h-1-\lambda}^{h-1}{[\rho_T-\rho(z)]\,\text{d}z}\right\}\mathbf{g},
 \end{equation}
 \noindent where $\rho(z)$ is the continuous density distribution of the fluid mixture, which might be independent from $\mu(z)$. The position of the swimmer as a function of time $h(t)$ can then be obtained by solving the ordinary differential equation
 \begin{equation}
     L_H\frac{{\rm d} h}{{\rm d} t}=U(h).
 \end{equation}
Note that this might not be solvable analytically for an arbitrary viscosity profile, but it is straightforward to do numerically.

\subsection{\label{subsec:tail_first} Head-first vs tail-first approach}
 The calculations above were all carried out in the case where the head of the swimmer crosses the interface first (pusher mode). 
Since the motion is dominated by viscosity, expressions for the speed when the swimmer approaches the interface with the tail first  may then be obtained by time reversal. Indeed, time reversal of Stokes flow corresponds to the map \mbox{$\{U,\Omega,h,\mu',\mu\} \mapsto\{-U,-\Omega,-h,\mu,\mu'\}$}, therefore the tail-first approach is obtained by evaluating Eq.~\eqref{eq:cont_2} at $h'=-h+\lambda+1$, where the translation $\lambda+1$ comes from the fact that $h'=0$ corresponds to the moment when the tail meets the fluid interface. Note that we need to be careful with the sign of the gravitational field, and to remember that in the experiments, the high viscosity fluid always sits at the bottom, so $\mathbf{F}_g\cdot\mathbf{e}_z<0$ when the swimmer crosses from high to low viscosity and vice-versa.
 
 \subsection{\label{subsec:parameter}Model predictions: Parameter dependence}
 
Before  comparing the model predictions with the experimental data, we  explore the impact of the different parameters of the problem on the swimming speed. One of the advantages of the model developed here is that it allows us to explore a wider set of conditions than those attainable experimentally. In this and the following section we will adopt the convention \mbox{$\mu'<\mu$} for clarity. This means that, when the swimmer moves up the gradient, \mbox{$\mu(h\rightarrow-\infty)\rightarrow\mu'$} and \mbox{$\mu(h\rightarrow\infty)\rightarrow\mu$}. When the swimmer moves down the gradient, we swap $\mu'$ and $\mu$. Furthermore, we will use the terminology defined in Table~\ref{tab:2} to refer to the different swimming conditions.

To simplify the interpretation of predictions, we first neglect  buoyancy; in such a case, the time-reversal symmetry between the head-first and tail-first is exact. The dimensional parameters we are left with are the sizes of the swimmer ($L_H$, $r_H$, $L_T$, $r_T$) and  the pitch  of the helical tail ($P_T$). Another length scale is the size of the transition region ($\Delta$) or, equivalently, the time at which the experiment is performed after the two-fluid mixture is set up. We will keep the proportions of the head fixed as well as the pitch and thickness of the helical filament so that we are left with two dimensionless parameters: $\delta=\Delta/L_H$ and $\lambda=L_h/L_H$, which quantify  the relative size of the transition region and the size of  the tail compared to the head of the swimmer. Finally, we will also consider variations in the viscosity ratio of the initial two mixture fluid, i.e.~$\mu/\mu'$.
 
 \begin{figure}
     \includegraphics[width=\textwidth]{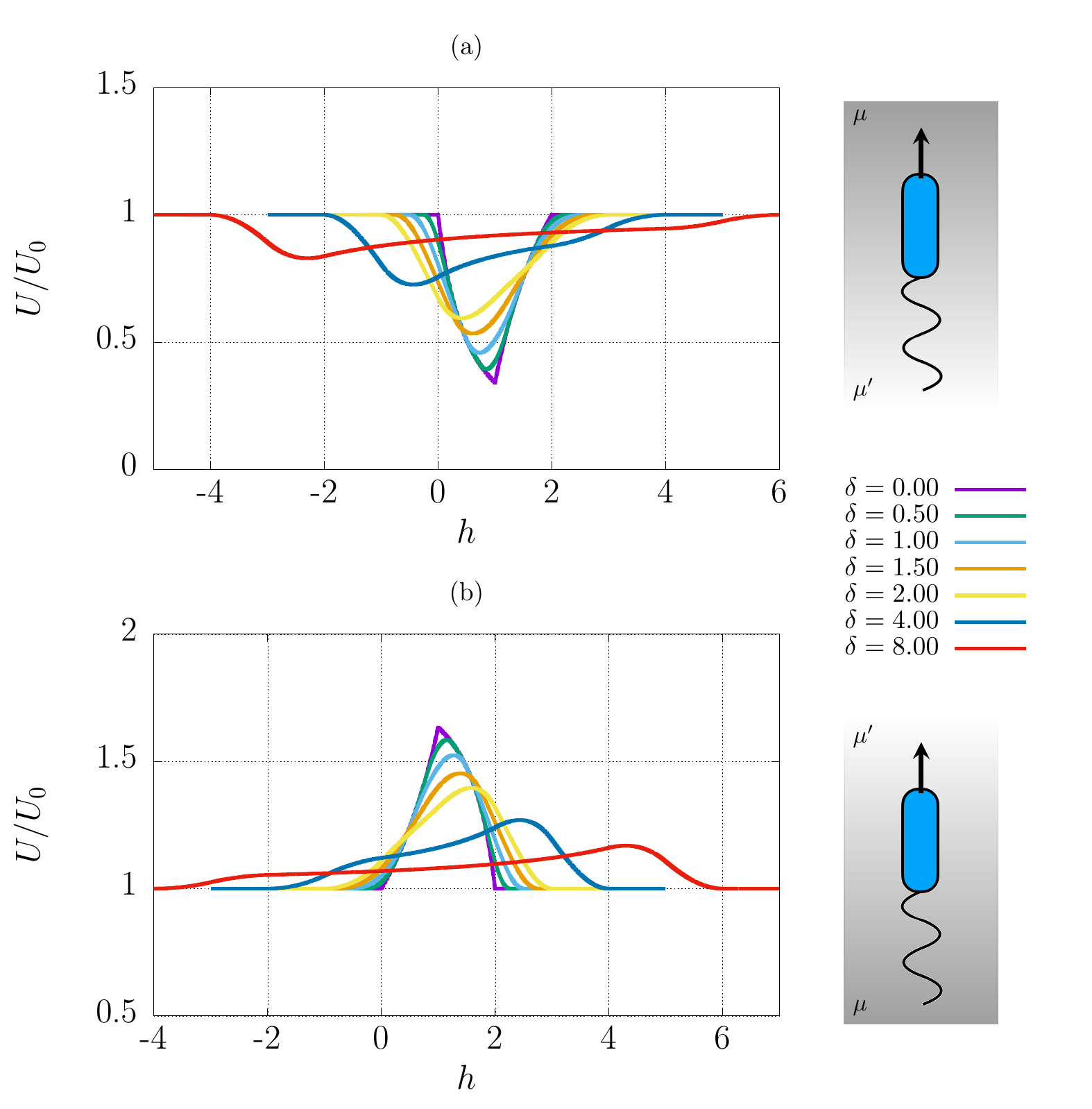}
     \caption{\label{fig:v_linear_c6}Swimming speed in a linear viscosity gradient for different widths of the transition region. (a) Case I, (b) Case III.          The viscosities of the semi-infinite fluids on either side of the interface  are given by the experimental values, i.e.~\mbox{$\mu'=\mu_-=0.55~\text{Pa}\cdot\text{s}$} and \mbox{$\mu=\mu_+=2.74~\text{Pa}\cdot\text{s}$}. The length of the tail and head are the same, i.e.~$\lambda=1$, and the swimmer is neutrally buoyant in both fluids. The diagrams on the right indicate the direction of motion, with dark and light gray representing high ($\mu$) and low viscosity ($\mu'$) respectively. The values are normalised by the terminal speed $U_0$.}
 \end{figure}
 
In Fig.~\ref{fig:v_linear_c6} we first show the swimming speed as a function of the distance between the head and the fluid-fluid interface (normalised by the initial, and terminal, speed $U_0$). We assume that the viscosity varies linearly between the two experimental values, \mbox{$\mu'=\mu_-=0.55~\text{Pa}\cdot\text{s}$} and \mbox{$\mu=\mu_+=2.74~\text{Pa}\cdot\text{s}$}, and we take the head and tail to have the same lengths, i.e.~$\lambda=1$. In panel (a) we consider Case I, and the fluid interface is located at $h=0$. The speed is constant when the swimmer is completely immersed on the low viscosity fluid. As the head crosses the interface, the drag increases but the propulsion stays the same, therefore the speed decreases. Once the tail meets the interface, the propulsion starts to increase thereby compensating the drag, and thus the speed increases until it plateaus back to a constant speed. Panel (b) shows the speed of the swimmer in Case III. Here the drag reduces as the head traverses the interface, hence the swimming speed increases until the tail meets the interface, when the propulsion  starts decreasing, compensating for the lower drag. This continues until the swimmer is completely immersed in the top fluid, at which point the speed reaches a new constant value.
 
The speed of the swimmer in Case II (tail-first) may be obtained by reflecting Fig.~\ref{fig:v_linear_c6}~(b) on the vertical axis $h=1$, with $h$ now measured from the tip of the tail to the interface. Similarly for Case IV, we reflect Fig.~\ref{fig:v_linear_c6}~(a). Notice that for $\lambda=1$, this reflection corresponds to the transformation \mbox{$h\rightarrow -h+1+\lambda$}. In general, for arbitrary $\lambda$, we need to reflect on the axis \mbox{$h=(1+\lambda)/2$} in order to obtain the swimming speed in the tail forward scenarios. Therefore, the behaviour of the swimming speed is reversed in the tail first approach: the swimming speed increases when it crosses from low to high viscosity and vice-versa.
 
\begin{figure}
     \includegraphics[width=\textwidth]{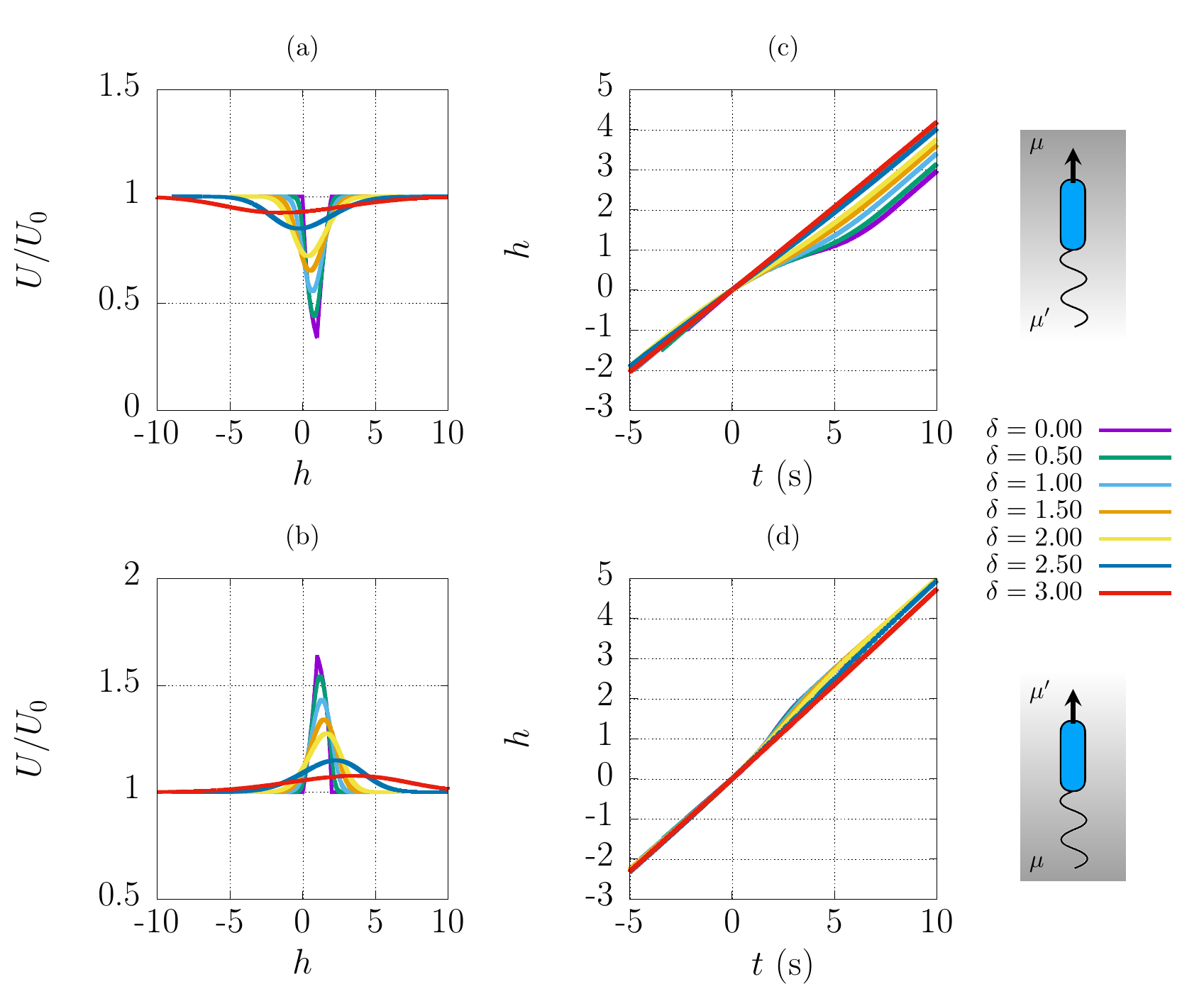}
     \caption{\label{fig:v_diff_c6}Swimming speed in a diffusive viscosity gradient for different widths of the transition region. (a) Case I, (b) Case III.
         The   parameters $\{\mu/\mu',\lambda\}$ are the same as in Fig.~\ref{fig:v_linear_c6}.}
\end{figure}

We next show in Fig.~\ref{fig:v_diff_c6}~(a) and (b)   analogous graphs to those in Fig.~\ref{fig:v_linear_c6} for a diffusive viscosity gradient with different values of the transition length scale \mbox{$\Delta=2\sqrt{2Dt}$}. The behaviour is seen to be qualitatively  the same as in the linear case. We also show the  position of the swimmer as a function of time for the same conditions in Fig.~\ref{fig:v_diff_c6}~(c) and (d). The evolution of the position is seen to not strongly depend on the width of the viscosity transition region, with the strongest variability occurring for Case I [panel (c)].
 
 \begin{figure}
     \includegraphics[width=\textwidth]{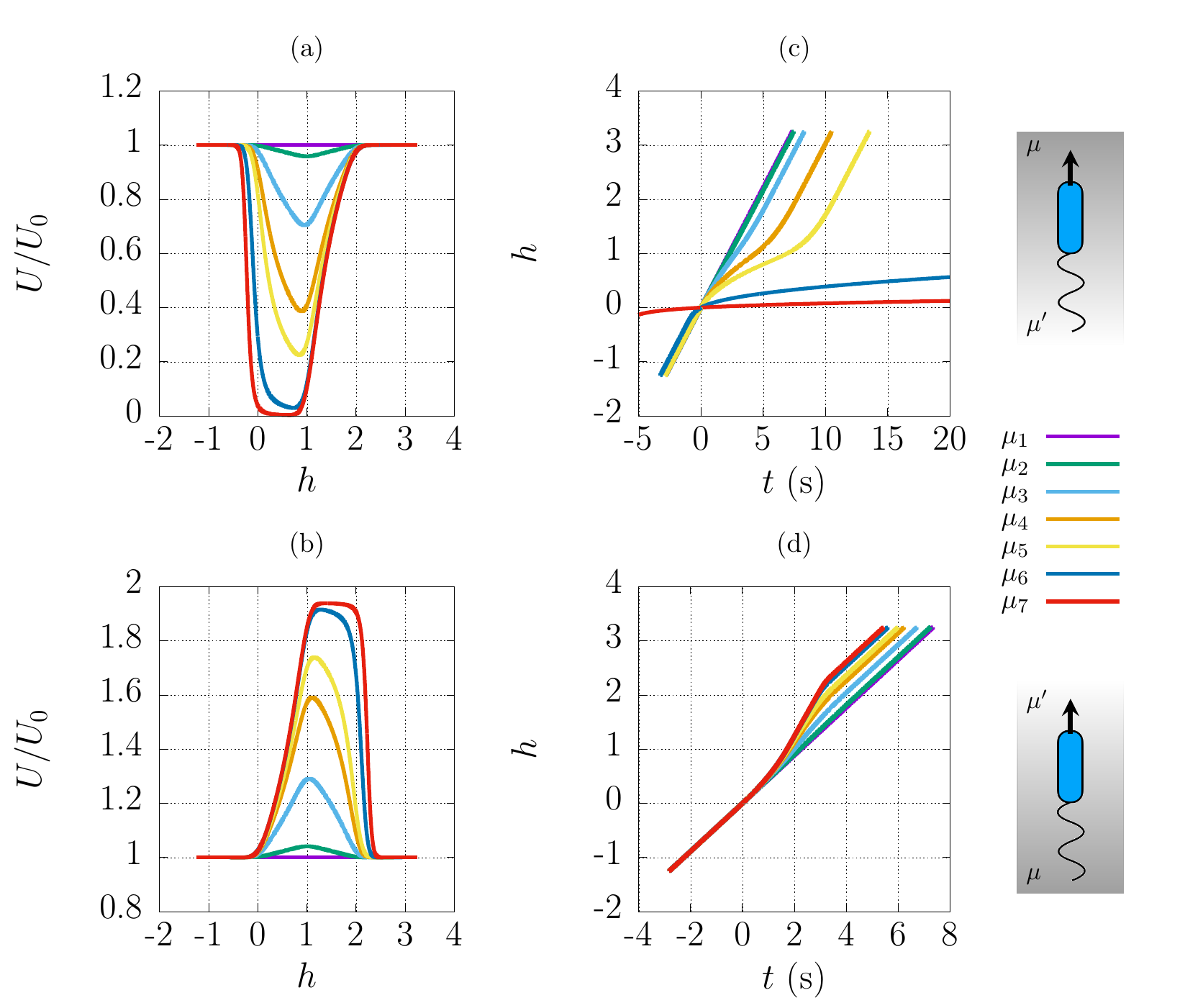}
     \caption{\label{fig:v_pos_diff_c6_mu}Swimming speed and position in a diffusive viscosity gradient for different values of the viscosity ratio: $\mu_i/\mu'=1.0$, $1.1$, $2.0$, $5.0$, $10.0$, $100.0$ and $1000.0$. (a), (c) Speed and position, Case I; (b), (d) Speed and position, Case III. The width of the transition region is $\delta=\Delta/L_H=0.25 $, the viscosity of the top fluid is \mbox{$\mu'=\mu_-=0.55~\text{Pa}\cdot\text{s}$} and $\lambda=1$.}
 \end{figure}
  
The behaviour of the swimmer dynamics with increasing values of the viscosity ratio ($\mu/\mu'$)  is shown in   Fig.~\ref{fig:v_pos_diff_c6_mu} (for values $\delta=0.25$ and $\lambda=1$).   As could have been expected,   the viscosity ratio greatly influences the amount by which the speed of the swimmer changes as it crosses the interface. In particular, when motion occurs  from low to high viscosity, it is possible for the swimmer to spend an arbitrarily long time crossing the gradient if the viscosity ratio is large (see Fig.~\ref{fig:v_pos_diff_c6_mu}~(a) and (c)).  In contrast, the effect on the dynamics when the swimmer crosses from high to low viscosity is less pronounced, with the time spent crossing the gradient decreasing by about 30$\%$ when the viscosity ratio is a thousand times larger.
 
\begin{figure}
     \includegraphics[width=\textwidth]{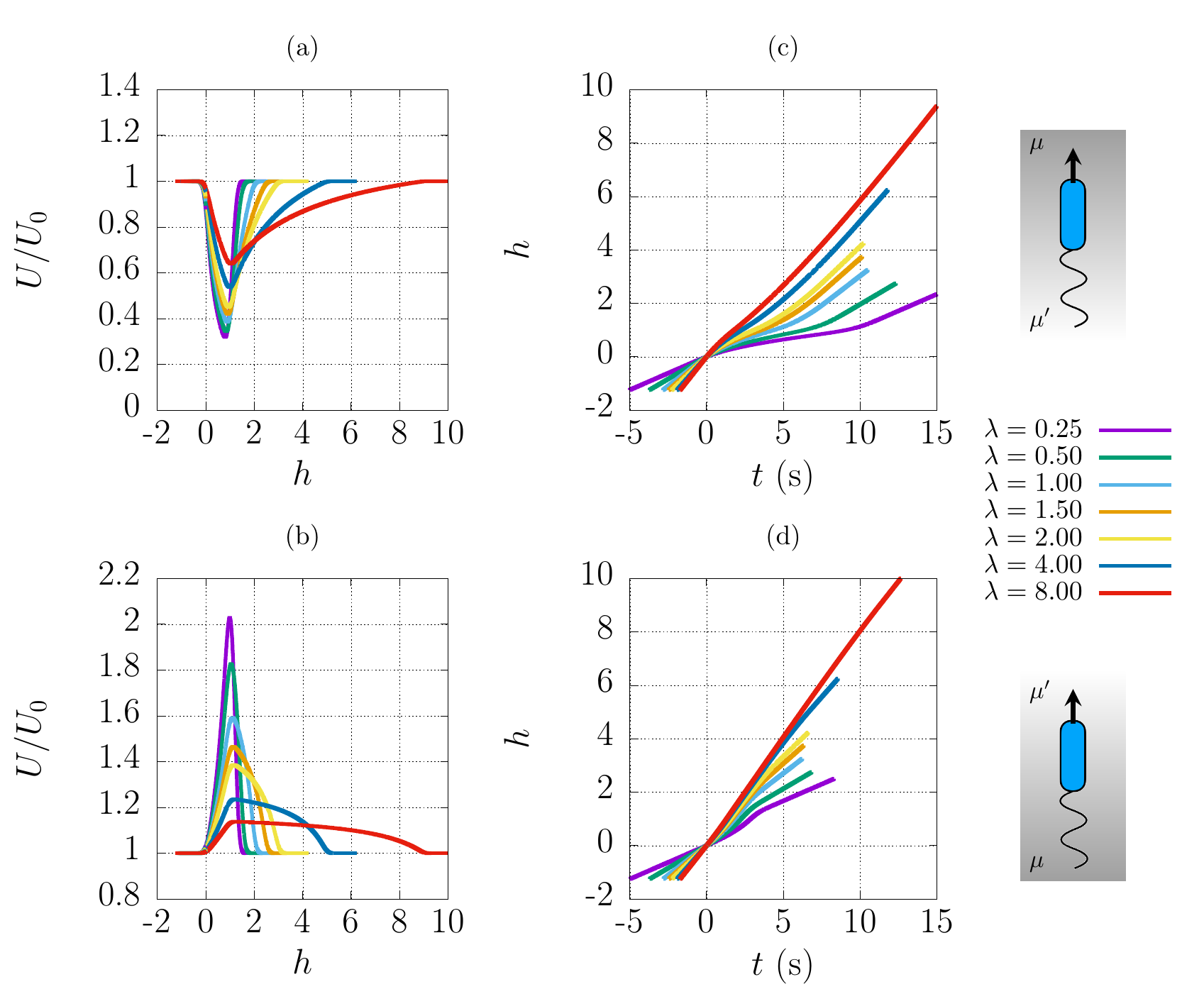}
     \caption{\label{fig:v_pos_diff_c6_l}Swimming speed and position in a diffusive viscosity gradient for different values of the tail to head size ratio, $\lambda$. (a), (c) Speed and position, Case I; (b), (c) Speed and position, Case III. The width of the transition region is $\delta = \Delta/L_H=0.25 $ and the viscosity ratio is the same as in Fig.~\ref{fig:v_linear_c6} (experimental values).}
\end{figure}

The impact of the dimensionless length of the tail ($\lambda$) is next plotted in Fig.~\ref{fig:v_pos_diff_c6_l}, for which we use the same size of the transition region and the viscosity values used in Fig.~\ref{fig:v_linear_c6}. The speed of the swimmer is seen to  increase with the length of its tail, as expected since it is the tail that  generates  propulsion. We further observe  that the time the swimmer takes to cross the interface decreases with $\lambda$ in both cases.
 
We  also observe that the speed changes less for swimmers with long tails; indeed, if the tail is much larger than the size of the transition region, then the propulsion remains almost the same during a crossing event. Therefore, we expect swimmers with short tails to be less efficient at crossing the interface.
 
\section{\label{sec:experiments} Comparison with experiments}
  
\subsection[Low to high viscosity]{\label{subsec:low-high_h-f} Positive viscosity gradient}
  
\begin{figure}
   \includegraphics[height=7cm]{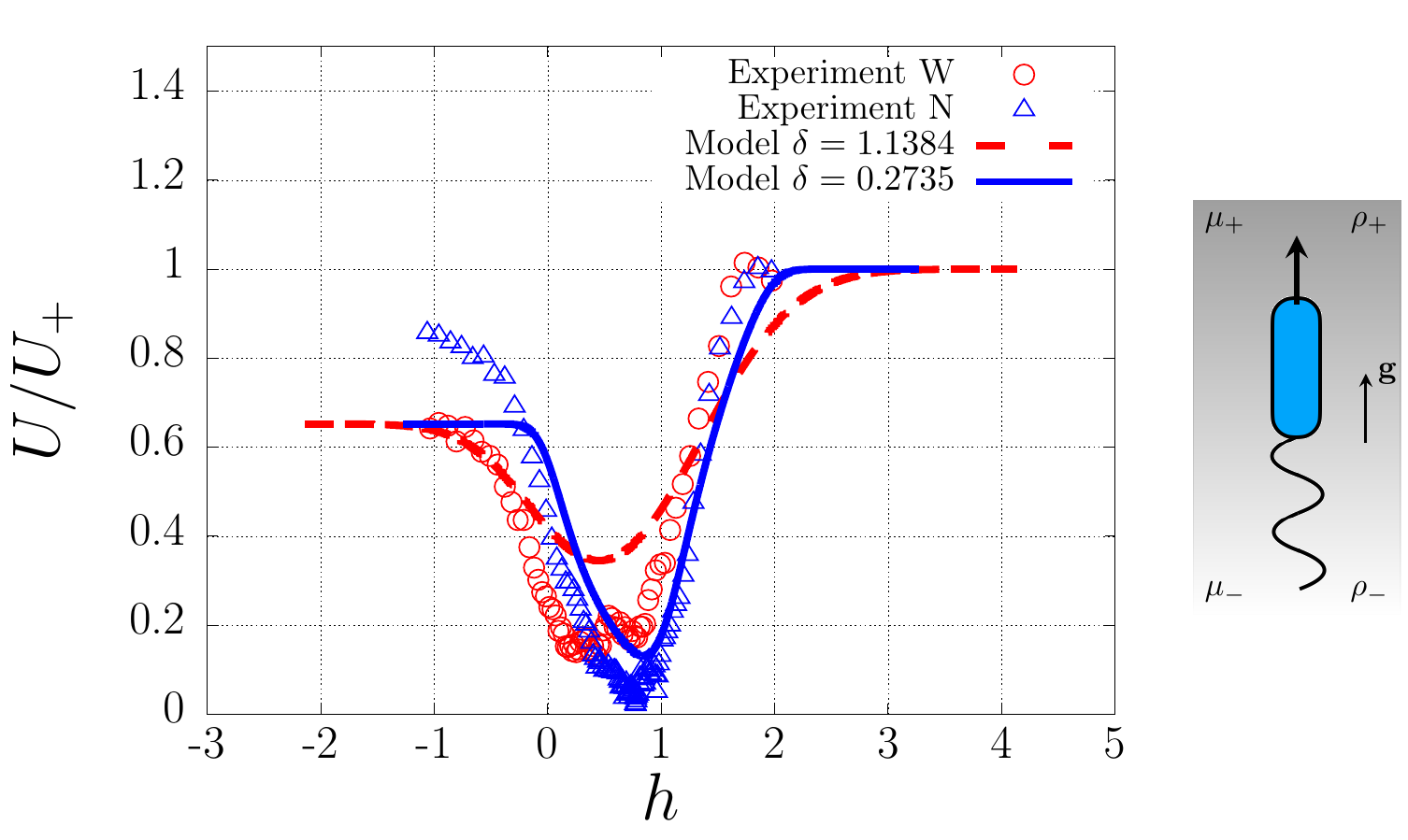}
   \caption{\label{fig:case_6_h-f_l-h}Comparison between experiments and data for Case I (i.e.~head-first (pusher) swimmer crossing a positive viscosity gradient): Speed of the swimmer as a function of the dimensionless position of the head $h$ with respect to the interface located at $h=0$. The speed is normalised by the speed in the fluid of high viscosity, $U_{+}=U(h\rightarrow\infty)$. We display the measurements early after the viscosity gradient has been set up ([N]arrow gradient): triangles (experiment) and blue solid line (model), and sixteen hours after ([W]ide gradient): circles (experiment) and red dashed line (model).}
\end{figure}

In this section we compare the predictions of our model to the experiments of Section~\ref{sec:results}.   

We begin with the situation where the swimmer crosses the interface from the low to the high-viscosity domain and we start by analysing Case I, in which the head approaches  the interface first. The swimmer moves at a constant speed when it is completely immersed in fluid $1$. When the head reaches the interface, based on the results from the previous section, we expect the speed to decrease due to an increase in drag experienced by the head. When the tail meets the interface, the propulsion should then start to increase until it compensates the higher drag, achieving a constant terminal speed. Indeed, both Eq.~\eqref{eq:cont_2} and the experimental data confirm this. 
We plot in Fig.~\ref{fig:case_6_h-f_l-h} the speed of the swimmer (normalised by the swimming speed in the high-viscosity fluid) as a function of the dimensionless position of the head $h$; the swimmer starts from the low-viscosity fluid (\mbox{$\mu'=\mu_-=0.55\,\text{Pa}\cdot\text{s}$}) and approaches the interface  head-first. We compare the experimental data against our model, Eq.~\eqref{eq:cont_2}, using the experimental parameters, i.e.~\mbox{$\lambda=1$}, \mbox{$\mu=\mu_+=2.74~\text{Pa}\cdot\text{s}$}, \mbox{$\rho'=\rho_-=1310~\text{kg/m}^3$}, \mbox{$\rho=\rho_+=1370~\text{kg/m}^3$} and an average density \mbox{$\rho_{swimmer}=1270~\text{kg/m}^3$} for the swimmer. The size of the transition region $\delta$ is obtained by the procedure described in Section~\ref{subsec:diffuse}.  With no additional fitting parameters, we observe that our model matches the experiments very well specially at early times [in the narrow viscosity gradient, indicated by (N)]. The model is able to predict also that the speed reduction decreases with an increase in the thickness of the fluid interface, $\delta$. 

  \begin{figure}
      \includegraphics[height=7cm]{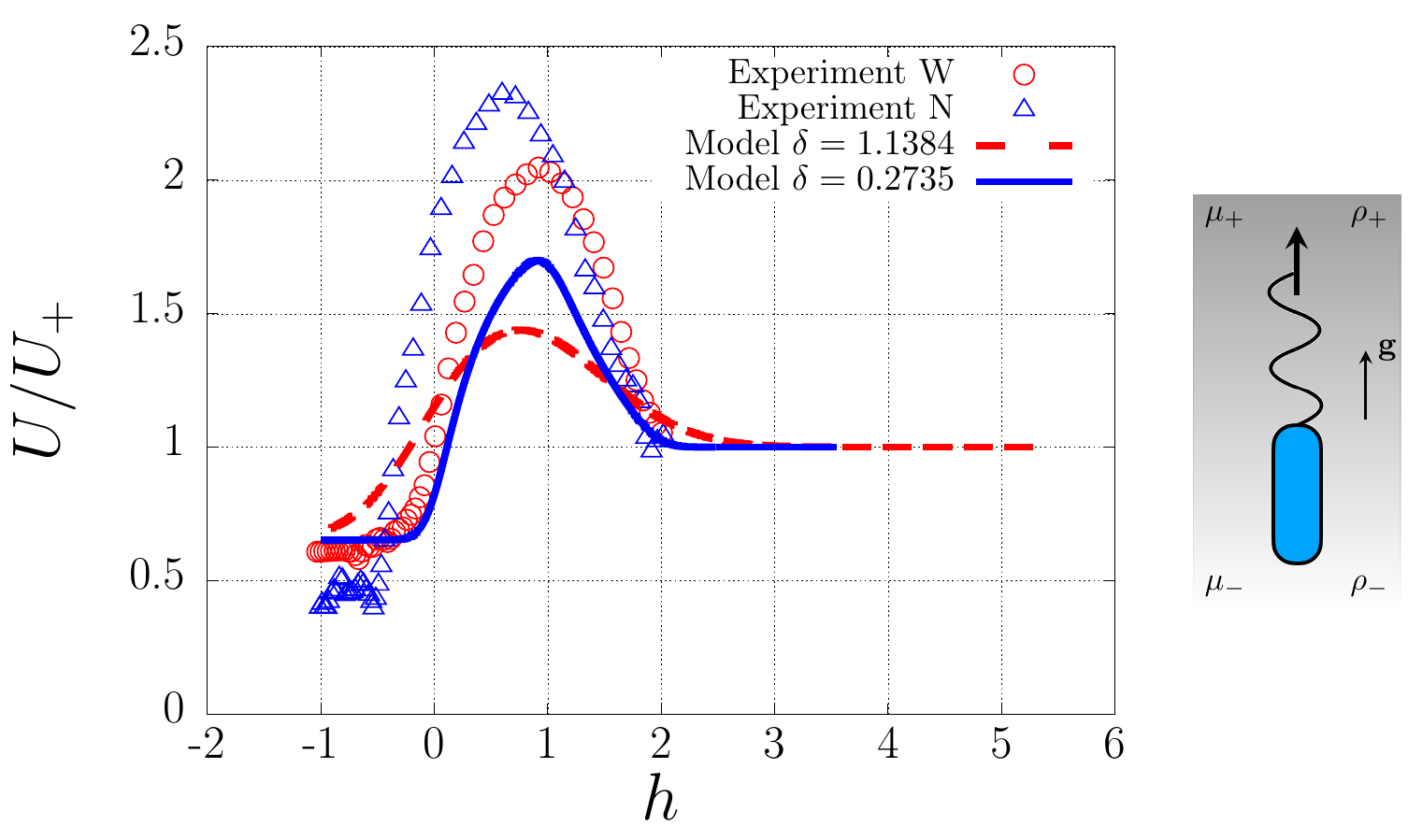}
      \caption{\label{fig:case_6_t-f_l-h}Comparison between experiments and data for  Case II (tail-first (puller) swimmer in a positive viscosity gradient). The speed of the swimmer is plotted as a function of the dimensionless position of the tail $h$ with respect to the interface located at $h=0$, and 
      the speed is normalised by that in the high-viscosity fluid,  $U_{+}=U(h\rightarrow\infty)$. 
      We show the measurements  soon after the viscosity gradient has been set up ([N]arrow gradient case): triangles (experiment) and blue solid line (model), as well as  sixteen hours after the start of the experiment ([W]ide gradient case): circles (experiment) and red dashed line (model).}
  \end{figure}

  In contrast to the head first approach, in Case II (tail-first motion) we expect the swimmer to speed up as it traverses the viscosity gradient, this as a result of an increase in propulsion. When the head meets the interface, the drag increases and the speed should decrease, until the swimmer achieves a constant speed. Both our model in Eq.~\eqref{eq:cont_2} and the experiments agree with this behaviour, as shown  in Fig.~\ref{fig:case_6_t-f_l-h}; we use the same values for the parameters $\lambda$ and $\rho_{swimmer}$ as in Figs.~\ref{fig:case_6_h-f_l-h} and \ref{fig:case_6_h-f_h-l}. The position $h$ is now measured from the tip of the tail to the interface. We swap the values of the viscosities and densities $\mu'=\mu_+$, $\mu=\mu_-$, $\rho'=\rho_+$ and $\rho=\rho_-$, to be consistent with reversibility and the speed is still normalised by $U_{+}$. Here, we also observe that the model can reproduce the experimental behaviour, especially  at early times. It can also capture the reduction of the increase in speed with the width of the transition region $\delta$.
  
 \subsection[High to low viscosity]{\label{subsec:high-low_h-f} Negative viscosity gradient}
 
 We now move on to the case where the swimmer crosses the interface from the high to the low-viscosity region, that is swimming down the gradient. To compare the results against our model, Eq.~\eqref{eq:cont_2}, we use the same set of the parameters: \mbox{$\mu'=\mu_+=2.74~\text{Pa}\cdot\text{s}$}, \mbox{$\mu=\mu_-=0.55\,\text{Pa}\cdot\text{s}$}, \mbox{$\rho'=\rho_+=1370~\text{kg/m}^3$}, \mbox{$\rho=\rho_-=1310~\text{kg/m}^3$}, \mbox{$\rho_{swimmer}=1270~\text{kg/m}^3$} and \mbox{$\lambda=1$}. The width of the transition region, $\delta$, is obtain as before by fitting Eq.~\eqref{eq:diff_2} to the experimental data.
 
\begin{figure}
      \includegraphics[height=7cm]{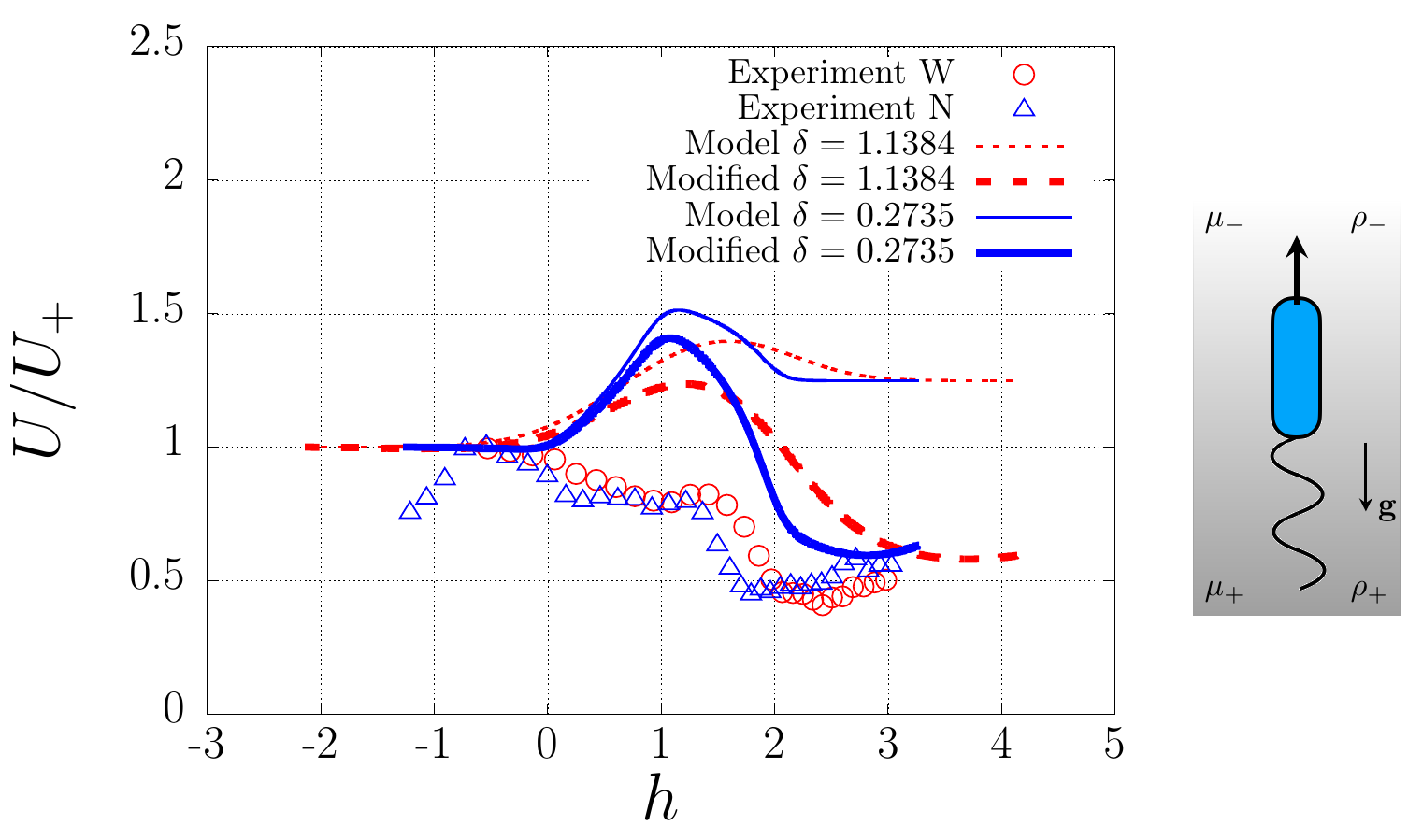}
      \caption{\label{fig:case_6_h-f_h-l} Comparison between experiments and data for  Case III (i.e.~head-first swimmer (pusher) crossing a negative viscosity gradient). We plot the dimensionless speed of the swimmer as a function of the dimensionless position of the head $h$ measured relative to  the interface located at $h=0$. The speed of the swimmer is  normalised by the speed in the high viscosity fluid, $U_{+}=U(h\rightarrow-\infty)$. We display the measurements early after the viscosity gradient has been set up ([N]arrow gradient): triangles (experiment) and blue solid line (model), and sixteen hours after ([W]ide gradient): circles (experiment) and red dashed line (model). Thin lines represent the original model not taking into account entrainment while the thick lines show the modified model with variable buoyancy and $\alpha_{max}=0.1$.}
 \end{figure}

In Case III (head-first) we expect to see a behaviour opposite to that of Case I. Again, the swimmer travels at constant speed when it is completely immersed in the high viscosity fluid. As predicted by our model, when the head crosses the interface we would expect the drag experienced by the swimmer to decrease, resulting in an increase in the swimming speed. Then, when the tail reaches the interface, the propulsion should decrease, compensating for the reduced drag, until the swimming speed reaches a constant value.  However, the experimental data show a completely different behaviour. We plot in Fig.~\ref{fig:case_6_h-f_h-l}   a comparison between the experimental data and the predictions of Eq.~\eqref{eq:cont_2} (theoretical predictions are shown in thin lines). In the  experiments, the swimmer seems to maintain a constant speed as the head crosses the interface. When the tail then meets the interface, the speed starts decreasing. It is only once the swimmer has fully crossed and is completely immersed in the low viscosity fluid that the speed starts to increase. 

\begin{figure}
    \centering
    \subfigure[]{\includegraphics[height=6cm]{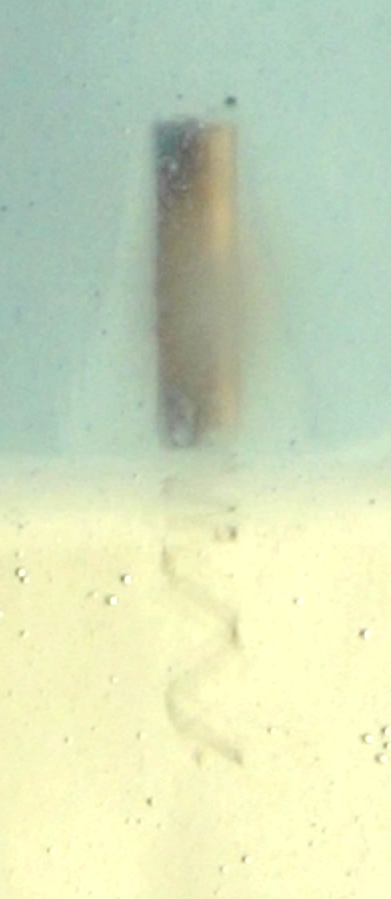}\quad
        \includegraphics[height=6cm]{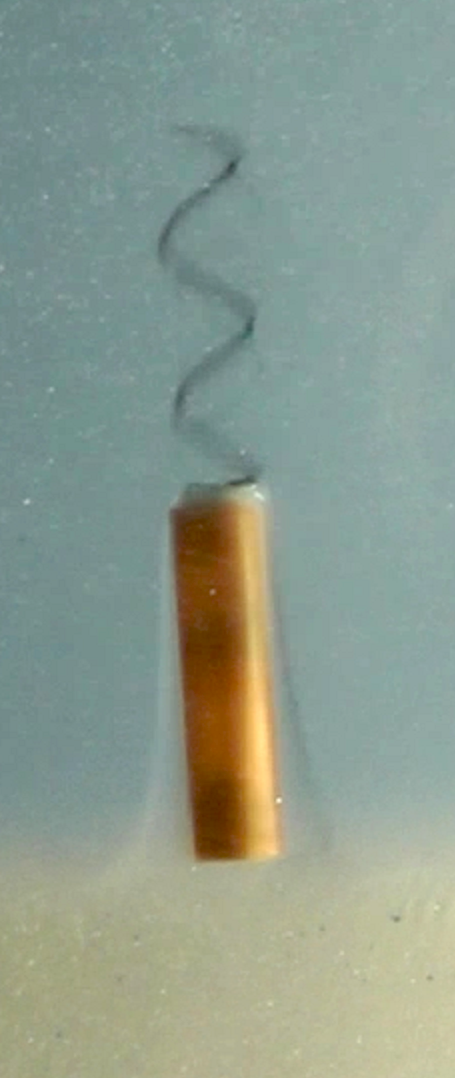}}\quad
    \subfigure[]{\includegraphics[height=6cm, clip=true]{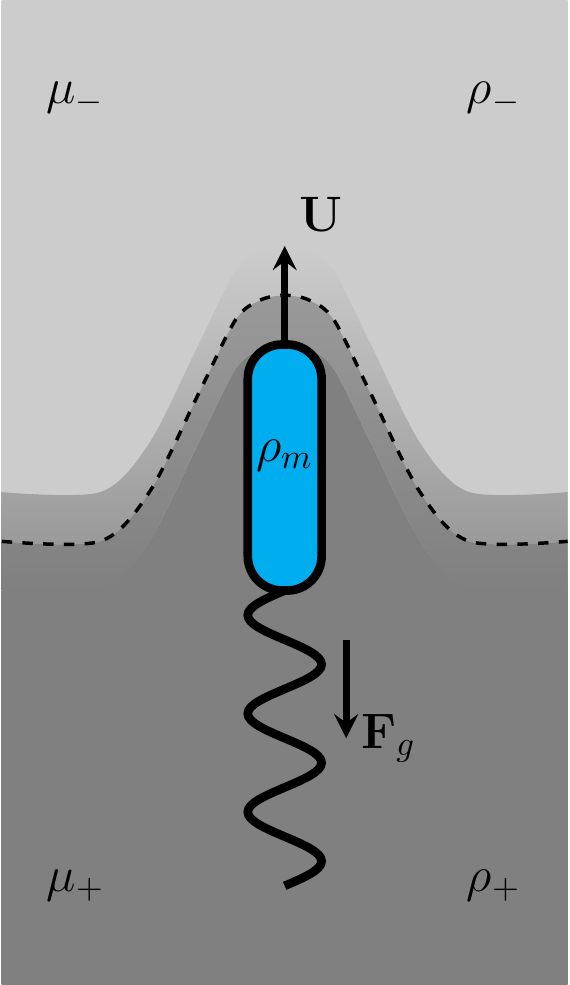}\quad
    \includegraphics[height=6cm, clip=true]{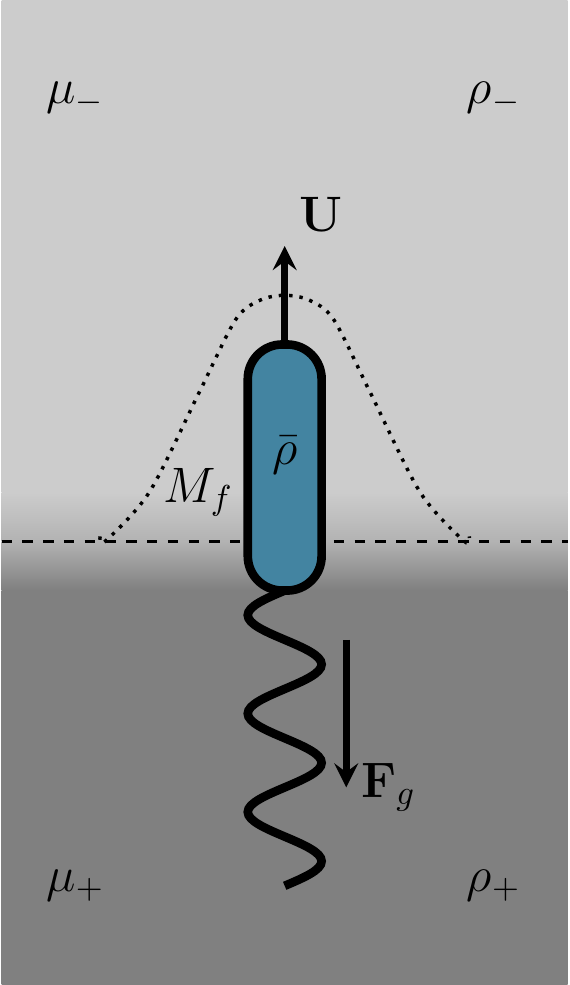}}
\caption{\label{fig:entrainment}Viscous entrainment of the high-viscosity fluid by the swimmer. (a) Experimental picture showing  the swimmer moving down the gradient and  entraining some of the high-viscosity fluid as it crosses the interface, regardless of its orientation relative to the interface. (b) The entrained fluid accounts for an increase in the apparent density of the swimmer from $\rho_{swimmer}=M_{swimmer}/V_{swimmer}$ to $\bar{\rho}=[1+\alpha(h)]\rho_{swimmer}$, where $\alpha$ depends on the  mass of fluid dragged along with the swimmer, $M_f=\rho V_f$, as given in Eq.~\eqref{eq:experiments_01}.}
\end{figure}

Based on experimental observations, we hypothesise that this counter-intuitive behaviour is due to the head of the swimmer entraining a significant amount of  high-viscosity fluid with it as it crosses into the low-viscosity region, thereby increasing its effective density and being slowed down. We show   experimental evidence of this entrainment in   Fig.~\ref{fig:entrainment} (a).

It is difficult to  precisely calculate the amount of fluid that the swimmer entrains. However, we can use our model to show that an increase in the effective swimmer density leads to theoretical results closer to what is observed experimentally. In order to do that, we assume that the swimmer has an average density $\rho_{swimmer}$ which increases by a height-dependent fraction $\alpha(h)$ as \mbox{$[1+\alpha(h)]\rho_{swimmer}$}. The increase is set explicitly by the relation  
  \begin{equation}
      \label{eq:experiments_01}
      \bar{\rho}(h)=[1+\alpha(h)]\rho_{swimmer}=\frac{M_{swimmer}+M_f(h)}{V_{swimmer}+V_f(h)},
  \end{equation}
where $\{M_{swimmer},V_{swimmer}\}$, $\{M_f,V_f\}$ are the masses and volumes of the swimmer and the entrained fluid (see Fig.~\ref{fig:entrainment}(b)). Therefore the maximum increase in density is obtained in the limit $1\ll V_f/V_{swimmer}$ and is given by \mbox{$\alpha_{max}=(\rho_f-\rho_{swimmer})/\rho_{swimmer}$}. Once the density reaches its maximum, the fluid slides-off and the density decreases. To reflect the fact that the amount of fluid that the swimmer drags may depend on the shape of the fluid interface, we set $\alpha$ to be a Gaussian function with variance \mbox{$[(1+\lambda)/2+\delta]^2$} and maximum $\alpha_{max}$ at $h=1+\lambda+\delta$. This means that: (i) the changes in apparent density are negligible before the swimmer meets the interface, (ii) the maximum increase in density occurs when the swimmer has fully crossed the interface, and (iii) most of the dragged fluid slides off after the swimmer has travelled the same distance it did before accumulating the maximum amount of entrained fluid. Although this approach is a  phenomenological way to account for the effect of the drift volume, it shows that an increase in the effective swimmer density plays an important role in the dynamics. We show in Fig.~\ref{fig:case_6_h-f_h-l} the predictions of the modified model with the increase in density as thick lines. This new  model is now able to capture  the qualitative features observed in experiments.
  
  \begin{figure}
      \includegraphics[height=7cm]{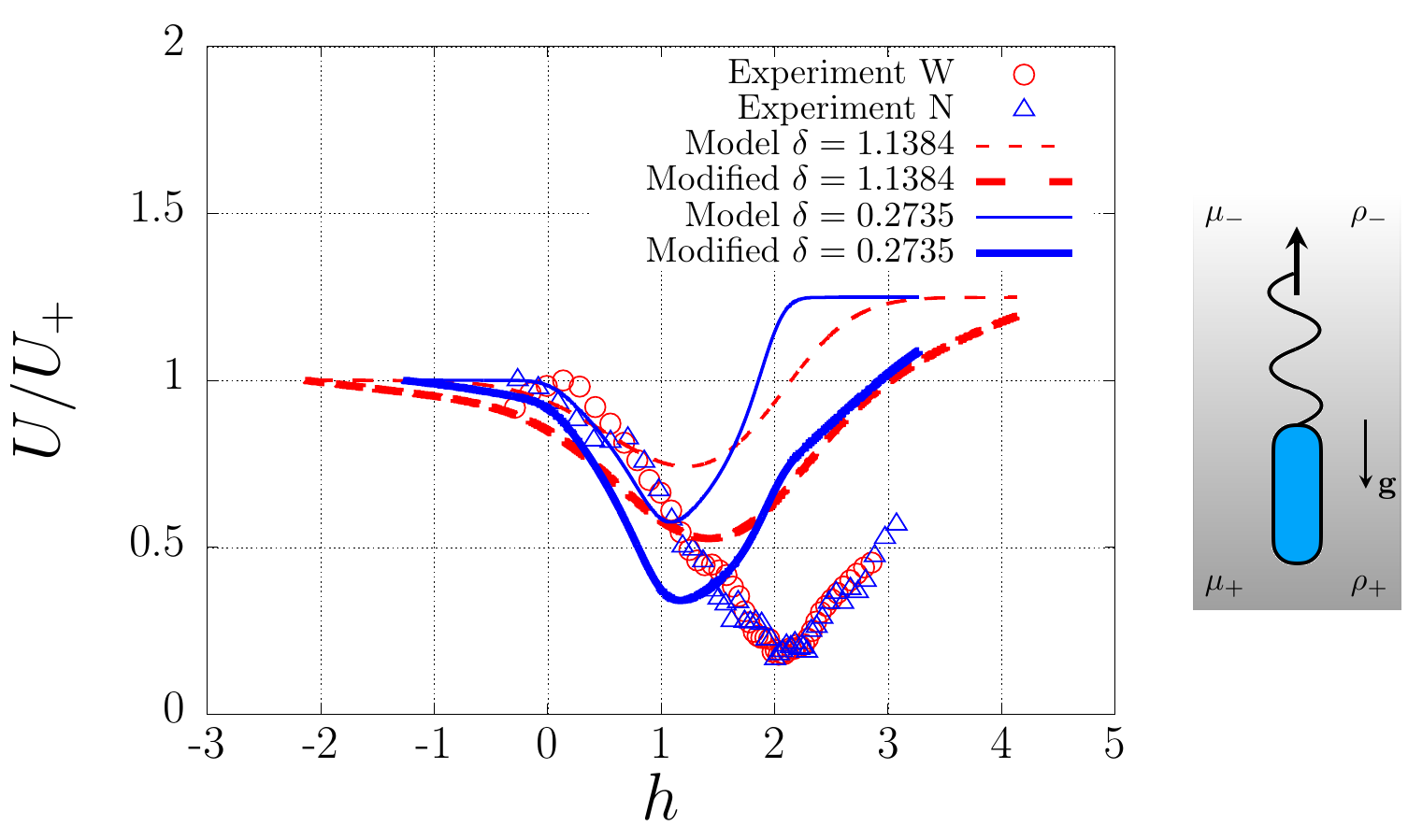}
      \caption{\label{fig:case_6_t-f_h-l}Comparison between experiments and data for  Case IV (tail-first swimmer (puller) crossing the viscosity gradients from high to low viscosity): Speed of the swimmer as a function of the dimensionless position of the tail $h$ with respect to the interface located at $h=0$. Values for the speed are normalised by the speed in the high viscosity fluid $U_{+}=U(h\rightarrow-\infty)$.  We display the measurements early after the viscosity gradient has been set up ([N]arrow gradient): triangles (experiment) and blue solid line (model), and sixteen hours after ([W]ide gradient): circles (experiment) and red dashed line (model). The thin lines show the predictions of  the original model while we plot in thick lines the modified model with variable buoyancy and $\alpha_{max}=0.1$.}
  \end{figure}
  
  We finally address the situation in Case IV with a swimmer approaching the interface  tail-first (puller case). Here we expect the swimming speed to slow down as the swimmer crosses the gradient  as a result of a decrease in propulsion. As soon as the head meets the interface the drag should decrease, compensating for the lower propulsion, until the speed reaches a constant value. However we can see in Fig.~\ref{fig:case_6_t-f_h-l} that, in the experiments, the swimmer does not speed up until it has completely crossed the viscosity gradient ($h=0$), unlike the predictions from the original model (thin lines). An increase in the effective density of the swimmer due to entrainment of the high-viscosity fluid might here also be at the origin of this result. We use the modified model  outlined above  and plot its predictions in Fig.~\ref{fig:case_6_t-f_h-l} as thick lines; we see that the new model is able to come closer to the experimental data. As for Case II, here $h$ is measured from the tip of the tail and we flip the values of the viscosities and densities to be consistent with reversibility, $\mu'=\mu_-$, $\mu=\mu_+$, $\rho'=\rho_-$ and $\rho=\rho_+$.

\section{\label{sec:conclusions} Conclusions}
    
In this paper, we   present a joint experimental-theoretical  study of the dynamics of synthetic magnetic helical swimmers moving across viscosity gradients between two miscible fluids. The viscosity gradients are seen to play a significant role in the swimming dynamics.  

For motion  up the viscosity gradient, there are two possible behaviours: first, for up the gradient motion, if the swimmer moves head-first (pusher mode), its speed reduces due to an increase in drag. On the other hand, the swimmer speeds up when it swims tail-first (puller mode), due to an increase in the viscous propulsion. When the swimmer moves instead from high to low-viscosity regions, the opposite behaviour is expected, i.e.~the swimming speed should increase if the swimmer moves head-first and decrease if it moves tail-first. However, we  observe in our experiments that the swimmer slows down in both cases. We hypothesise that buoyancy forces resulting from entrainment of the high-viscosity fluid are responsible for such counter-intuitive behaviour: as the swimmer traverses the gradient, it drags a large amount of fluid with it, increasing its apparent mass and slowing it down. We show evidence of this mechanism by modifying our model to include a buoyant term that increases as the swimmer advances.

Since we only focus on swimming motion parallel to the viscosity gradient in this paper, our model cannot tackle the issue of viscotaxis for single swimmers. However, our results suggest that, regardless of the viscous entrainment, it is always harder for a pusher-like swimmer to swim up the gradient and that the opposite is true for a puller-like swimmer. Therefore, in addition to the reorientation of  chiral swimmers  in viscosity gradients~\cite{liebchen2018viscotaxis}, 
 our results point to  both positive and negative collective viscotaxis as  being not only possible but   governed solely by the  motility pattern of the cells. 
 
 Specifically, let us consider   microorganisms which perform a run-and-tumble dynamics, as  is the case for  many bacteria~\cite{Taute2015, Taktikos2013}, and therefore  swimmers that repeatedly stop their motion to  change direction. 
For simplicity, we can assume that the swimmer's mode is always the same, i.e.~that it always remains a puller or a pusher. For example the bacterium \textit{E.~coli} remains a pusher during its swimming motion. If the motility pattern of the swimmer has a large positive directional persistence (i.e.~the swimming direction after a reorientation event is close to the  previous direction), then pusher-like swimmers would be predicted to statistically accumulate in regions of high viscosity (collective positive viscotaxis), because individual swimmers would spend more time in regions were they swim slower. The opposite situation would happen for puller-like swimmers (negative collective viscotaxis). In contrast, if the directional persistence is negative (i.e.~reorientation angles larger than 90$^\circ$ on average), then pusher swimmers would exhibit negative  collective viscotaxis while 
pullers would display positive viscotaxis. 

The situation is more complex for bacteria such as \textit{H.~pylori} that can switch between swimming modes~\cite{Antani2020} or \textit{V.~alginolyticus} that exhibits a bi-modal motility pattern with two different persistence parameters~\cite{Taute2015}. In the case of \textit{H.~pylori}, persistence is negative and  the cell switches between pusher and puller modes during its locomotion. Using our results, we predict that a swimmer with this type of motility would accumulate in regions of high viscosity. This, in turn, would be  advantageous for the cell as it would tend to spend longer times in the  high-viscosity mucus layer that protects the stomach, ultimately leading to penetration and colonization of the stomach wall. The reorientation towards, or away from, the gradient might of course modify the collective viscotactic effect, and therefore further investigation  would be  necessary to draw definite conclusions. On the other hand, our results also indicate that a healthy mucus layer will do its function if it remains sharp: if the bacteria remain at the viscosity gradient for too long, they would exhausts their available energy trying to cross the interface. The understanding of this process may be helpful in trying to understand prevention or remediation of gut infection and inflammation~\cite{Hansson2012}.
    
The idealized system considered in our paper was aimed to emulate biological processes, in particular the one by which bacteria are capable of penetrating mucus layers or membranes to cause infections. Even in the simplified situation considered in our paper the process is seen to exhibit rich dynamics. We hope that this first study will motivate further work on swimming in viscosity-stratified fluids.
    
\section*{Acknowledgements}

CEL and JGG contributed equally to this work. We thank K. Dekkers for his help while conducting the experiments. JGG acknowledges the support of DGAPA-UNAM for postdoctoral financial support. This project has received funding from the European Research Council (ERC) under the European Union's Horizon 2020 research and innovation programme  (grant agreement 682754 to EL).


\end{document}